\documentclass[manuscript,screen,nonacm]{acmart}

\pdfoutput=1

\usepackage{times}
\usepackage{helvet}
\usepackage{courier}
\usepackage{graphicx}
\usepackage{subfig}
\usepackage{url}

\usepackage{amsmath,amsfonts} 
\usepackage[ruled,vlined,noend]{algorithm2e}
\usepackage{xcolor}
\usepackage{multicol}
\usepackage{multirow}
\usepackage{listings}
\usepackage[utf8]{inputenc}
\usepackage{bm}
\usepackage{gensymb}
\usepackage{soul}
\usepackage{ragged2e}
\usepackage{enumitem}
\usepackage[para]{footmisc} 
\usepackage{etoolbox}

\newcommand{\etal}{\emph{et~al.}}
\newcommand{\ie}{i.e.,}
\newcommand{\eg}{e.g.,}

\newcommand{\covid}{COVID19\xspace}

\newcommand{\mysubsubsection}[1]{\vspace{4pt} \noindent \textbf{#1}}

\definecolor{darkgreen}{RGB}{0,100,0}
\definecolor{forestgreen}{RGB}{34,139,34}
\definecolor{lightgrey}{RGB}{204, 204, 204} 
\definecolor{darkgrey}{RGB}{51, 51, 51} 
\definecolor{black}{RGB}{0, 0, 0} 
\definecolor{red}{RGB}{204,0,51} 

\AtBeginDocument{%
  \providecommand\BibTeX{{%
    \normalfont B\kern-0.5em{\scshape i\kern-0.25em b}\kern-0.8em\TeX}}}

\providetoggle{arXiv}
\settoggle{arXiv}{true}

\copyrightyear{2021}
\acmYear{2021}
\acmConference[]{}{}
\acmBooktitle{}
\acmPrice{}
\acmDOI{10.1145/xxxx.xxxx}
\acmISBN{978-1-4503-8312-7/21/04}

\author{Yelena Mejova}
\email{yelenamejova@acm.org}
\affiliation{
  \institution{ISI Foundation}
  \city{Torino}
  \country{Italy}
}

\author{Nicolas Kourtellis}
\email{nicolas.kourtellis@telefonica.com}
\affiliation{
  \institution{Telefonica Research}
  \city{Barcelona}
  \country{Spain}
}

\begin{document}

\title{YouTubing at Home: Media Sharing Behavior Change as Proxy for Mobility Around COVID-19 Lockdowns}
\pagestyle{plain} 

\begin{abstract}
Compliance with public health measures, such as restrictions on movement and socialization, is paramount in limiting the spread of diseases such as the severe acute respiratory syndrome coronavirus 2 (also referred to as \covid).
Although large population datasets, such as phone-based mobility data, may provide some glimpse into such compliance, it is often proprietary, and may not be available for all locales.
In this work, we examine the usefulness of video sharing on social media as a proxy of the amount of time Internet users spend at home.
In particular, we focus on the number of people sharing YouTube videos on Twitter before and during \covid lockdown measures were imposed by 109 countries.
We find that the media sharing behavior differs widely between countries, in some having immediate response to the lockdown decrees -- mostly by increasing the sharing volume dramatically -- while in others having a substantial lag.
We confirm that these insights correlate strongly with mobility, as measured using phone data.
Finally, we illustrate that both media sharing and mobility behaviors change more drastically around mandated lockdowns, and less so around more lax recommendations.
We make the media sharing volume data available to the research community for continued monitoring of behavior change around public health measures.
\end{abstract}

\maketitle

\section{Introduction}

The \covid pandemic has brought drastic changes in people's lives around the world.
With respect to online activities and usage of the Internet, people have turned to social media and other networking and messaging platforms for socialization, networking, sharing and consumption of content~\cite{anshika2020twitterexcess-emotions-covidlockdown,bottger2020covid19-internetreaction-imc}.
In particular, during the first \covid lockdown (Q2-Q3 2020), popular social platforms such as Twitter~\cite{iqbal2020twitterstats}, Facebook~\cite{iqbal2020facebookstats}, and YouTube~\cite{iqbal2020youtubestats}, saw an increase in daily and monthly active users~\cite{chaffey2020worldstats-social-media,fischer2020worldstats-social-media2}, with YouTube and Facebook-related content being among the top shared~\cite{bottger2020covid19-internetreaction-imc,kemp2020worldstats-social-media3}.
Such large changes in social media use may be not only due to increased interest, but indeed due to the physical constraints of the restrictive measures taken by the governments worldwide.

Currently, lively research is being done around misinformation that is possibly spreading among the above channels, affecting health-related beliefs and behaviors of their audience~\cite{cinelli2020socialmediausage,yang2020covid19}. 
However, in the unique circumstances of physical lockdowns, media consumption and sharing may be a sign of another health-related behavior: social and physical isolation.
Compliance with the social distancing measures is critical in controlling the spread of \covid~\cite{ecdc2020considerations}, and alternative data sources have been proposed to track it, including mobile phone GPS and traffic congestion data~\cite{kuiper2020intelligent}.
Social media may be another such resource. 
Already, previous studies of social media have found daily and weekly periodicity, indicating shifts in check-in and sharing behavior around different activities~\cite{hasan2013understanding}. 
Small-scale studies are already being published for using social media posts to track individuals who may be associated with the spread of the disease~\cite{bisanzio2020use,zeng2021spatial}. 
In this work, we propose to use a large language- and space-agnostic collection of YouTube video sharing on Twitter as another tool for measuring population behavior around the social distancing measures. 

Beyond monitoring the activity of users on separate platforms, studying cross-platform posting offers unique insights into users' behavior, as it indicates higher engagement on multiple platforms.
Given that such behavior is more time-consuming, and requires (higher) users' attention to view the video on YouTube and share it on Twitter, it could reveal tendencies of users with respect to daily schedules such as working vs. resting hours, etc.
Further, it is an act of social or ``communal'' video co-consumption, which can provide insights into community or group-focused video sharing and diffusion.
Finally, technically Twitter stream provides an easier way to monitor engagement of users with YouTube videos, and to model those users through their posting history, than building crawling engines for both YouTube and Twitter and handling API-related constraints for both platforms.

Concretely, in this paper, we use a dataset of 390 million Twitter posts mentioning YouTube videos spanning the period from July 2019 to September 2020. 
After geo-locating the users of these posts, we select 109 countries which have had a \covid-related social distancing recommendation or mandate. 
We find that, indeed, the posting trend shifts dramatically around government measures, although the precise timing may differ, depending on the peculiarities of each government's handling of the pandemic. 
We also find that these changes in media sharing have a strong negative relationship with the phone-based mobility data, suggesting social media may provide valuable insights, especially where such mobility data is not available.

In particular, we make the following contributions:
\begin{itemize}
    \item We are the first to study cross-platform video posting behavior and how \covid-related lockdowns may have impacted it.
    \item We study this behavior worldwide for 109 countries, and provide case studies of countries with different governmental actions around the pandemic.
    \item We investigate if the country-specific changes in media sharing behavior can be used as a proxy for the detection of reduced mobility of its residents.
    \item Finally, we provide to the research community a unique dataset with daily aggregated per-country media sharing statistics in order to encourage reproducibility and future research of health-related behaviors\footnote{\url{https://github.com/ymejova/yt-tw-covid}}.
\end{itemize}

\section{Background \& Related Work}

\mysubsubsection{YouTube on Twitter.}
Twitter, as a social media platform, is frequently used to disseminate various types of content produced and published in other platforms, such as news, photos, videos, etc.
Such cross-platform data have been used to better understand user engagement with YouTube content, predict the extent of its diffusion, as well as to profile its viewers.
For example, Christodoulou \etal~\cite{christodoulou2012youtubediffusiontwitter} studied Twitter users and how their properties impact the popularity and diffusion of YouTube videos.
Several papers have studied the problem of finding the best Twitter users to promote YouTube videos to increase engagement and viewership of said videos~\cite{deng2014cross-net-recommendation, yan2014cross-net, yan2015cross-net2}.
Furthermore, other papers investigated the phases that a video goes through with respect to popularity (proxied by viewership), by building models to predict popularity of videos using power-law fitting~\cite{yu2015lifecyle}.
In addition, characteristics of YouTube videos shared on Twitter~\cite{abisheva2014watches}, as well as complex graph features~\cite{yu2014twitter} were employed to model video popularity.
In this work, we consider a large dataset of YouTube links diffused via Twitter posts across over a hundred countries around the world, crucially capturing the sharing behavior before and during the rise of \covid in the early 2020. 

\mysubsubsection{Impact of \covid on usage of online platforms.}
\covid has had a great impact on many aspects of our offline and online lives.
In relation to the focus of this paper, \covid has affected the intensity and frequency with which we use online platforms, including social media and digital devices.
Recent studies have measured these changes by focusing on different aspects of our online world.
For example, Lutu \etal~\cite{lutu2020covid19-traffic-imc} performed a characterization of the impact that the pandemic had on the traffic processed by a mobile network operator, finding that mobile traffic was reduced due to users' reduced mobility during lockdown.
Also, Bottger \etal~\cite{bottger2020covid19-internetreaction-imc} studied how the Facebook Edge Network reacted to \covid and the traffic induced by users, which was found to be higher during and after lockdown was imposed in different countries; in many cases this increase was due to higher video consumption.
Finally, Feldmann \etal~\cite{feldmann2020covid19-lockdowneffect} studied how being under lockdown affected various aspects of internet traffic around the world.
On the other hand, very recent studies looked into \covid-related information spreading through various social media platforms.
For example, Cinelli \etal~\cite{cinelli2020socialmediausage} studied how (mis)information related to \covid spreads on Twitter, Instagram, YouTube, Reddit and Gab, as well as differences in the diffusion of information from questionable and reliable sources on each platform.
Li \etal~\cite{li2020youtube-videos-covid19} studied popular YouTube videos related to \covid and identified that over 25\% of such videos contained misinformation regarding \covid.
In this work, not only do we find that the focus on \covid-related videos increases around the onset of lockdown measures, but also that sharing of other content also markedly increases, pointing a behavior shift beyond attention to the ongoing events. 
Thus, we consider the media sharing as a proxy of offline behavior, and specifically we compare this data to mobility.

\mysubsubsection{Social media and mobility.}
In the previous decade, the proliferation of mobile technology has provided researchers with unprecedented fine-grained data on human mobility, often coming from phone records or large internet platforms, which has been combined with more traditional sources such as air traffic and public transit data.
Geo-tagged social media posts then have been used as a way to enrich the existing mobility information, for example by annotating movement by content associated with it \cite{wu2015semantic} or explaining traffic events in a city \cite{wu2016interpreting}. 
Quercia \etal~\cite{quercia2014shortest} use Flickr photos as proxies of beauty to measure ``happiness'' of trajectories through a city.
However, going beyond annotation, attempts have been made to derive mobility data directly from the social media activity, such as international and national travel flows networks \cite{barchiesi2015quantifying,beiro2016predicting} or local commuting patterns \cite{mcneill2017estimating}. 
This information then can be used to track disease \cite{kraemer2018inferences}, migration patterns of refugees \cite{hubl2017analyzing} or as a testbed for building mobility prediction models \cite{feng2018deepmove}.
Recently, attempts have been made to use Twitter data for tracking the spread of \covid. 
For instance, Bisanzio \etal~\cite{bisanzio2020use} correlated the posting behavior of 161 users with \covid cases in locales they visited, while Zeng \etal~\cite{zeng2021spatial} used geo-tagged tweets to estimate mobility of the U.S. state of South Carolina.
However, the size of the dataset presented in this paper allows for a much larger view of the interaction between \covid-related measures and media consumption.
In the pages below, we both quantify the relationship between media sharing and various stages of lockdowns, and discuss several case studies illustrating the complex nature of such data.
Also, we provide additional visualizations and aggregated data to the research community.

\begin{table}[t]
    \begin{minipage}{0.58\linewidth}
    \centering
    \small
    \begin{tabular}{lr}
    \toprule
    \textbf{Twitter/YouTube collection} & \\
    \midrule
    Duration                &   459 days (06/19/2019-09/20/2020) \\
    Posts (tweets)          &   390,002,678                      \\ 
    Unique Locations        &   8,436,027                        \\
    Countries               &   249                              \\
    Languages               &   66                               \\ 
    \midrule
    \multicolumn{2}{l}{\textbf{\covid Policy data (Oxford's Coronavirus Response)}} \\
    \midrule
    Duration                &   276 days (01/01/2020-10/01/2020) \\
    Countries               &   185                              \\
    Lockdown states         &   3                                \\ 
    \midrule
    \multicolumn{2}{l}{\textbf{Mobility data (Apple Mobility Trends)}} \\ 
    \midrule
    Duration                &   231 days (01/14/2020-09/01/2020) \\ 
    Countries               &   63                               \\
    Mobility states         &   driving, walking, transit        \\ 
    \bottomrule
    \end{tabular}
    \caption{Datasets used in our analysis, before filtering took place.}
    \label{tab:datasets}
    \end{minipage}
\hfill
    \begin{minipage}{0.38\linewidth}
    \centering
    \small
    \begin{tabular}{llr}
    \toprule
    \textbf{Rank} & \textbf{Country} & \textbf{\% of tweets} \\ 
    \midrule
    1       &   USA           &   25.24           \\
    2       &   Japan         &   14.47           \\
    3       &   Great Britain &   5.19            \\
    4       &   Brazil        &   5.11            \\
    5       &   France        &   4.04            \\
    6       &   India         &   3.53            \\
    7       &   Spain         &   2.76            \\
    8       &   Italy         &   2.60            \\
    9       &   Canada        &   2.16            \\
    10      &   Mexico        &   1.91            \\ 
    \bottomrule
    \end{tabular}
    \caption{Top 10 countries by number of posts.}
    \label{tab:top10countries-by-volume}
    \end{minipage}
\end{table}

\section{Data}
\label{sec:data}

In this section, we describe the datasets used in the present study, along with preprocessing and selection steps.
First, we detail our efforts to build the Twitter \& YouTube dataset (Sec.~\ref{sec:data-twyt}).
Then, we briefly describe the extraction of lockdown dates from the Oxford \covid regulations data (Sec.~\ref{sec:data-covid19}).
Finally, we conclude with a description of the Apple mobility dataset (Sec.~\ref{sec:data-mobility}).
The various datasets used are summarized in Table~\ref{tab:datasets}.

\subsection{Twitter \& YouTube Data}
\label{sec:data-twyt}

\textbf{Collection.}
This dataset was collected using the Twitter Streaming API with queries ``youtube'' and ``youtu.be'' (a popular shortened URL for the platform).
Besides the query, no other constraint was applied.
The collection took place from June 19, 2019 to September 20, 2020, capturing at least a half of year before \covid is discovered, and several months of worldwide lockdowns in 2020.
The captured information includes tweet text, user information, metadata about posting time, and link to the YouTube video.
This collection resulted in 390,002,678 tweets.

\noindent\textbf{Preprocessing.}
We then geo-locate the users in this dataset using the Location field of their profiles. 
We use the GeoNames location database to match the user-specified free-text location strings to a populated location.\footnote{The optimized code for geolocation can be found at \url{https://sites.google.com/site/yelenamejova/resources}}
We are able to locate 4,055,292 out of 8,436,027 unique location strings present in the data, which covers 165,520,903 tweets (42.4\% of all).

\begin{figure}
    \includegraphics[width=0.8\linewidth]{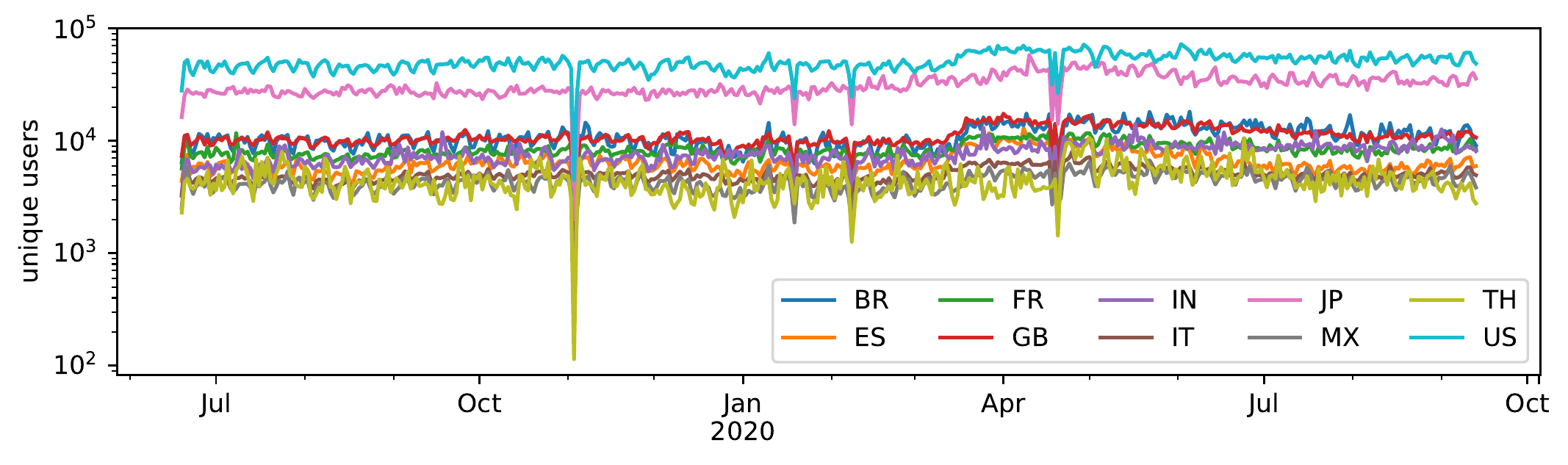}
    \caption{Daily number of unique users for the top 10 countries across time (June 2019 - October 2020.}
\label{fig:top10countries_stats}
\end{figure}

Figure~\ref{fig:top10countries_stats} shows daily statistics on unique users of the top 10 countries, sorted by volume of posts (also Table~\ref{tab:top10countries-by-volume}).
We can see a handful of points in time when, due to technical difficulties, the collection was disrupted.
Also, we can already observe that 1) there is a weekly periodicity of users' behavior in sharing YouTube videos on Twitter, 2) there was a shift in posting frequency between March and May 2020, and 3) USA and Japan have significantly higher posting activity across time compared to the other countries.

Next, we consider three metrics for data analysis: raw number of posts per day, unique number of users, and unique videos (figures are omitted due to space).
As expected, all three metrics have skewed distributions across countries.
For example, the median for daily number of posts, users and videos is 82, 51 and 71, respectively, but the mean is 1491, 892 and 948, respectively.
Therefore, we apply an activity threshold and keep countries with activity (in unique users) above the median of all countries, and across all available dates.
After applying this filter, from 249 countries, we are left with 125 countries.
Furthermore, the three available metrics we have at hand (\ie~number of posts, unique users, and unique videos) are highly correlated with each other: average Spearman correlation between users and videos across all countries is 0.74, and between users and posts is 0.85.
Given that the number of posts or videos involved in this sharing activity can be skewed by abnormal or extreme posting by any particular user, for the rest of the study, we focus on unique users posting per day.
Finally, we linearly interpolate the values for the days in which the collection went down.
We dub the resulting time series as the $T$ dataset.

\subsection{\covid Response Data}
\label{sec:data-covid19}

\textbf{Collection.}
\label{data-collection}
We use the University of Oxford's Coronavirus Government response tracker~\cite{oxford2020covid,oxford2020covidgithub} to gather the indicator variables for the levels of lockdown in the above countries.
It covers the time period from January 1, 2020 to October 2, 2020.
We use the field ``C6: Stay at home requirements'', which ``Records orders to ``shelter-in-place'' and otherwise confine to the home''.
This field has an ordinal scale of several levels: L0: no measures, L1: recommend not leaving house, L2: require not leaving house with exceptions for daily exercise, grocery shopping, and 'essential' trips, and L3: require not leaving house with minimal exceptions. 
Using the above scale, we define \emph{\textbf{lockdown}} ($L$) as the period of time that a country was \emph{requiring} its residents to stay at home unless (minimal) exceptions apply (\ie~L2 or L3).
We also consider the days under L1 as recommended lockdown, and we use this distinction when appropriate.

\noindent
\textbf{Preprocessing.}
The data regarding the lockdown and other related information on \covid were reported for 185 countries, covering all 125 countries in our data.
For each country, we find the starting date $D_{L2}$ when it entered in \emph{lockdown} mode, \ie~L2 or L3 (from here on, when referring to L2 we include both states), and the ending date $D_{L2end}$, \ie~when the country entered L0 or L1.
In case more than one starting and ending dates are detected, 
we apply a buffer time $L2_B = 7$ to merge consecutive lockdown periods (we experimented with $L2_B$ up to 28, but results changed minimally). 
The median country went through $\sim$52 days of lockdown, which lasted for up to 107 days for 75\% of the countries.
About 10\% of countries have remained in a lockdown mode for the whole duration of observation ($\sim$200 days), while 41 countries (\ie~33\%) had 0 days in lockdown. 
The latter countries either did not define any stay at home guidelines (L0, 9 countries) or did not provide any \covid data (5 countries), and where subsequently removed.
Finally, we removed 2 countries which had enforced fewer than 30 days of lockdown. 
These steps resulted in two lists of countries: 82 that had a (L2) lockdown, and a more inclusive one of 109 countries that had at least a lockdown recommendation (L1).

\subsection{Mobility Data}
\label{sec:data-mobility}

\textbf{Collection \& Preprocessing.}
Finally, in order to explore the relationship between media sharing ($T$) and mobility ($M$), we utilize the Mobility Trends Report published by Apple, which ``reflects requests for directions in Apple Maps''\footnote{\url{https://covid19.apple.com/mobility}}.
The report offers normalized mobility activity in three categories: driving, walking, and transit.
For our analysis, we use the walking estimates as the finest of granularity levels.
In the end, out of the 109 countries which have passed the activity and \covid lockdown thresholds, 56 had mobility data (signal $M$) provided by the Mobility Trends Report.

\section{Media Sharing About \covid}

\begin{figure}[t]
    \centering
    \subfloat[Daily users mentioning \covid]{\includegraphics[width=0.49\linewidth]{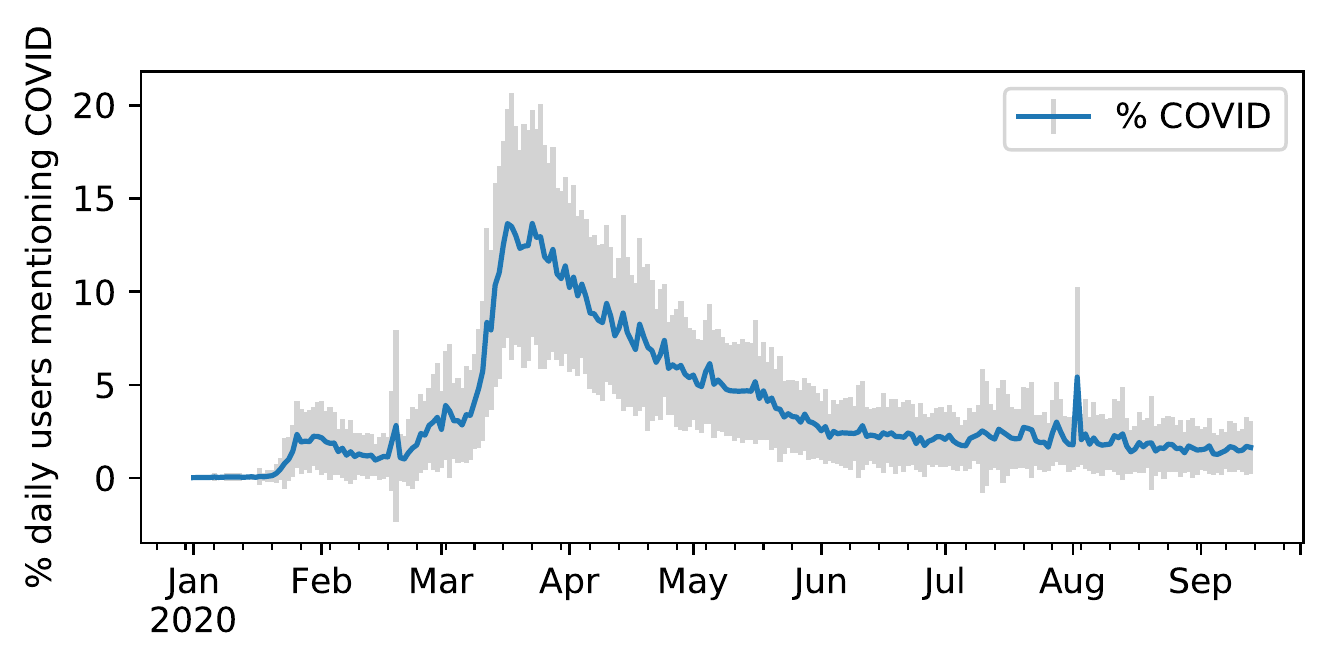}}
    \hspace{0.2cm}
    \subfloat[Normalized volume]{\includegraphics[width=0.49\linewidth]{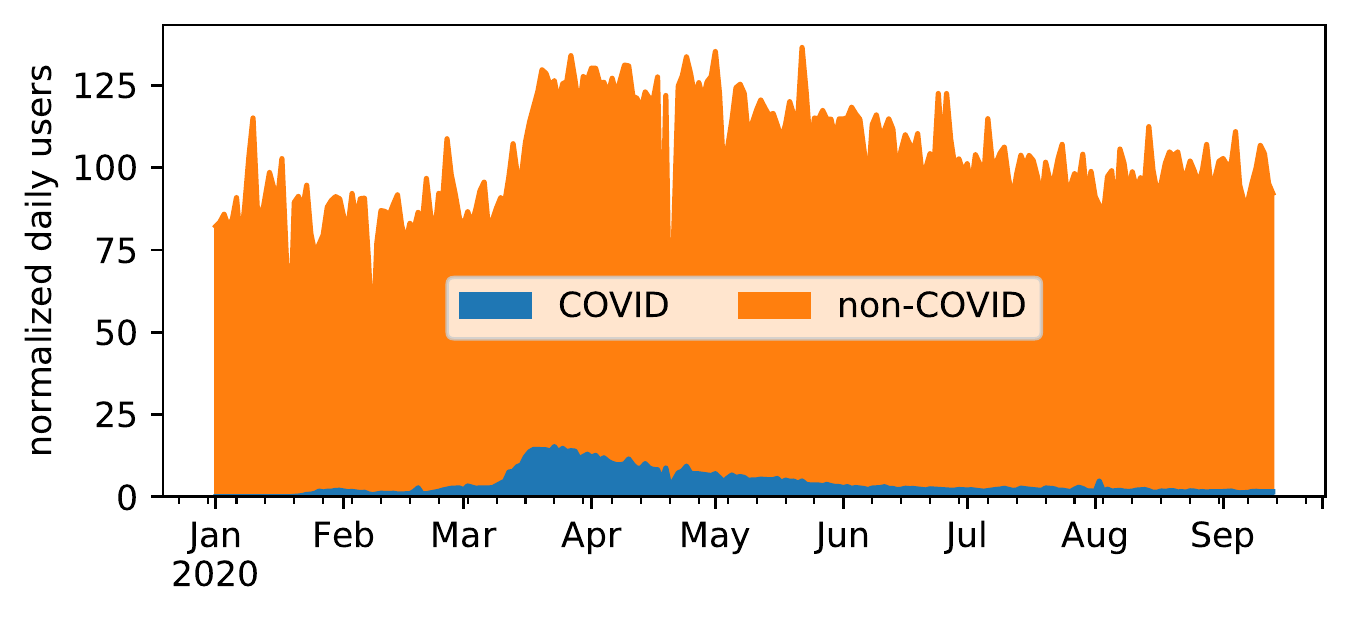}}
    \caption{(a) Portion of daily unique users who are mentioning \covid-related keywords, macro-averaged over countries. Gray bars indicate standard deviation per day.
    (b) Normalized volume of tweets (by volume on first day), which do/do not mention \covid-related keywords, macro-averaged over countries.}
    \label{fig:percCovid}
\end{figure}

We begin by examining the extent to which the increase in shared media ($T$) can be directly attributed to conversations about the pandemic, or if other topics were also at play.
We use a dictionary of 590 keywords defined by Twitter for its multilingual \covid stream\footnote{\url{https://developer.twitter.com/en/docs/labs/covid19-stream/filtering-rules}}, since these keywords indicate interest in the topic of the pandemic.
Figure~\ref{fig:percCovid}(a) shows the country macro-average of the daily unique users in our dataset who are mentioning \covid-related keywords.
Such spike on Twitter and Facebook-related activity has also been reported in recent works~\cite{yang2020covid19}.
On average, the maximum contribution of \covid-related posting is $\sim$15\%, and appears to happen in late March 2020, when most lockdowns started across many countries.
The spike tends to be higher for countries having fewer users, with Spearman $\rho = -0.362$ ($p < 0.001$) between average volume of tweets vs. maximum contribution of \covid-related posting.

Interestingly, Figure~\ref{fig:percCovid}(b) shows that the volume increase around the second half of March is not only due to the posting around \covid.
In this figure, we normalize the volume of posting in each country by dividing it by the volume on the first day and multiplying by 100, such that the volume increase is standardized.
We find that, despite a rise in \covid-related volume (in blue), there is additional volume (in orange) not attributable to the interest in pandemic.

To check whether topics other than \covid are mentioned in these posts, we take the U.S. as an example (which also has the highest volume per country), and compute the top terms mentioned in the two months before and two months during the lockdown.
The top terms, shown in Table~\ref{tab:my_label}, show that, even though \covid-related keywords emerge in the topics during the lockdown (\eg~\emph{coronavirus} and \emph{covid}), others remain, including \emph{music}, mentions of President \emph{Trump}, and other general terms like \emph{love}, \emph{check}, and \emph{day}.
Thus, we postulate that the increased media sharing behavior is not wholly explained by the increased interest in the pandemic, but includes the habitual use of the platform.

\begin{table}[t]
    \centering
    \small
    \begin{tabular}{l|r|r||l|r|r}
    \toprule
    \textbf{Rank} &
    \textbf{Popular Word Before $L$} &
    \textbf{Popular Word During $L$} &
    \textbf{Rank} &
    \textbf{Popular Word Before $L$} &
    \textbf{Popular Word During $L$} \\ \midrule
    1   &   via (28.13\%)   & via (27.03\%)     &   11  &   like (1.18\%)   & \textbf{covid (1.56\%)} \\
    2   &   youtube (5.25\%)& youtube (5.13\%)  &   12  &   love (1.14\%)   & got (1.51\%)\\
    3   &   video (5.10\%)  & video (4.85\%)    &    13  &   one (1.04\%)    & trump (1.42\%)\\
    4   &   amp (2.51\&)    & amp (2.37\%)      &    14  &   full (1.00\%)   & like (1.05\%)   \\
    5   &   new (2.35\%)    & live (2.20\%)     &    15  &   de (0.95\%)     & one (1.01\%)   \\
    6   &   official (2.31\%)& new (2.02\%)     &    16  &   \textbf{coronavirus (0.95\%)} &love (1.80\%) \\
    7   &   watch (2.11\%)  & watch (2.12\%)    &    17  &   song (0.94\%)   & de (0.98\%)   \\
    8   &   live (2.09\%)   & official (2.29\%) &    18  &   check (0.91\%)  & song (0.98\%)   \\
    9   &   trump (1.83\%)  & \textbf{coronavirus (1.80\%)}&    19  &   time (0.82\%)   & check (0.93\%)   \\
    10  &   music (1.71\%)  & music (1.56\%)    &    20  &   day (0.81\%)    & day (0.92\%)   \\ \bottomrule
    \end{tabular}
    \caption{Popular words in tweets with YouTube links.
    Popularity shown in \% out of top 100 words.
    \vspace{-0.6cm}}
    \label{tab:my_label}
\end{table}

\section{Media Sharing Around \covid Lockdowns}

\begin{figure}[t]
    \centering
    \subfloat[United States]{\includegraphics[width=0.258\linewidth]{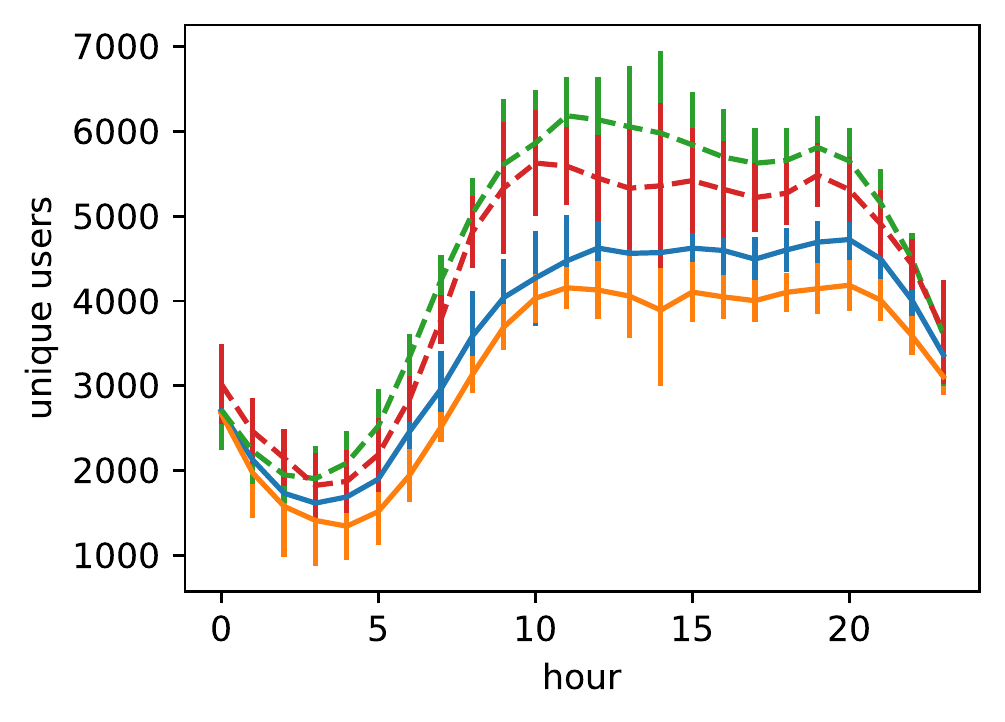}}
    \subfloat[Spain]{\includegraphics[width=0.24\linewidth]{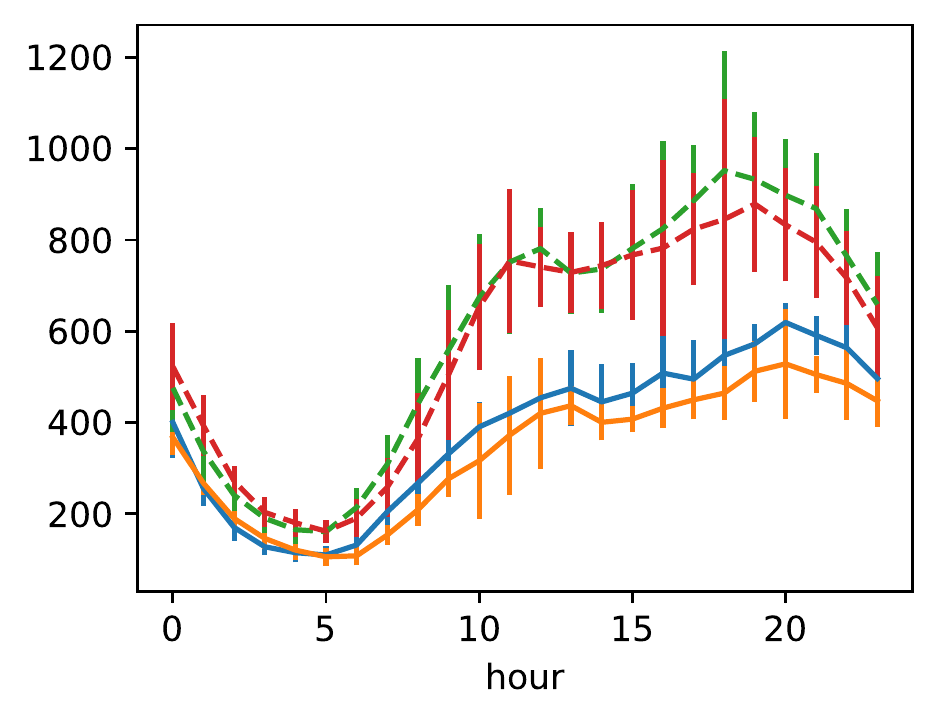}}    
    \subfloat[Brazil]{\includegraphics[width=0.24\linewidth]{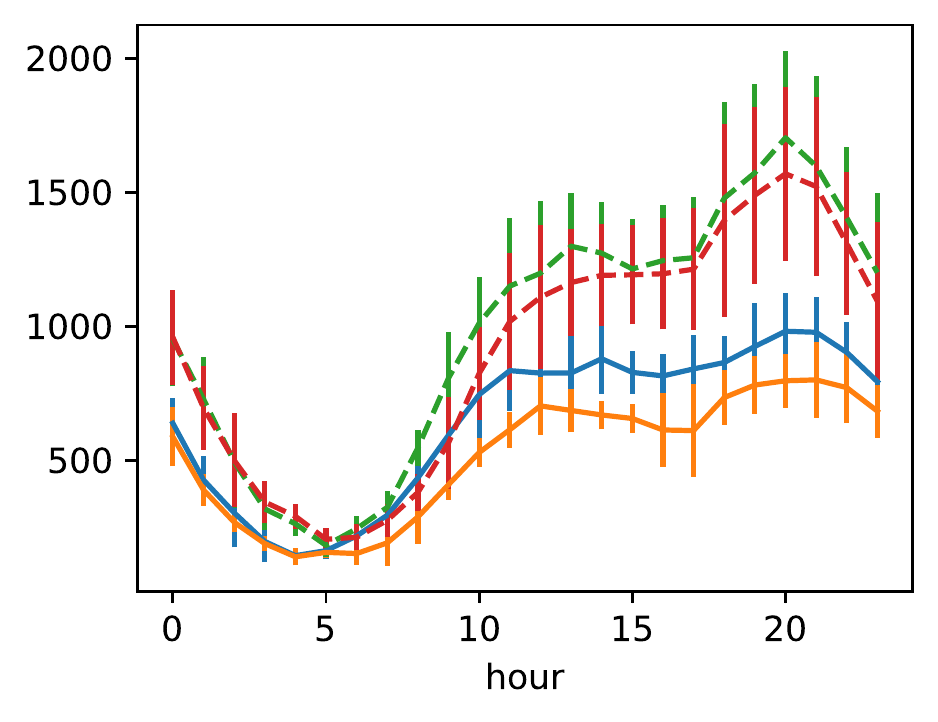}}
    \subfloat[Japan]{\includegraphics[width=0.24\linewidth]{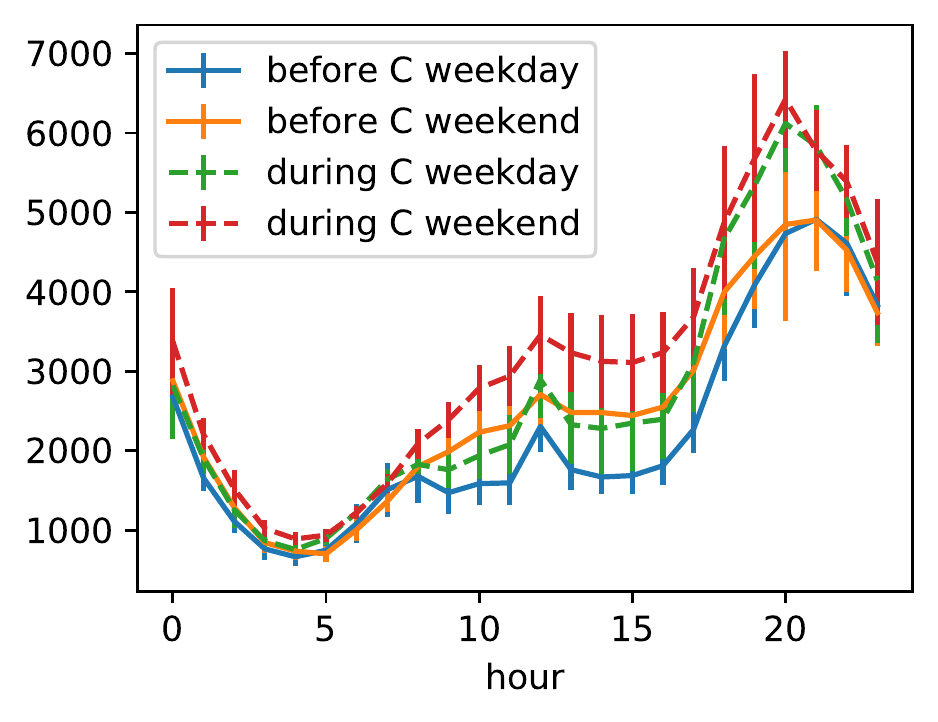}}
    \caption{Hourly volume (by unique posting users) of signal $T$ in four countries, 2 months before and 2 months after lockdown announcements ($D_{L2}$) in each country. Bars indicate standard deviations.}
    \label{fig:hourly}
\end{figure}

\subsection{Hourly Sharing Behavior Change Before and During Lockdown}

Focusing specifically on the posting behavior before and after lockdown was announced ($D_{L2}$) by the respective governments of the studied countries, we first examine the hourly periodicity of the posting behavior.
Figure~\ref{fig:hourly} shows examples of unique users posting during the hours of the day in four countries, separately for weekdays and weekends, and before and during \covid lockdowns.
Here, we consider 2 months before and 2 months after the lockdown announcements, average the posting volume over the days, and show standard deviation in vertical bars.
Not only do we see a marked rise in posting volume during L2 lockdowns, but also a shift in posting patterns, though differently for each country.
For instance, in U.S., the posting rate in the morning increases, compared to pre-\covid levels, whereas in Brazil it elevates especially around evening.
Also, the difference in posting between weekdays and weekends becomes less clear during lockdown (\eg~in Spain or Brazil), suggesting a similar media consumption on weekdays and weekends.

\subsection{Sharing Behavior Change at the Lockdown Start}

Next, we examine the change in the media sharing behavior ($T$) around the lockdown dates.
As described in Sec.~\ref{data-collection}, we define the lockdown dates at several levels.
Specifically, we consider two scenarios: (1) an L1 lockdown (with its starting date signified as $D_{L1}$), when only a recommendation to stay at home is promoted, and (2) an L2 lockdown (with its starting date signified as $D_{L2}$), when only essential trips are allowed.
To compare the behavior change to these days, we employ change point detection on signal $T$, using the pruned exact linear time (PELT) algorithm~\cite{wambui2015power} that detects change points through minimising a cost function over possible numbers and locations, as implemented in the python \texttt{ruptures} library~\cite{truong2020selective}.
We apply this algorithm to $T$ data from each country separately, taking the 2 months before their lockdown date ($D_{L1}$ or $D_{L2}$), and 2 months after the date, or until the studied lockdown is over (whichever comes first).
Dates of detected trend change are signified as $D_{TC(T)}$.
Out of the 109 countries having L1 and/or L2, we detected a change point for 100 countries, and for the 82 countries having L2 only, we detected change point for 77 countries.

\begin{figure}[t]
    \centering
    \subfloat[]{
    \includegraphics[width=0.33\linewidth]{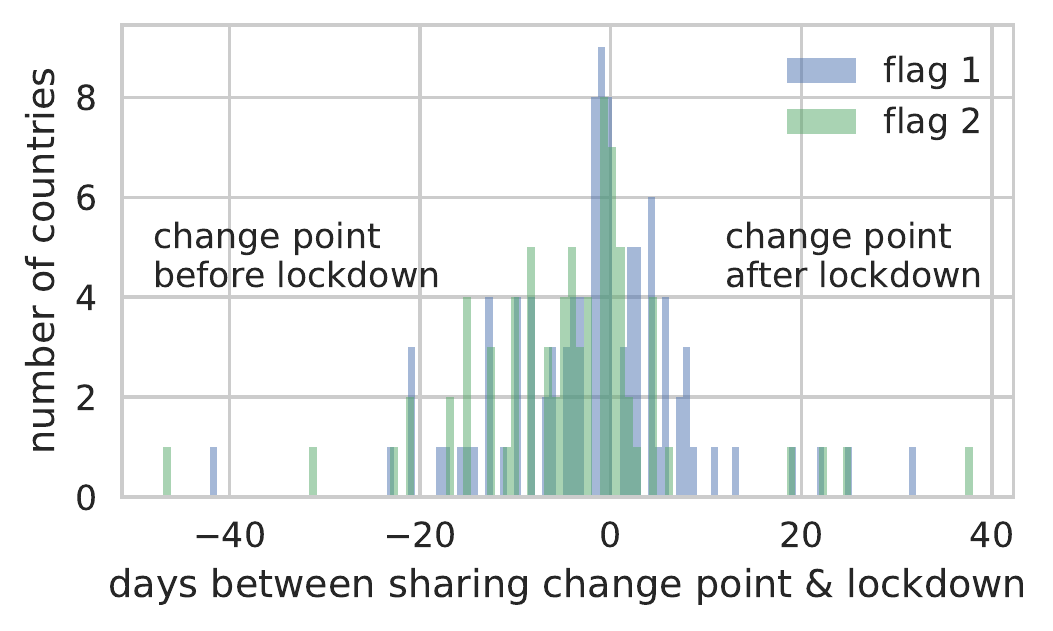}}
    \subfloat[]{
    \includegraphics[width=0.33\linewidth]{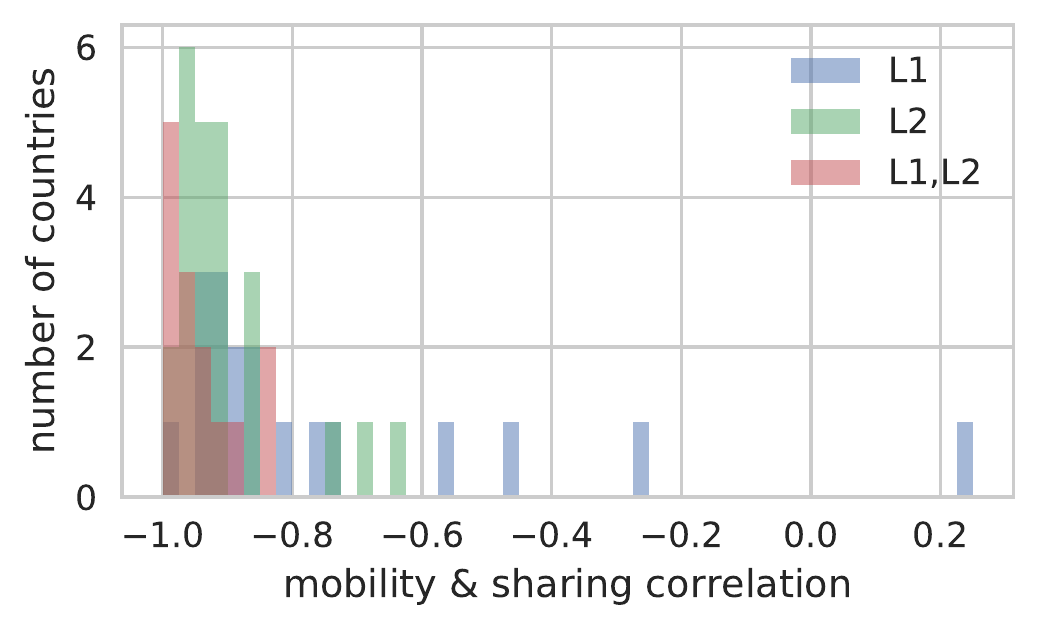}}
    \subfloat[]{
    \includegraphics[width=0.34\linewidth]{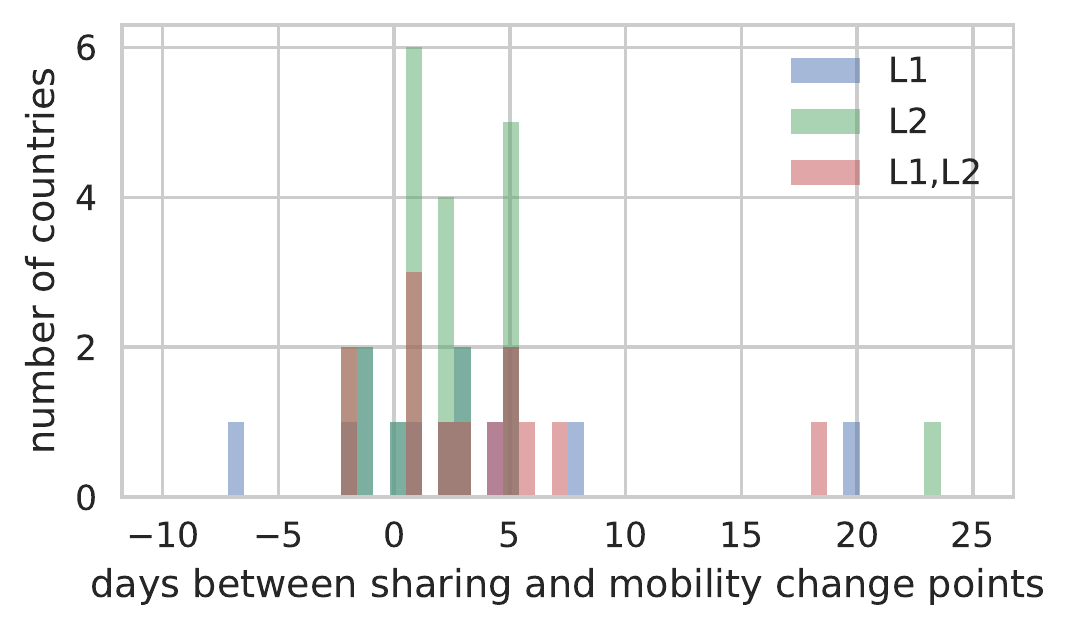}}
    \caption{Number of countries (y-axis) having (a) a difference (x-axis) between the change point in media posting behavior ($T$) and the starting date of L1 or L2 lockdowns (countries with both $D_{L1}$ and $D_{L2}$ are considered separately for each group).
    (b) Pearson correlation $r$ (x-axis) between a 7-day averaged mobility ($M$) and media sharing ($T$), by country lockdown situation (L1 vs. L2 vs. L1,L2).
    (c) Difference in days (x-axis) between change points of mobility ($D_{TC(M)}$) and media sharing ($D_{TC(T)}$), by country lockdown situation (L1 vs. L2 vs. L1,L2).}
    \label{fig:histChangeLockdown}
\end{figure}

Figure~\ref{fig:histChangeLockdown}(a) shows the frequency distribution of countries having a difference between the detected change point and the lockdown dates (\ie~$D_{TC(T)}$~-~$D_{L1}$ or $D_{TC(T)}$~-~$D_{L2}$), such that points below 0 indicate countries in which change point happened before the lockdown, and above 0 after lockdown.
For both definitions of lockdown (L1 or L2), about half of the time the change points happened within 5 days of the lockdown (51\% for L1 and 52\% for L2).
And, as expected, for more strict lockdown definition (L2), more change points happened before 0 (64\%) than for less strict (51\%).

\begin{figure}
\centering
    \subfloat[United States]{\includegraphics[width=0.49\linewidth]{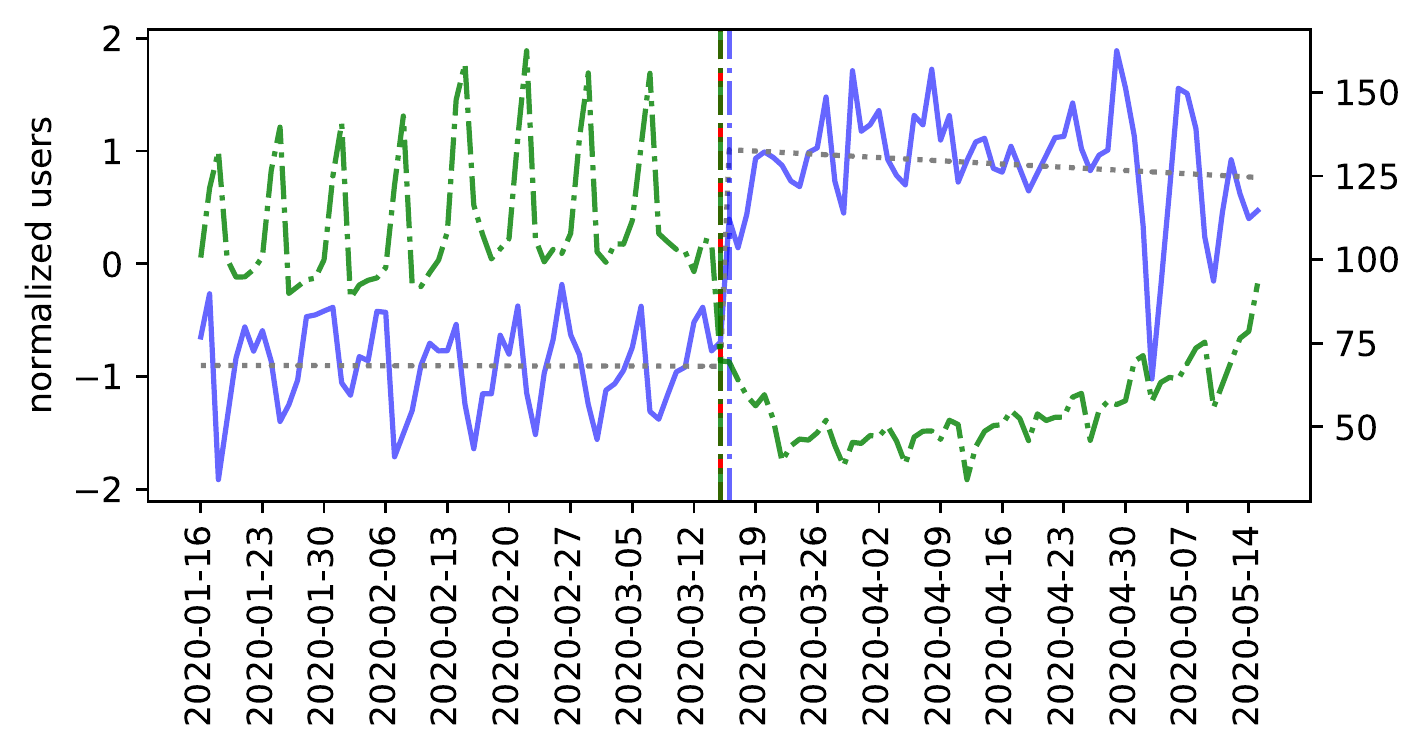}} 
    \vspace{0.1cm}
    \subfloat[Italy]{\includegraphics[width=0.49\linewidth]{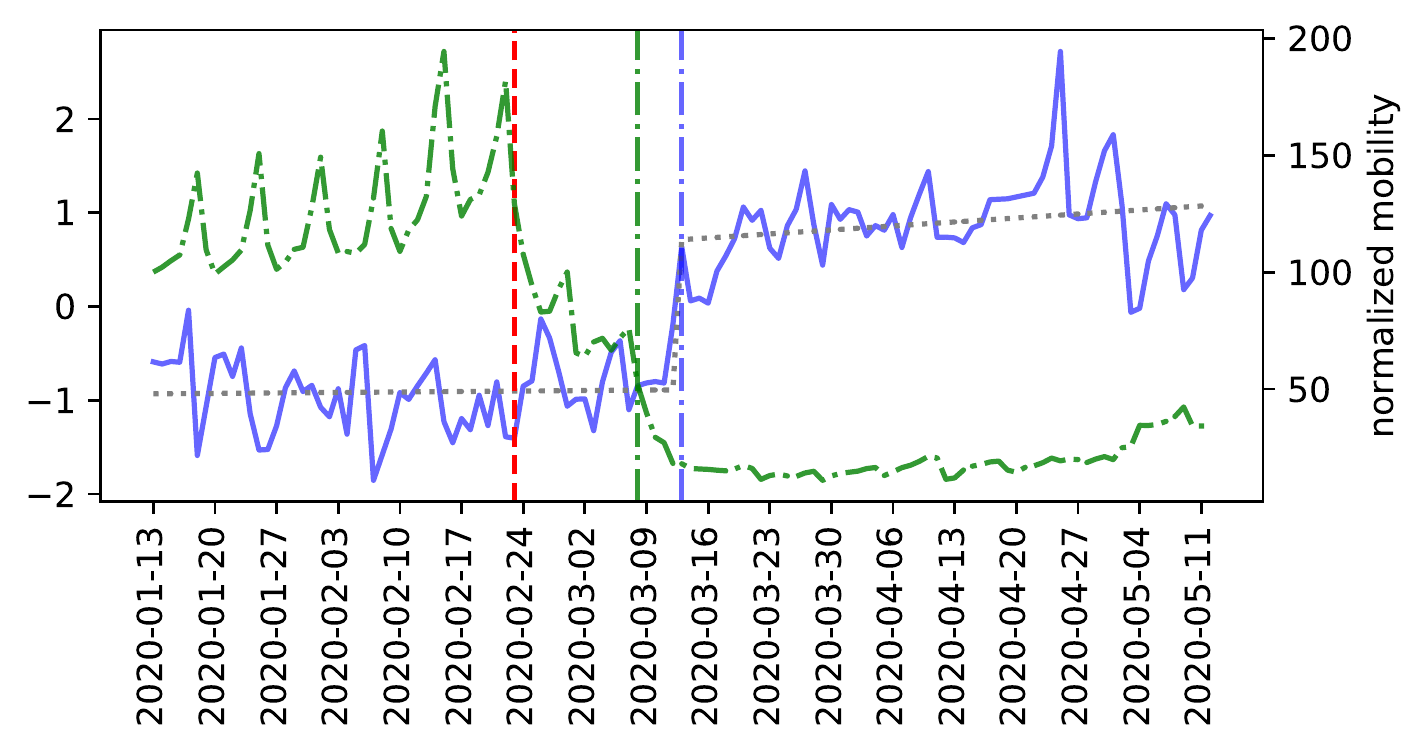}} \\
    
    \subfloat[Japan]{\includegraphics[width=0.49\linewidth]{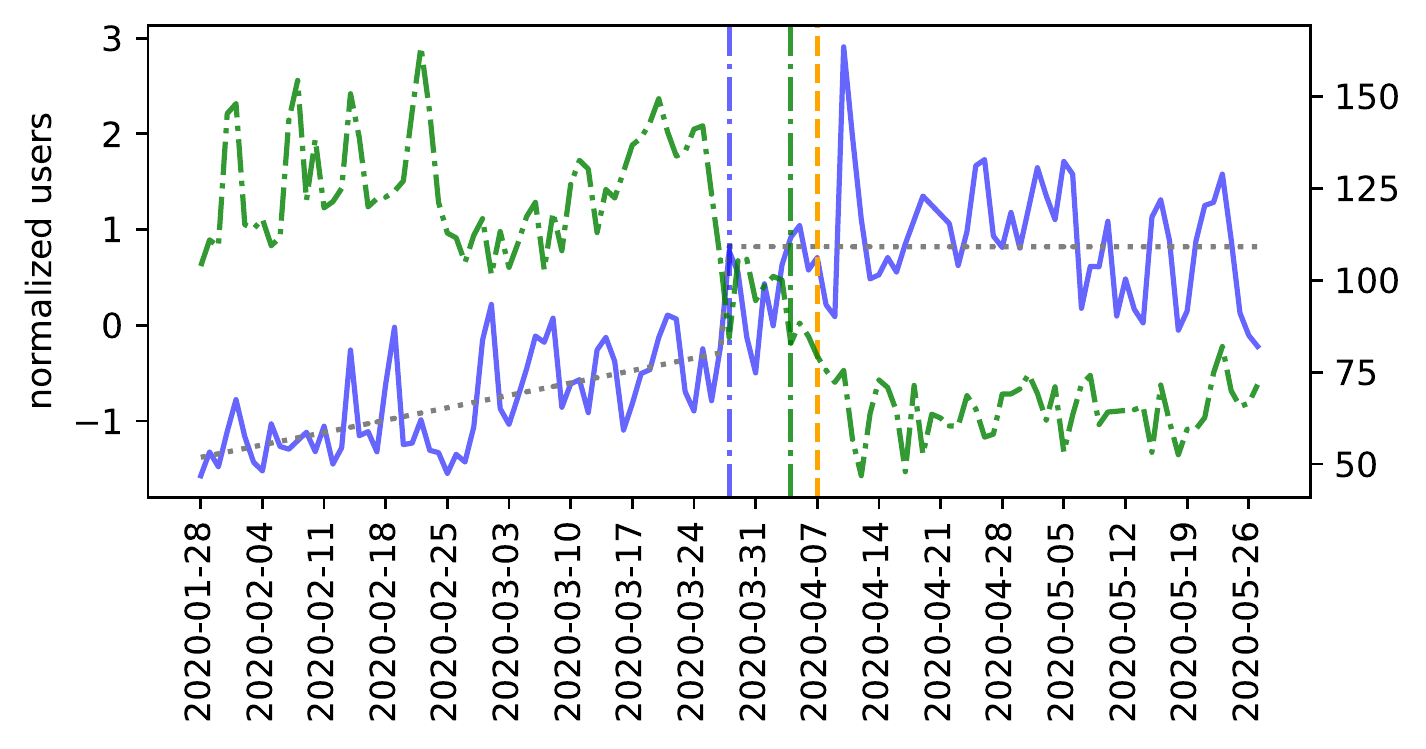}}
    \vspace{0.1cm}
    \subfloat[Brazil]{\includegraphics[width=0.49\linewidth]{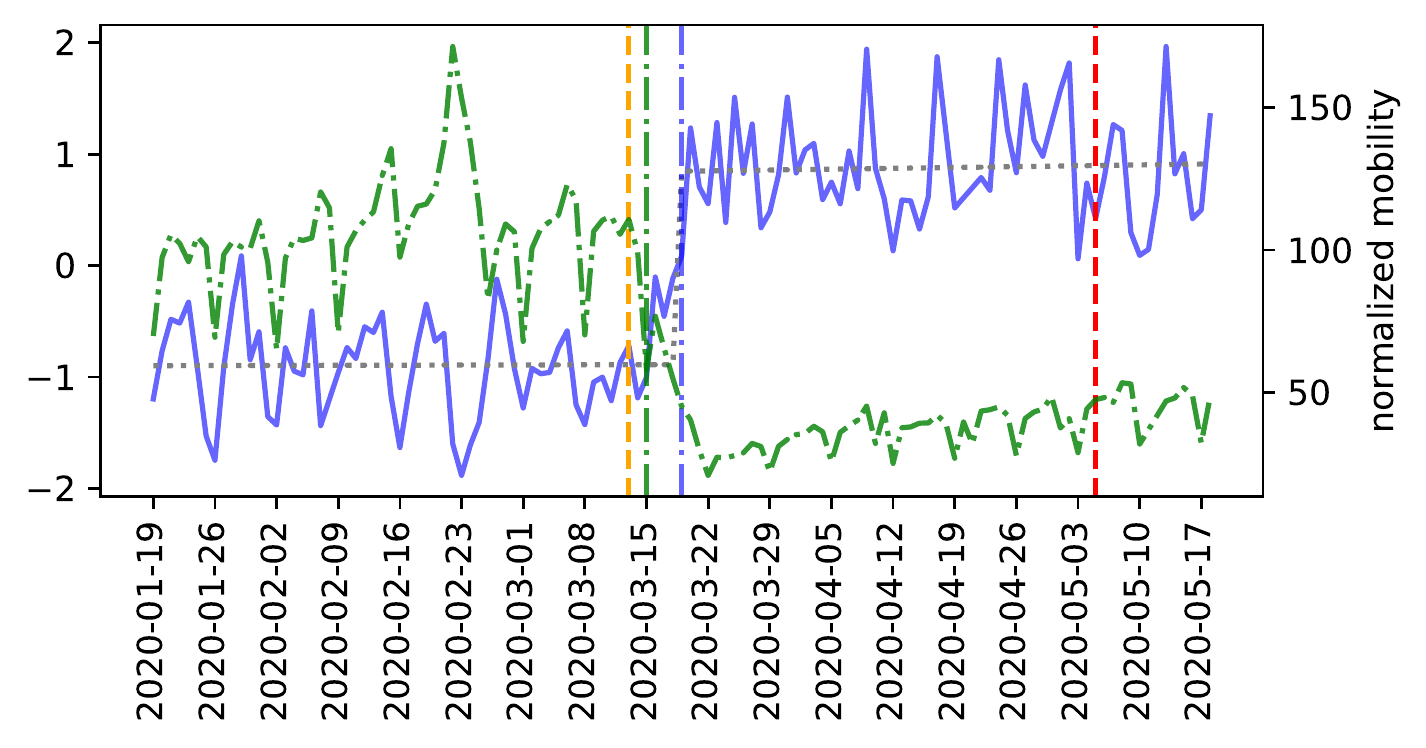}}
    \caption{Media sharing volume (blue) and mobility (green) in four countries.
    Vertical lines: L1 lockdown date $D_{L1}$ (orange), L2 lockdown date $D_{L2}$ (red), media sharing change point $D_{TC(T)}$ (blue), mobility change point $D_{TC(M)}$ (green).
    In grey, Interrupted Time Series model for media sharing $T$ around its change point $D_{TC(T)}$.} 
\label{fig:exampleChangePoints}
\end{figure}

Figure~\ref{fig:exampleChangePoints} shows four countries exemplifying the case in which the change in behavior is detected, on the same day as the announcement (United States), before the lockdown announcement (Japan \& Brazil), and after it (Italy).
\iftoggle{arXiv}{
  The readers are welcome to examine the plots for all countries at the end of this paper (Appendix A).}{}
The blue line shows the normalized sharing volume (in unique users), while the lockdown dates are shown in vertical lines -- orange for L1, and red for L2.
We explain the other lines in the next sections.
United States (as well as Serbia, New Zealand, Spain, and others) show changes in posting behavior that correspond precisely to the lockdown announcements.
In the case of the U.S., on March 16, President Trump announced ``15 Days to Slow the Spread'' with a series of guidelines on physical distancing and self-isolation, and recommendations to close schools and businesses \cite{mcgraw2020inside}. 
Italy's lockdown happened gradually, first quarantining 50,000 people from 11 different municipalities in the northern region on February 22, with gradual subsequent closures of regions \cite{dusi2020coronavirus}, until the Prime Minister Giuseppe Conte announced a country-wide quarantine on March 9 \cite{bbc2020coronavirus}.
Similarly, in Japan, Prime Minister Shinzo Abe requested that all Japanese elementary, junior high, and high schools close on February 27 \cite{tokyo2020coronavirus}, and the state of emergency declaration eventually expanded to include every prefecture within the country on April 16 \cite{Mainichi2020japan}. 
Brazil displays a notable shift in behavior around the time of the first \covid related death on March 17 \cite{bergamo020brasil}, without any official lockdowns until some of the northern cities announced the measures on May 7 \cite{globo2020belem}. 
The sudden or gradual shifts in the media sharing behavior, thus, visibly happen around the measures being taken at the time of the collection gathering, but not always, as in the case of Brazil.
\iftoggle{arXiv}{We encourage the reader to explore these graphs for all countries, shown at the bottom of the paper.}{We encourage the reader to explore these graphs for all countries in the companion website\footnote{\url{https://github.com/ymejova/yt-tw-covid/tree/main/plots_users_mobility}}.}

\subsection{Media Sharing as a Proxy for Mobility} 

Thus, we find that the media sharing behavior changes around the dates of lockdown enforcement.
We hypothesize that this change reflects the physical mobility of users, and the fact that they have more time to spend watching and sharing media online.
To explore the relationship between media sharing and mobility, we utilize the Apple mobility dataset ($M$) as described in Sec~\ref{sec:data}.
Recall that 56 countries that have passed our selection filters appeared in this dataset.

Figure~\ref{fig:histChangeLockdown}(b) shows the Pearson correlation $r$ between mobility $M$ and media sharing $T$, by the lockdown situations that happened in each country: having only L1, right away into L2, and having first L1 and later L2 (L1,L2).
Here, we smooth both trends using a 7-day average, in order to remove weekly periodicity (that was obvious in Figure~\ref{fig:top10countries_stats}).
Most correlations are strongly negative, and all those less than $-0.6$ are significant at $p < 0.0001$. 
Interestingly, the correlation is the weakest for countries which had only L1 lockdown (the rightmost points on the graph are Slovakia ($r = 0.23$, $p = 0.011$), Cambodia ($r = -0.25$, $p = 0.011$), Thailand ($r = -0.46$, $p < 0.0001$), and Hong Kong ($r = -0.56$, $p < 0.0001$)).
This strongly suggests that mobility is inversely proportional to media sharing behavior, as captured in the $T$ dataset.
Further, Figure~\ref{fig:histChangeLockdown}(c) shows the difference in days between the change point detected in the mobility and media sharing (both raw, un-smoothed). 
Out of 50 countries that had a change point detected for both mobility and media sharing, 44 are within 7 days, and for most countries the change in media sharing change follows the change in mobility.
The three countries having largest difference in change points (ticks in the far right of the plot) are Philippines (23), South Korea (20), and Singapore (18) -- all countries close to China that took early precautions against the virus.

If we return to the countries in Figure~\ref{fig:exampleChangePoints}, we now can examine the media sharing volume (in blue) in the contemporary context of the mobility score (green), and their respective change points (vertical lines in their respective colors).
In the case of the United States, the change happens all at once, and the change points, thus, happen within 1 day of each other.
The change points are a bit further apart in Brazil at 4 days, Italy at 5, and Japan at 7.

Some could argue that the identified cross-platform sharing changes were more predominant in countries with highly connected Internet users.
However, when we take the correlation of mobility and media sharing $r_{T,M}$ computed above, and correlate it to the internet penetration in the corresponding countries (data extracted from~\cite{2020internet-penetration}), we find no significant relationship ($r = -0.18$, $p = 0.16$).
This suggests that, once the countries have a minimum volume of Twitter usage, the signal can be used as a proxy for mobility irrespective of internet penetration. 
This is encouraging, since only 56 out of 109 originally selected countries had mobility data available, so that an alternative proxy is especially valuable for those locales where the major technology service providers do not release their data.

\section{Media Sharing Behavior Change: Interrupted Time-Series Modeling} 

To understand whether human behavior changes significantly around the time of COVID-related lockdowns, we employ Interrupted Time Series analysis (ITS)~\citep{bernal2017interrupted}, which aims to estimate the effect of an intervention which has a well-defined starting time.
Specifically, we examine both the time series of media sharing volume ($T$), as well as the mobility data ($M$), in order to gauge the extent of the behavior change.
We employ Ordinary Least Squares (OLS) regression to model the behavior time series of the two signals using two variables: $P$ signifying time passage in days and $X_t$, an indicator signifying the beginning of the intervention period.
For example, modeling change in the media sharing volume around a lockdown date, in following equation:
\vspace{-0.1cm}
\[ y_t = \beta_0 + \beta_1 P + \beta_2 X_t + \beta_3 P X_t \]

\noindent $y_t$ is the volume of media sharing at the time $t$, $\beta_0$ is the baseline volume at the beginning of the time series, $\beta_1$ is the baseline change in volume over time before the lockdown date, $\beta_2$ is the change in volume at the lockdown date, and $\beta_3$ is the trend (slope) change following the lockdown date.
In this paper, we focus specifically on the $\beta_2$, the change in behavior around the intervention.
Visually, the ITS models are plotted in grey in Figure~\ref{fig:exampleChangePoints} for the four countries.

To capture the complexity of the lockdown situations, we again examine three groups of countries: those that only had L1, those that only had L2, and those that went from L1 to L2 (L1=>L2).
Note that for the last group of countries, we can test both lockdown dates.
Additionally, since in the previous sections we found that the actual behavior change often happened some days before or after the official lockdown dates, we decided to perform ITS with 4 different types of dates: two lockdown dates ($D_{L1}$ or $D_{L2}$), and two trend change points ($D_{TC(T)}$ or $D_{TC(M)}$), on the groups of countries that each applies.
As in the previous section, we select 2 months before and 2 months after the selected date, and here consider only the 54 countries with mobility data.

Table~\ref{tab:its_rsquared_coeffs} shows statistics of ITS models for various scenarios.
It reports the mean ($\bar{\beta}$) and standard deviation ($\sigma_\beta$) of the intervention point coefficient, and the model fit ($\bar{R^2}$), aggregated over relevant countries ($n$).
First, we observe that the $\bar{\beta}$ coefficients are positive across all models. Note that we z-normalize the sharing volume ($T$), making it comparable across countries.
Second, strongly negative $\bar{\beta}$ coefficients for mobility indicate a sharp decrease around the interruption dates, echoing the negative relationship between sharing volume ($T$) and mobility ($M$) we have previously established.
Third, we find the model fit to be the worst for countries that had L1 lockdowns, indicating that the \textit{behavior change, both in media sharing and in mobility, are affected the least by policies merely recommending people to stay home}.
Further, we find similar $\bar{\beta}$ values and model fit for the two levels of lockdowns for the L1=>L2 countries, indicating that for such scenarios it may be difficult to discern the effect of L1 policy, as it is usually quickly followed by a more stringent one (for 75\% of the countries within 2 weeks).

Finally, we make two methodological observations.
We confirm that the extent of change of behavior (both in media sharing and mobility) is greater at change points than at lockdown dates, suggesting that \textit{factors other than those captured by lockdown policies -- including other communication by the authorities -- may have had an effect on behavior}.
The better fit of the ITS models in the case of mobility points to the \textit{quicker effect on this variable, and a less sudden one on media sharing, pointing to a gradual behavior change when it concerns media consumption during emergency measures}.
The implications in terms of attention to public messaging, adjustment in daily routine, and mental health maintenance are exciting future directions of research.

\begin{table}[t]
    \centering
    \small
    \begin{tabular}{l|rrrr}
\toprule
\textbf{Model} & $\bar{\beta}$ & $\sigma_\beta$ & $\bar{R^2}$ & $n$ \\
\midrule
media sharing \& L1 date (L1) & 1.126 & 0.832 & 0.477 & 18 \\
media sharing \& L2 date (L2) & 1.349 & 0.598 & 0.650 & 24 \\
media sharing \& L1 date (L1=>L2) & 1.158 & 0.685 & 0.652 & 14 \\
media sharing \& L2 date (L1=>L2) & 1.044 & 0.579 & 0.651 & 14 \\
media sharing \& sharing change point (L1) & 1.667 & 0.439 & 0.580 & 15 \\
media sharing \& sharing change point (L2) & 1.659 & 0.453 & 0.701 & 23 \\
media sharing \& sharing change point (L1=>L2) & 1.697 & 0.370 & 0.756 & 13 \\
media sharing \& mobility change point (all countries) & 1.298 & 1.084 & 0.623 & 54 \\
\midrule
mobility \& L1 date (L1) & -48.767 & 30.435 & 0.751 & 18 \\
mobility \& L2 date (L2) & -64.630 & 30.556 & 0.804 & 24 \\
mobility \& L1 date (L1=>L2) & -58.422 & 31.459 & 0.790 & 14 \\
mobility \& L2 date (L1=>L2) & -44.808 & 25.672 & 0.760 & 14 \\
mobility \& mobility change point (L1) & -65.221 & 21.345 & 0.817 & 16 \\
mobility \& mobility change point (L2) & -82.002 & 25.618 & 0.878 & 24 \\
mobility \& mobility change point (L1=>L2) & -53.372 & 82.838 & 0.865 & 14 \\
mobility \& sharing change point (all countries) & -67.582 & 29.810 & 0.831 & 51 \\
\bottomrule
    \end{tabular}
    \caption{Interrupted Time Series models with the mean ($\bar{\beta}$) and standard deviation ($\sigma_\beta$) of the intervention point coefficient, and average model fit ($\bar{R^2}$), aggregated over relevant countries ($n$).
    Model specification indicates data being modeled (media sharing or mobility) and kind of interruption point, as well as lockdown scenario (L1, L2, or L1=>L2).}
    \label{tab:its_rsquared_coeffs}
\end{table}

\section{Discussion \& Conclusions}

This paper provides a large-scale evidence of individual behavioral response to the social distancing interventions taking place around the world in the early 2020. 
We find that media sharing behavior increased in the majority of countries during the \covid lockdowns. 
However, \covid-related activity only accounts for 15\% of the overall volume of posts. In fact, users shared with each other content on a wide range of topics apart from \covid. 
Whether this change in behavior extends beyond volume, and impacts the media consumption qualitatively is a fascinating question. 
We are open to collaborating with researchers (as making full dataset is against Twitter Terms of Service) on exploring, for instance, the change in quality of information being shared, the kind of health advice people turn to in an emergency situation, and how users may maintain social ties and communities around co-consumption of online media.

We also established a strongly negative relationship between media sharing and mobility, however it is not certain that this relationship will endure.
We find the spike in interest about \covid dissipates after several months, whereas the sharing volume remains elevated over time. 
This data provides an opportunity to study psychological adaptation during a crisis, and test theories around, for instance, \emph{emotional evanescence} as individuals adjust their emotional reactions \cite{wilson2003making}, and the role of media in post-traumatic stress disorder after major events.

As far as the online behavior can be viewed as a proxy of mobility, it may be especially useful in the understanding of public compliance with the public health measures, and its implications on the spread of disease. 
In the post-hoc analysis of the effectiveness of government measures, online data may provide a window into individual behaviors that signify online and offline socialization, as well as interest in public health messages, which may relate to actions individuals take that may raise or lower the risk of disease contraction.
Further, country-specific cultural aspects may play a role, as well as trust in or disobedience to authority, and individual's relation to society.

There are several important limitations to our study.
Not only are we limited by the users of two specific internet platforms (albeit some of the most popular ones), the adoption of these platforms is non-uniform around the world, allowing us to examine only some countries. Analysis of local platforms, such as Sina Weibo in China, may provide a better coverage.
Further, as we have found, definitions of ``lockdown'' are highly varied, as each country presents a unique combination of local guidance, social support systems, and public messaging, which makes comparison across countries approximate.
Finally, a plethora of other signals should be considered when modeling public health-related behaviors beyond lockdown measures, and future work in understanding health belief formation and the changing of action affordances will provide a fuller picture of this emergency.

In order to support repeatable and extensible research, we make country-specific daily data and additional visualizations available at \url{https://github.com/ymejova/yt-tw-covid}.

\section{Acknowledgments}

The research leading to these results received partial funding from the EU H2020 Research and Innovation programme under grant agreements No 830927 (Concordia).
These  results reflect only the authors' view and the Commission is not responsible for any use that may be made of the information it contains.

\bibliographystyle{ACM-Reference-Format}
\bibliography{references}

\appendix{
\section{Media sharing, mobility, lockdowns, and ITS model for select countries}}

\centering
\includegraphics[width=0.49\linewidth]{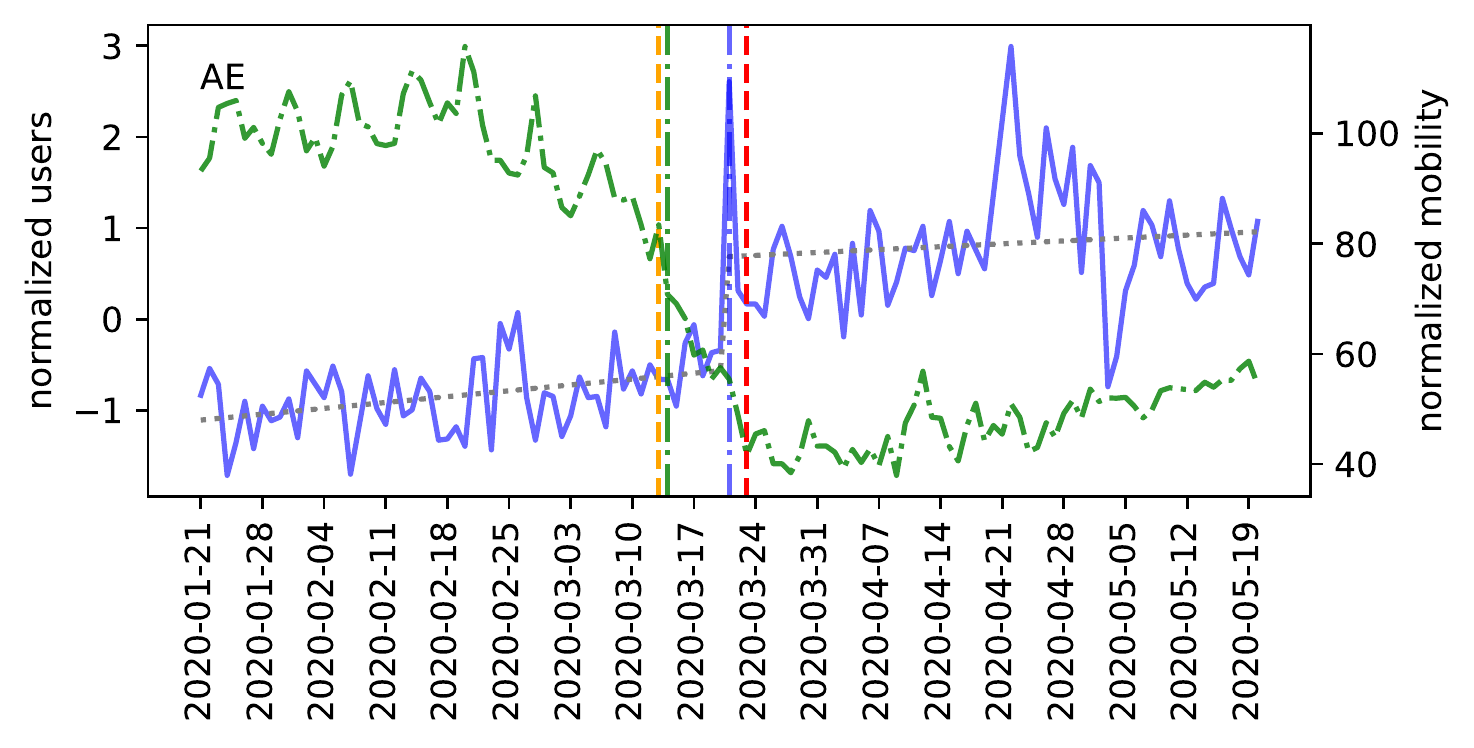}
    \vspace{0.1cm}
\includegraphics[width=0.49\linewidth]{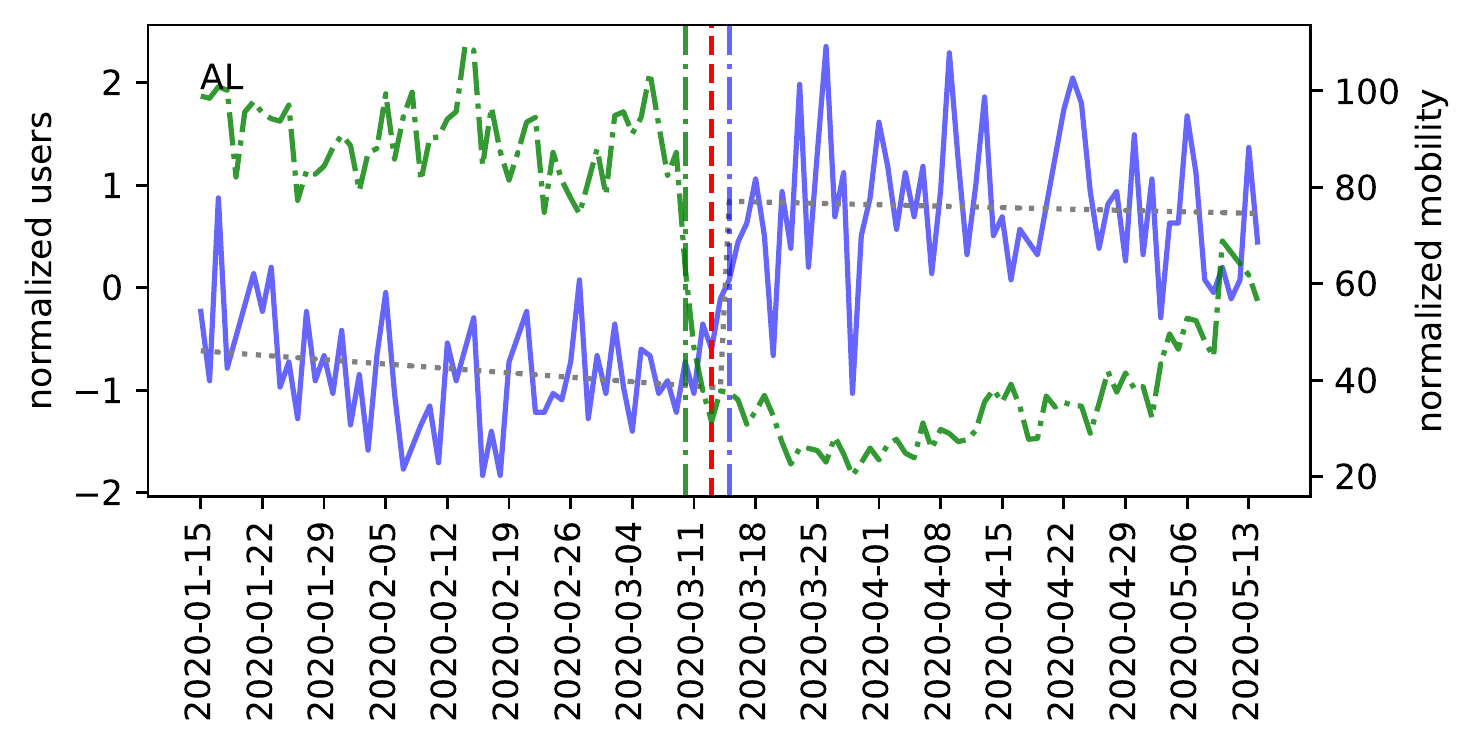} \\

\includegraphics[width=0.49\linewidth]{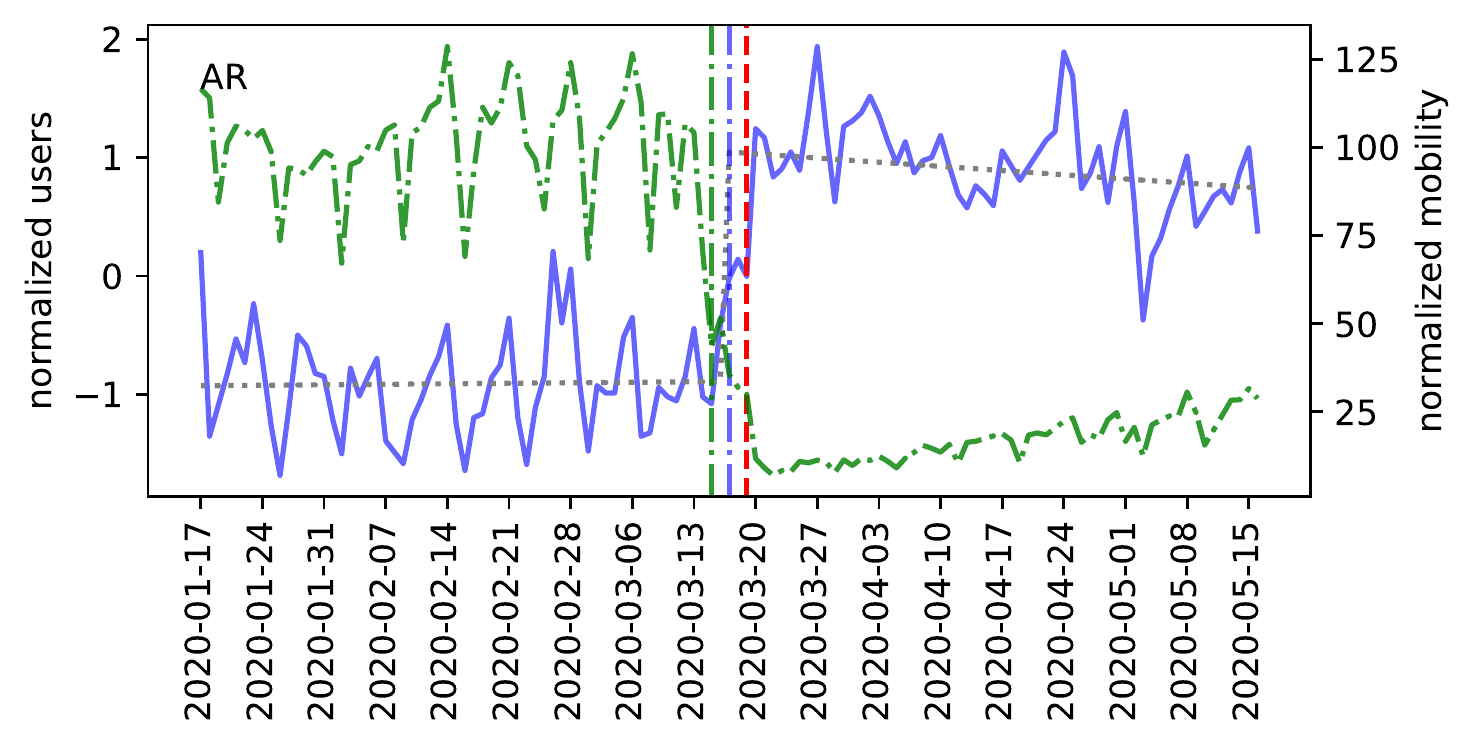}
    \vspace{0.1cm}
\includegraphics[width=0.49\linewidth]{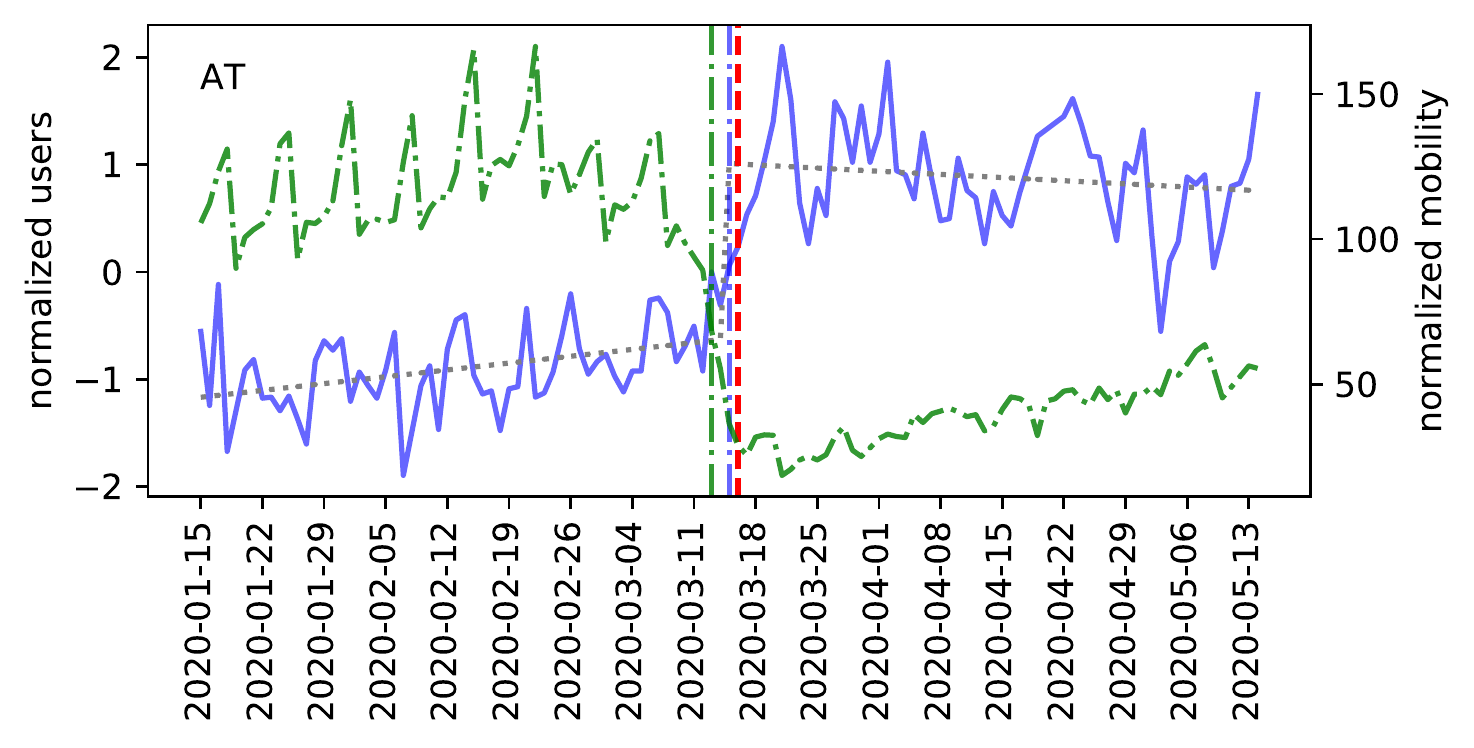}\\

\includegraphics[width=0.49\linewidth]{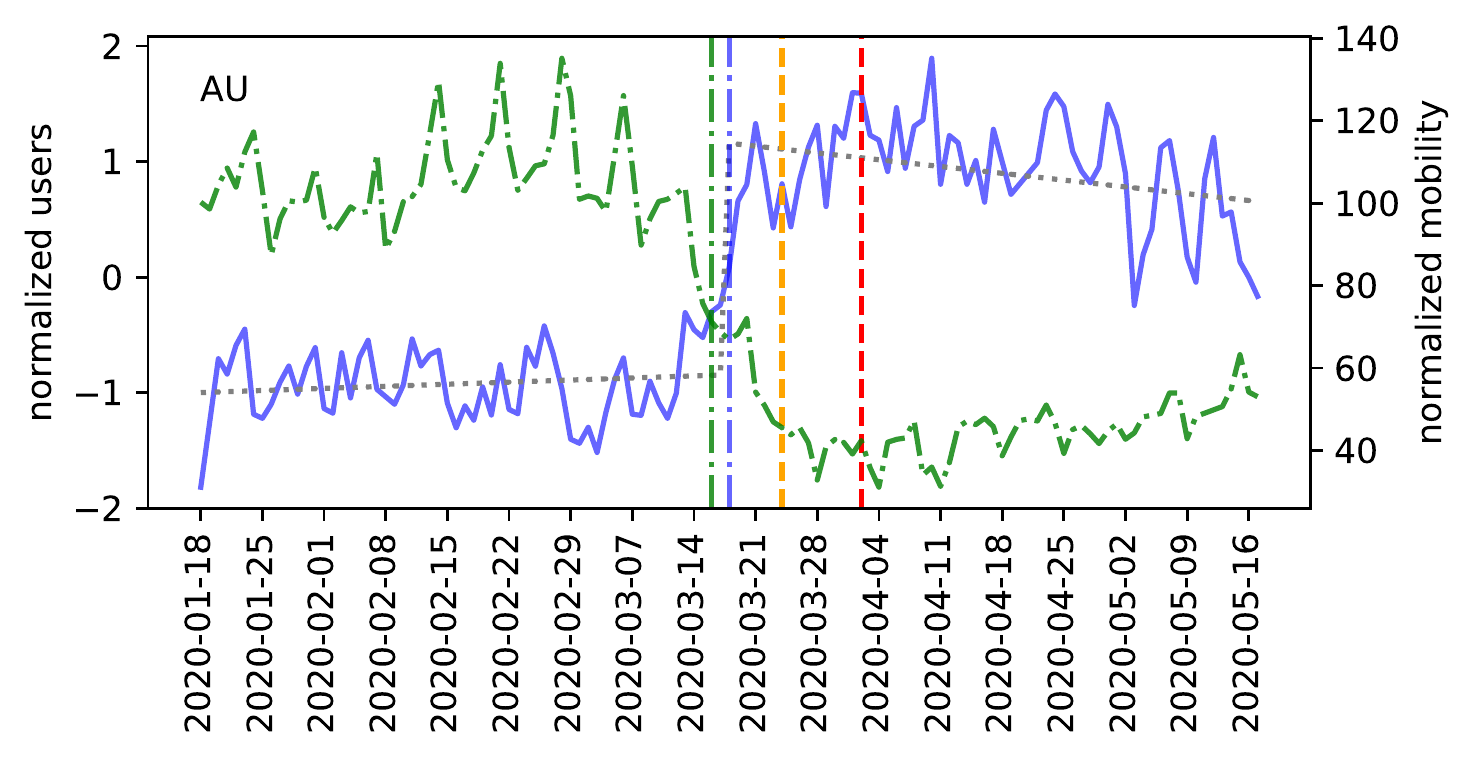}
    \vspace{0.1cm}
\includegraphics[width=0.49\linewidth]{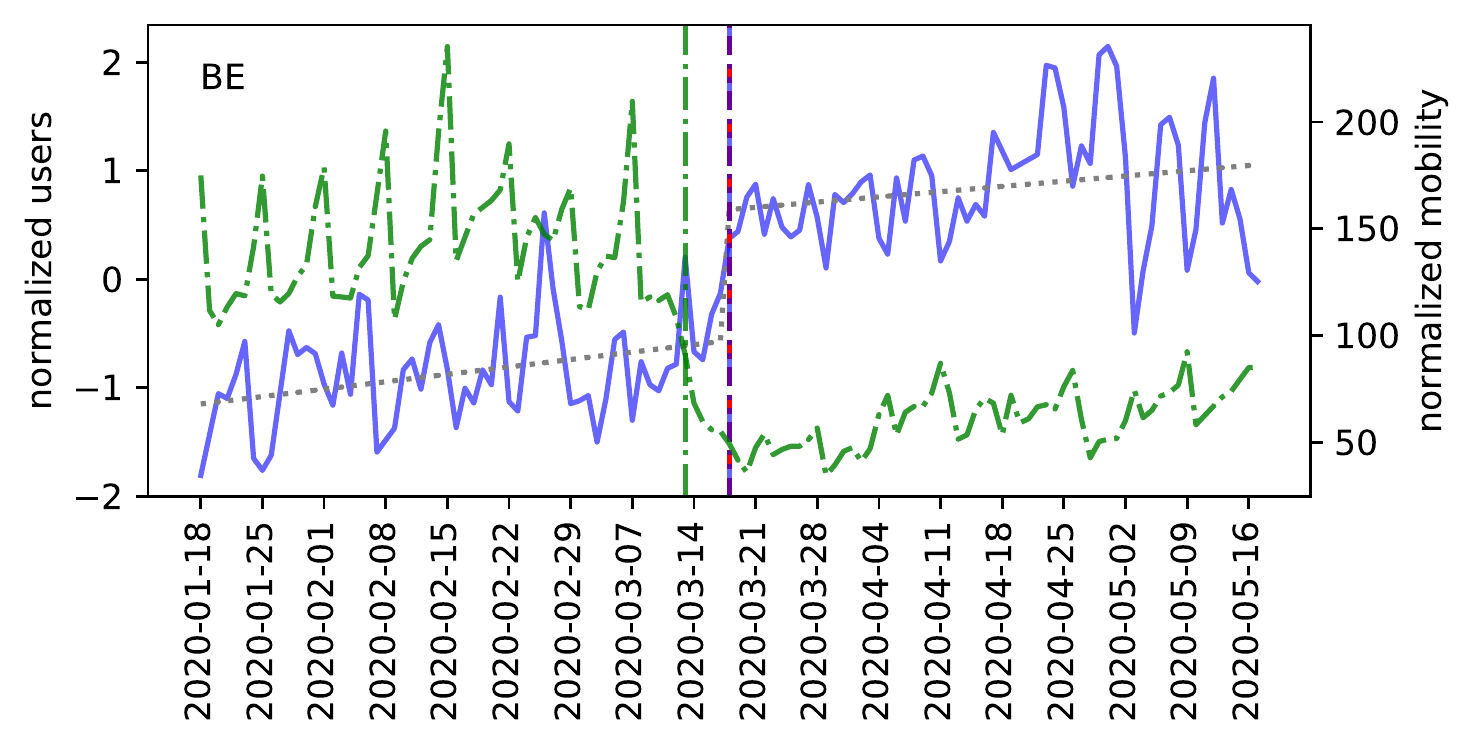}\\

\includegraphics[width=0.49\linewidth]{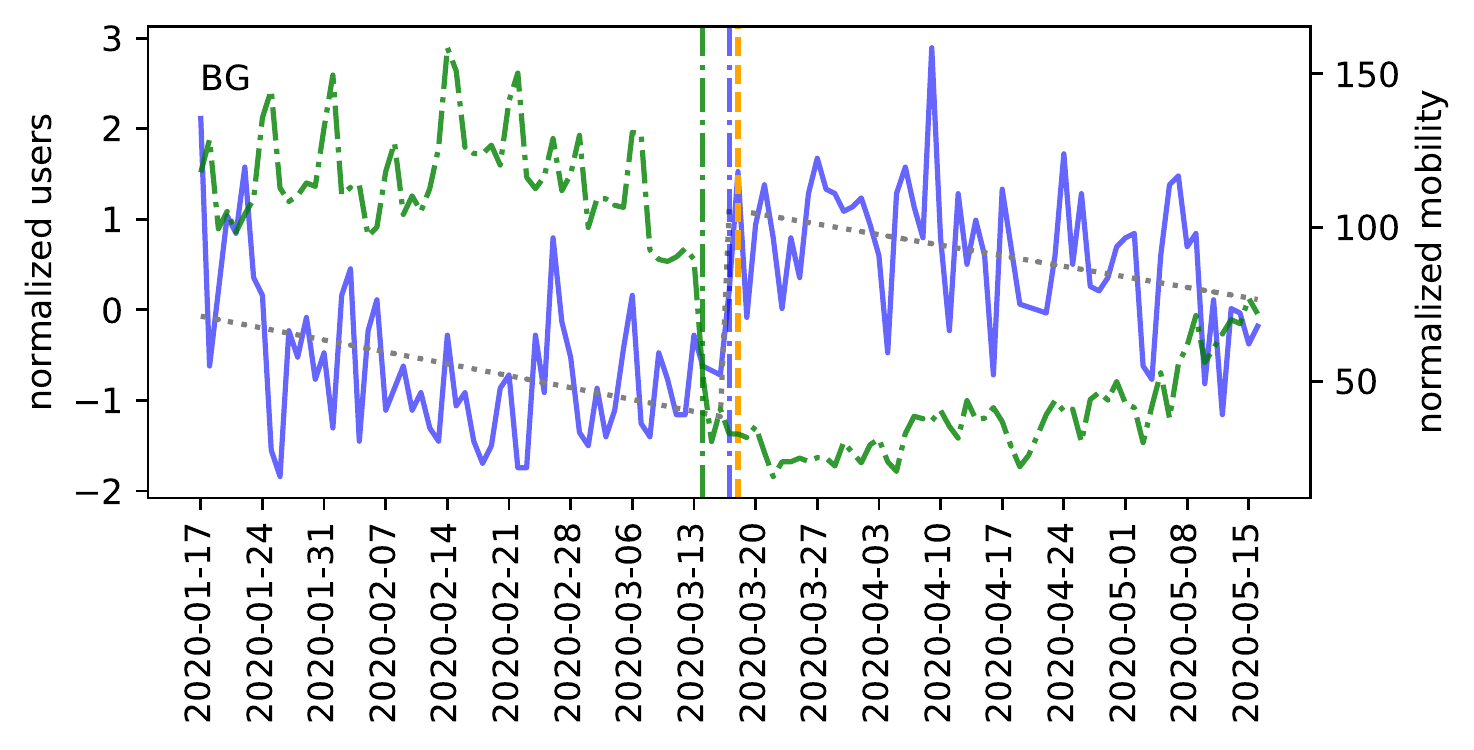}
    \vspace{0.1cm}
\includegraphics[width=0.49\linewidth]{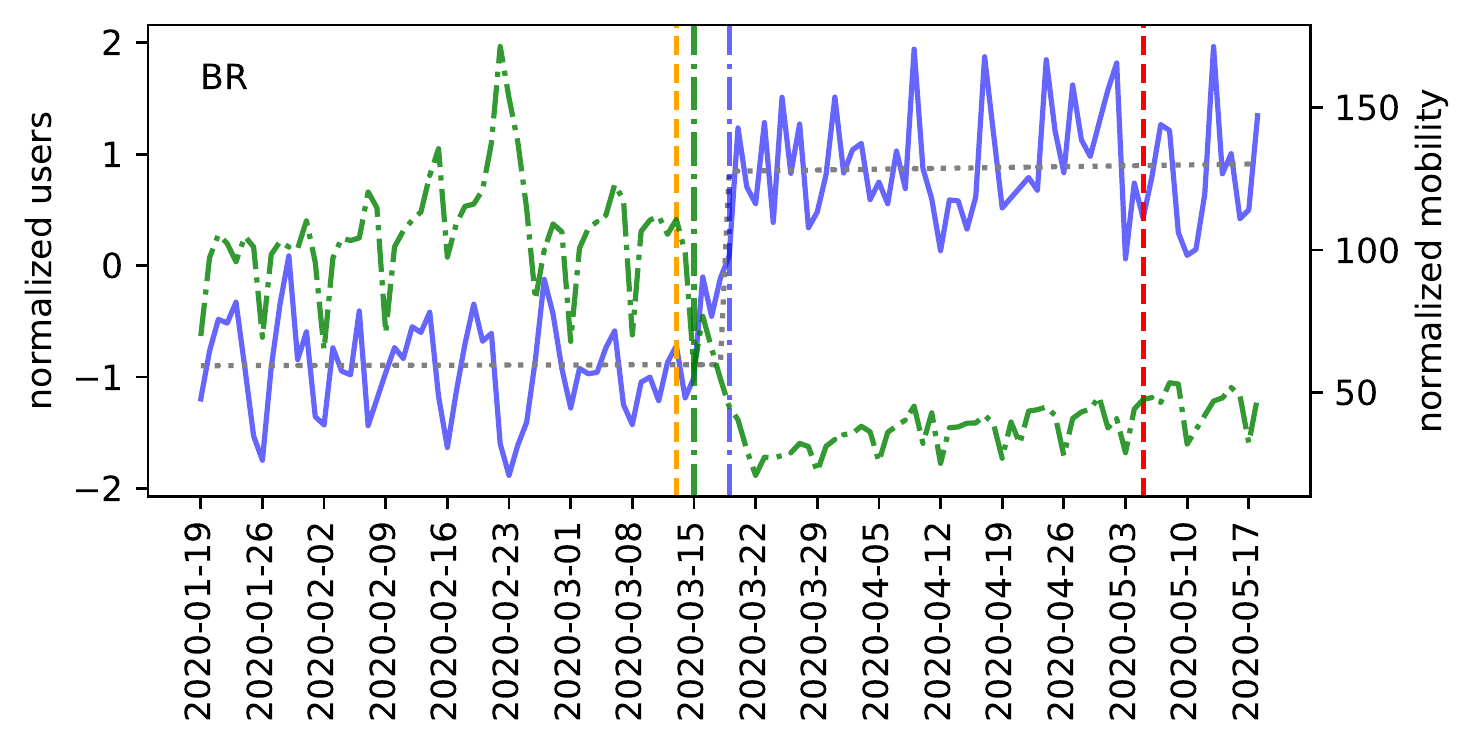}\\

\includegraphics[width=0.49\linewidth]{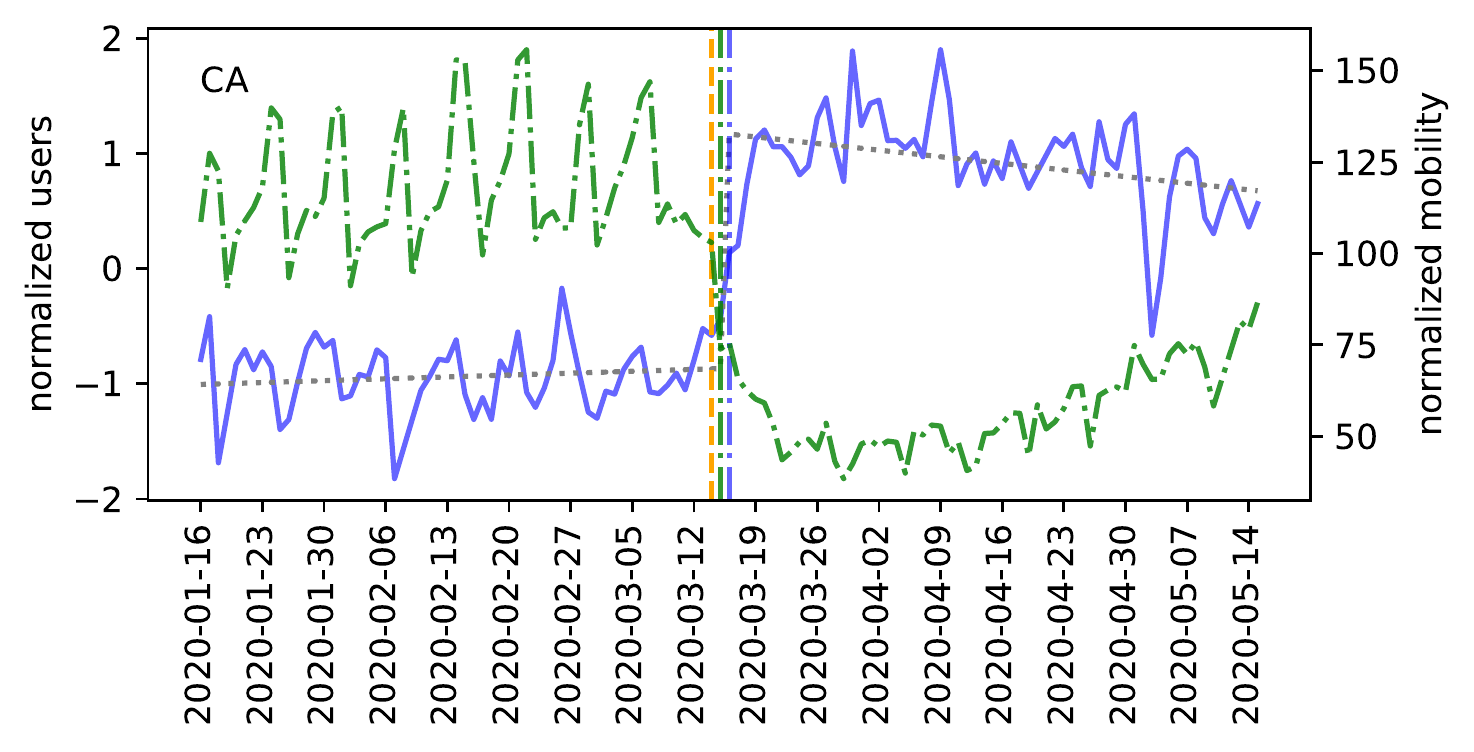}
    \vspace{0.1cm}
\includegraphics[width=0.49\linewidth]{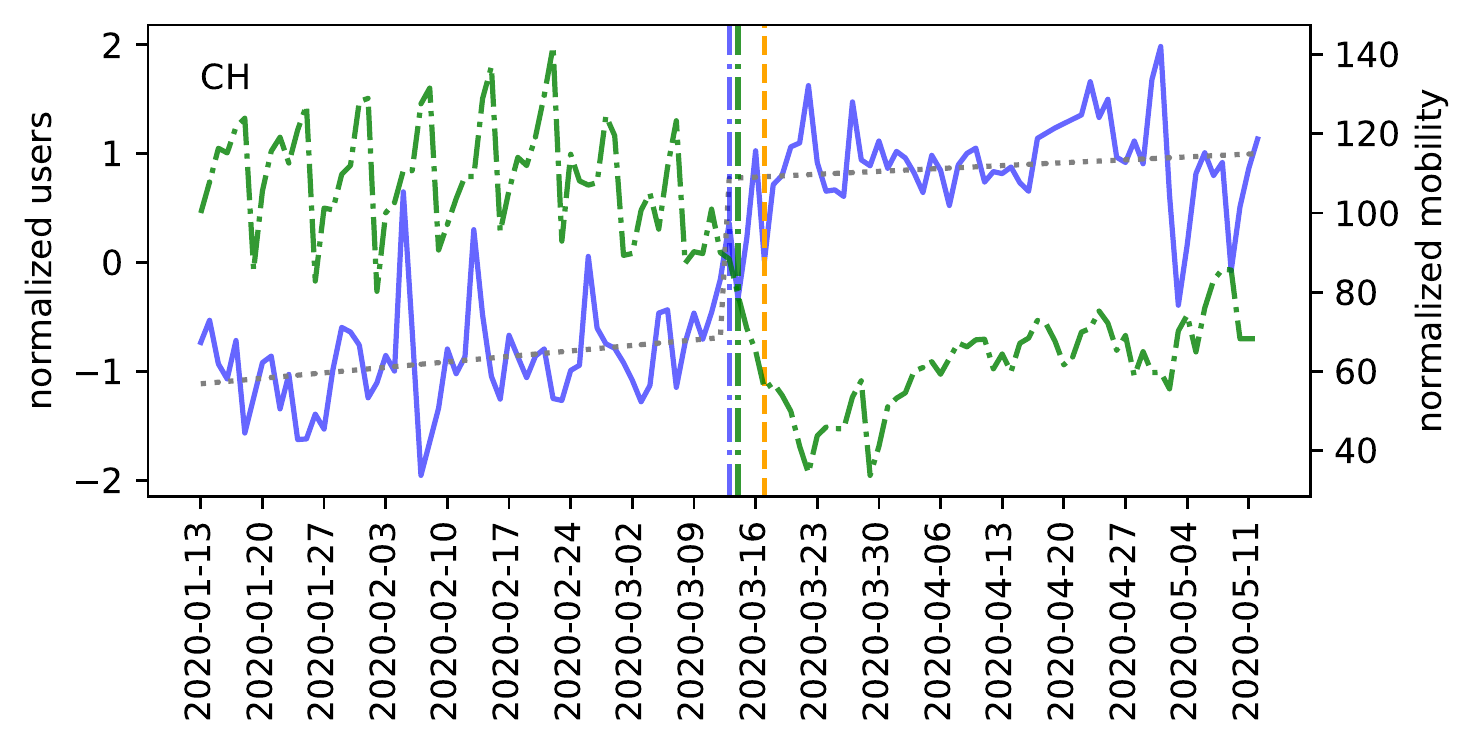}

\includegraphics[width=0.49\linewidth]{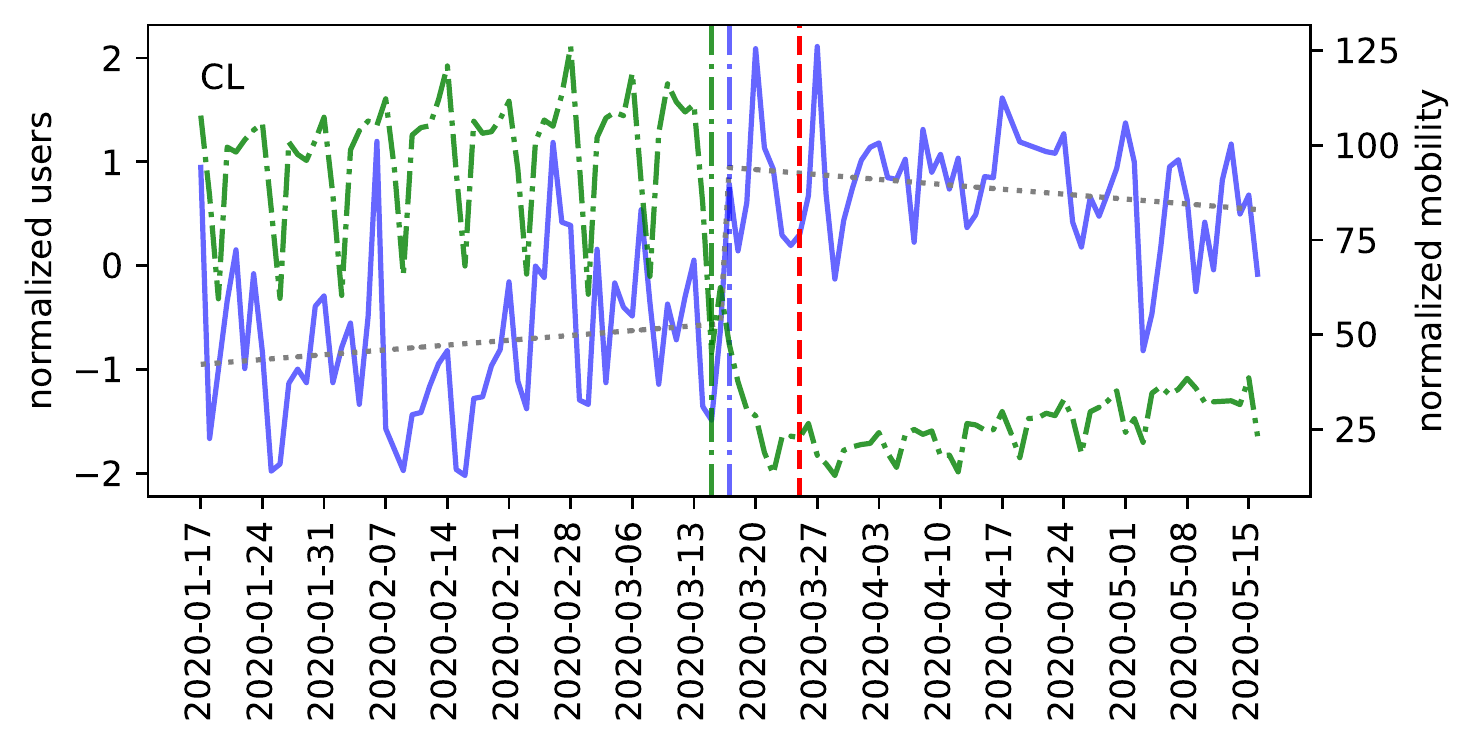}
    \vspace{0.1cm}
\includegraphics[width=0.49\linewidth]{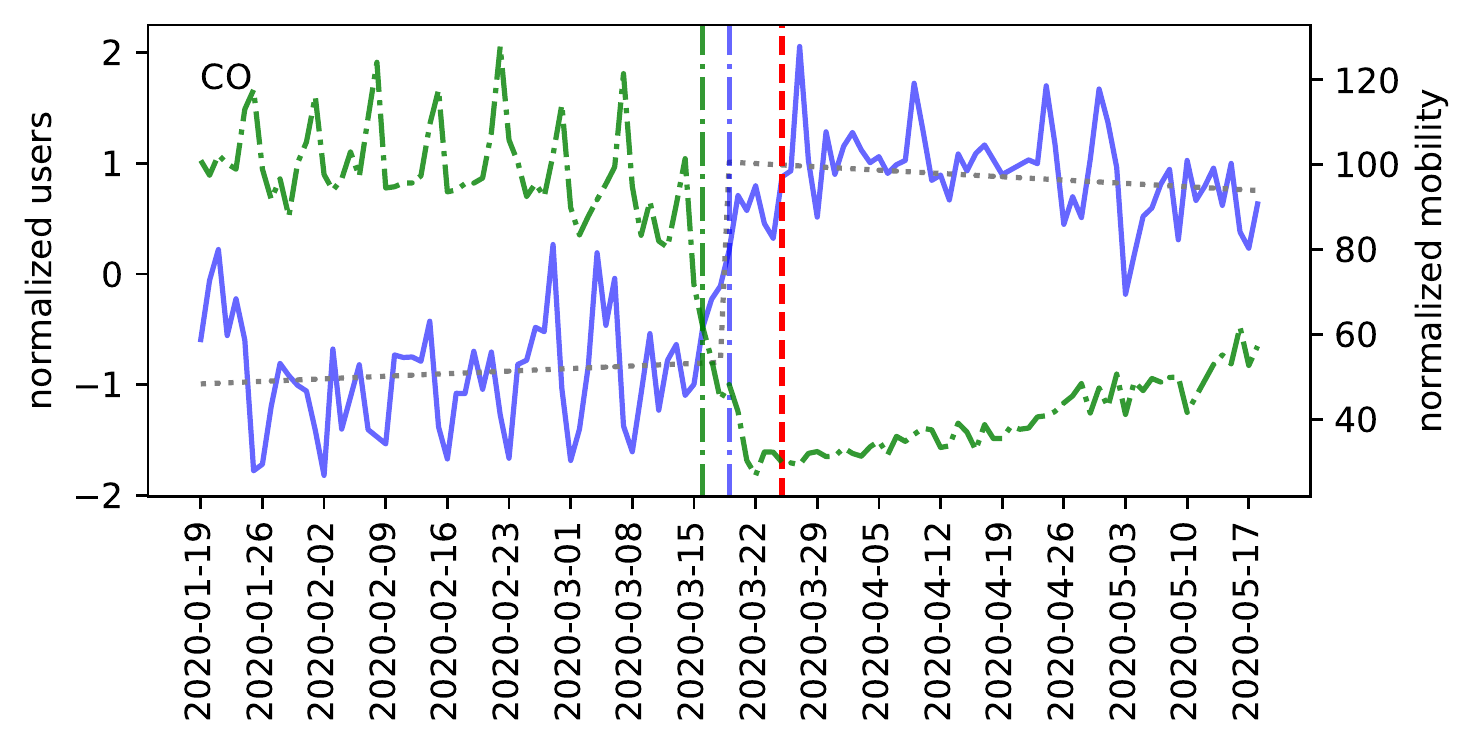} \\

\includegraphics[width=0.49\linewidth]{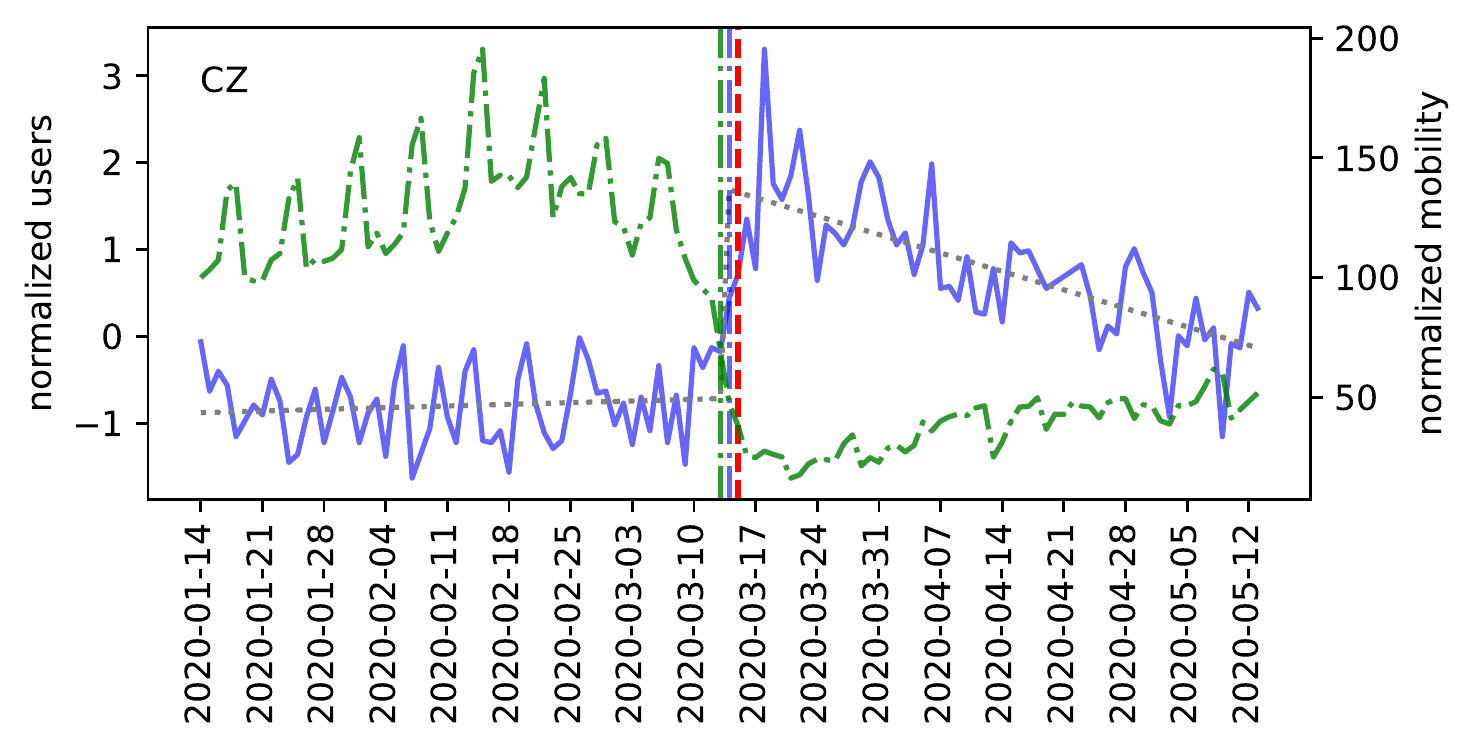}
    \vspace{0.1cm}
\includegraphics[width=0.49\linewidth]{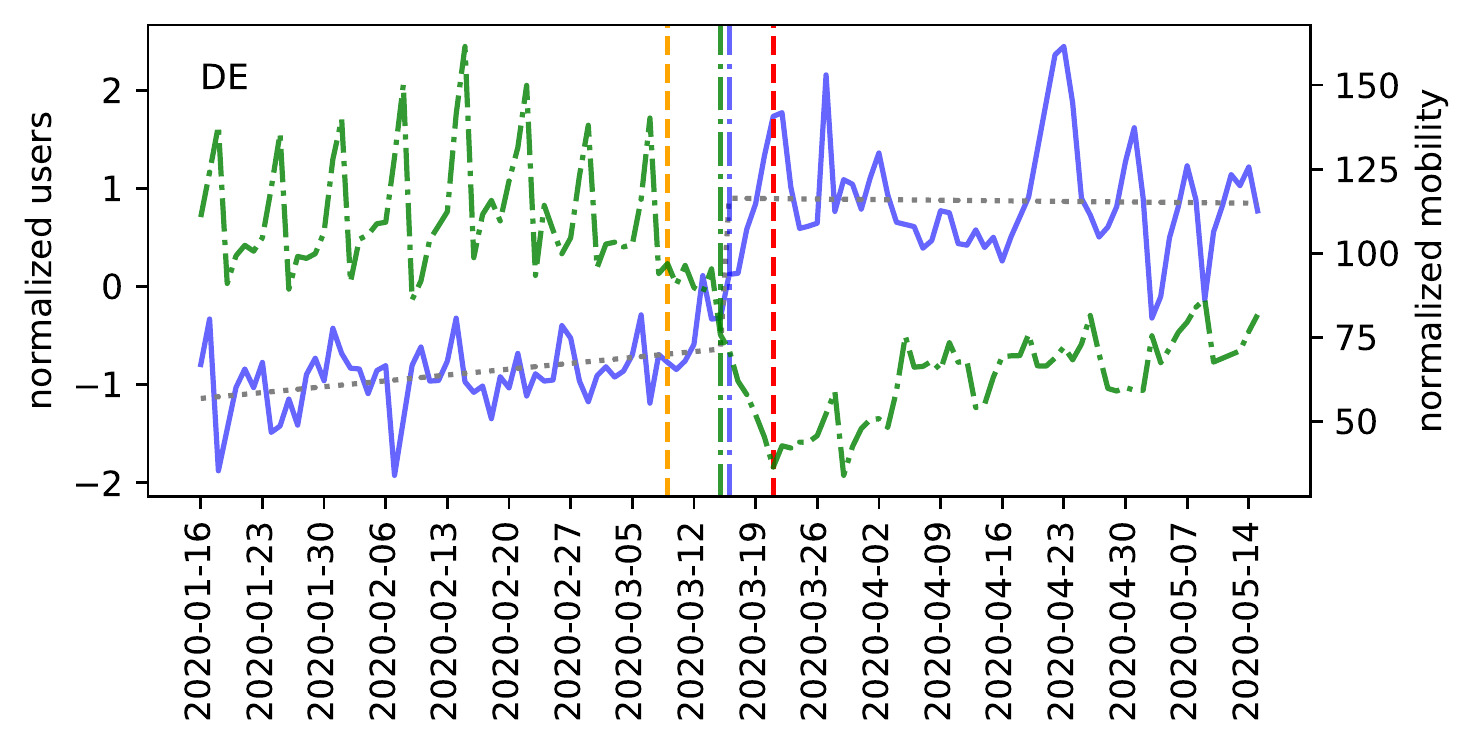}\\

\includegraphics[width=0.49\linewidth]{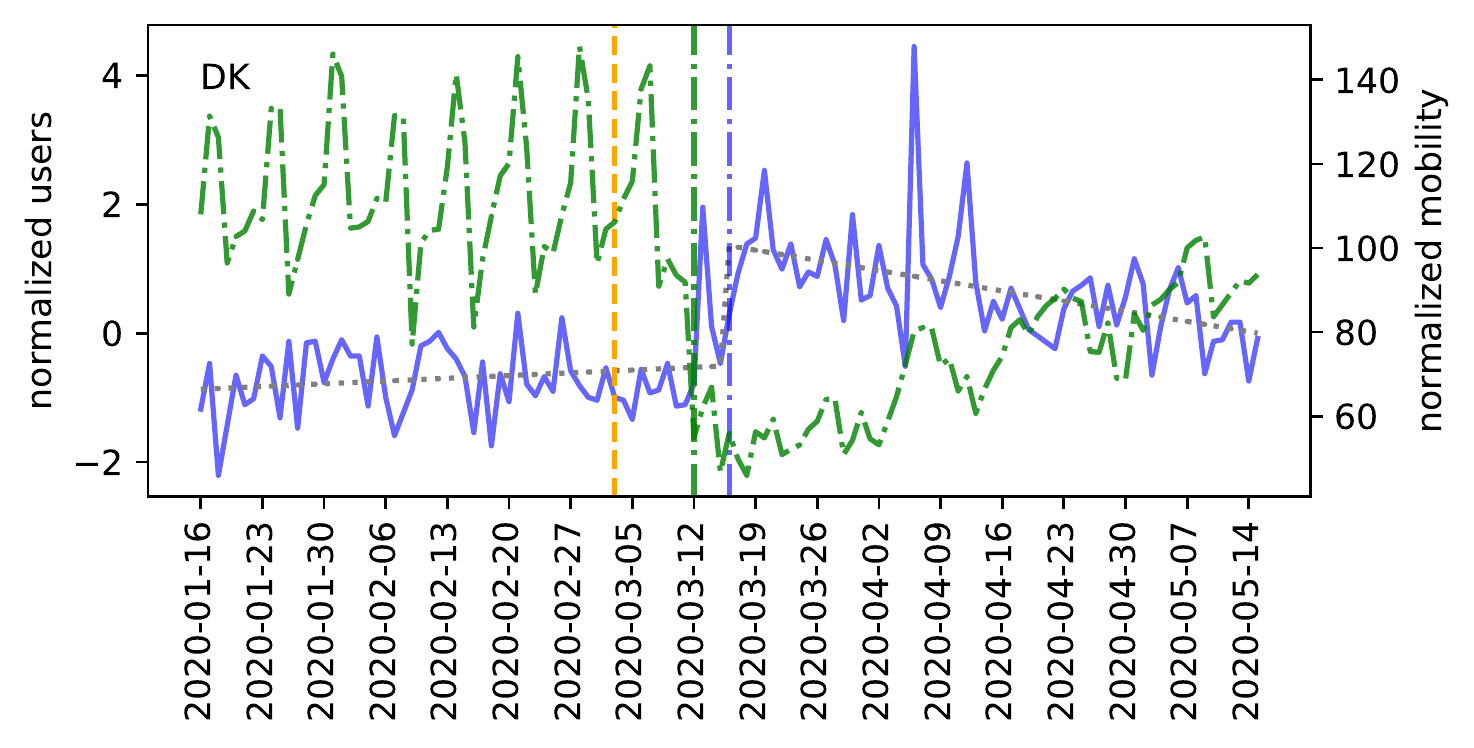}
    \vspace{0.1cm}
\includegraphics[width=0.49\linewidth]{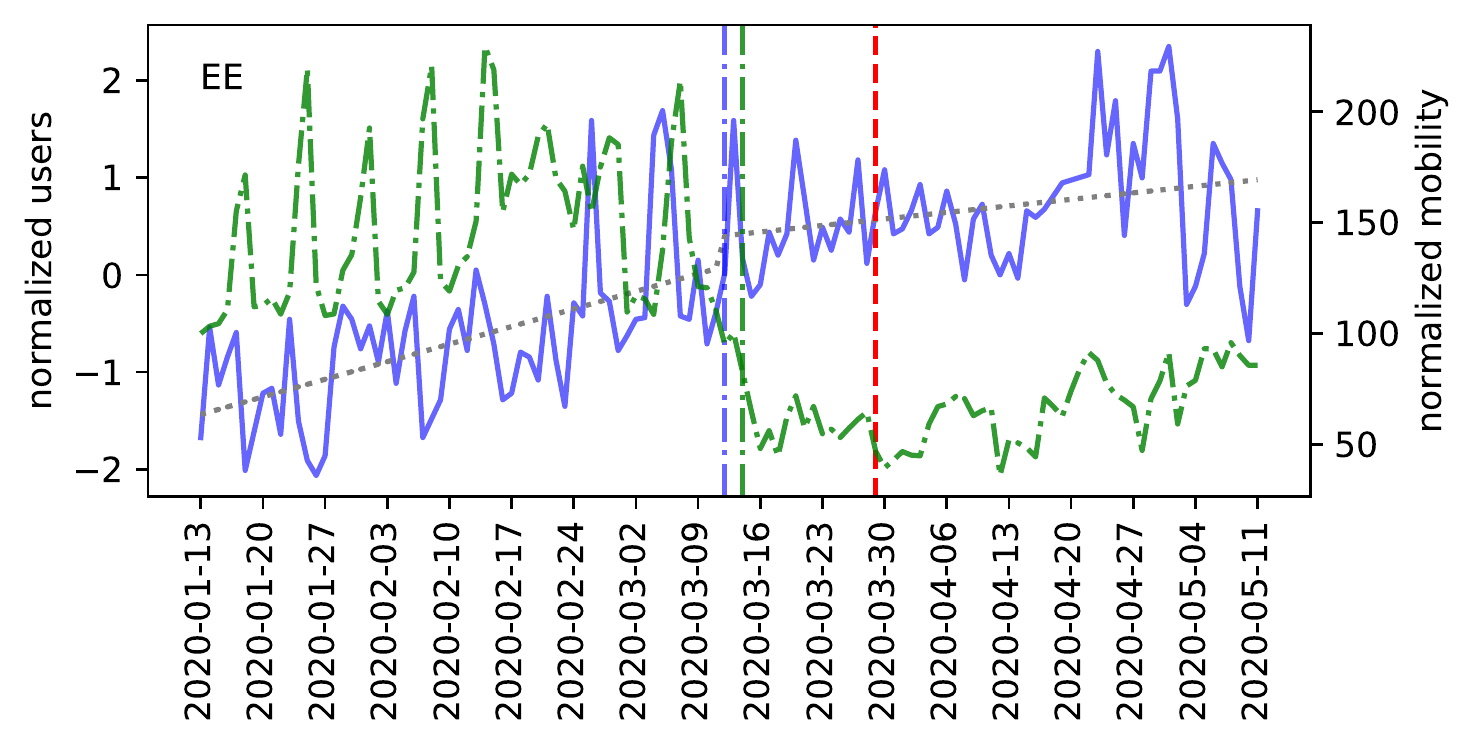}\\

\includegraphics[width=0.49\linewidth]{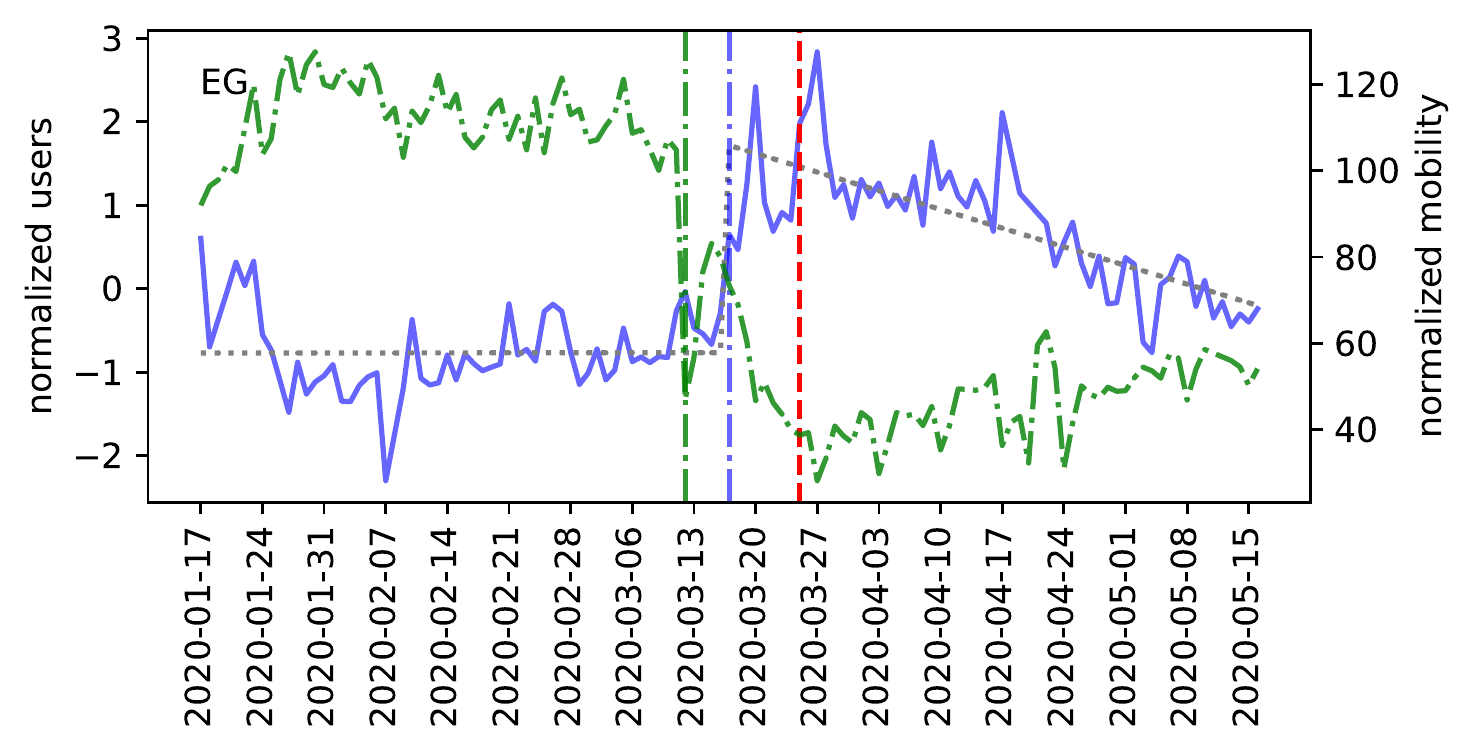}
    \vspace{0.1cm}
\includegraphics[width=0.49\linewidth]{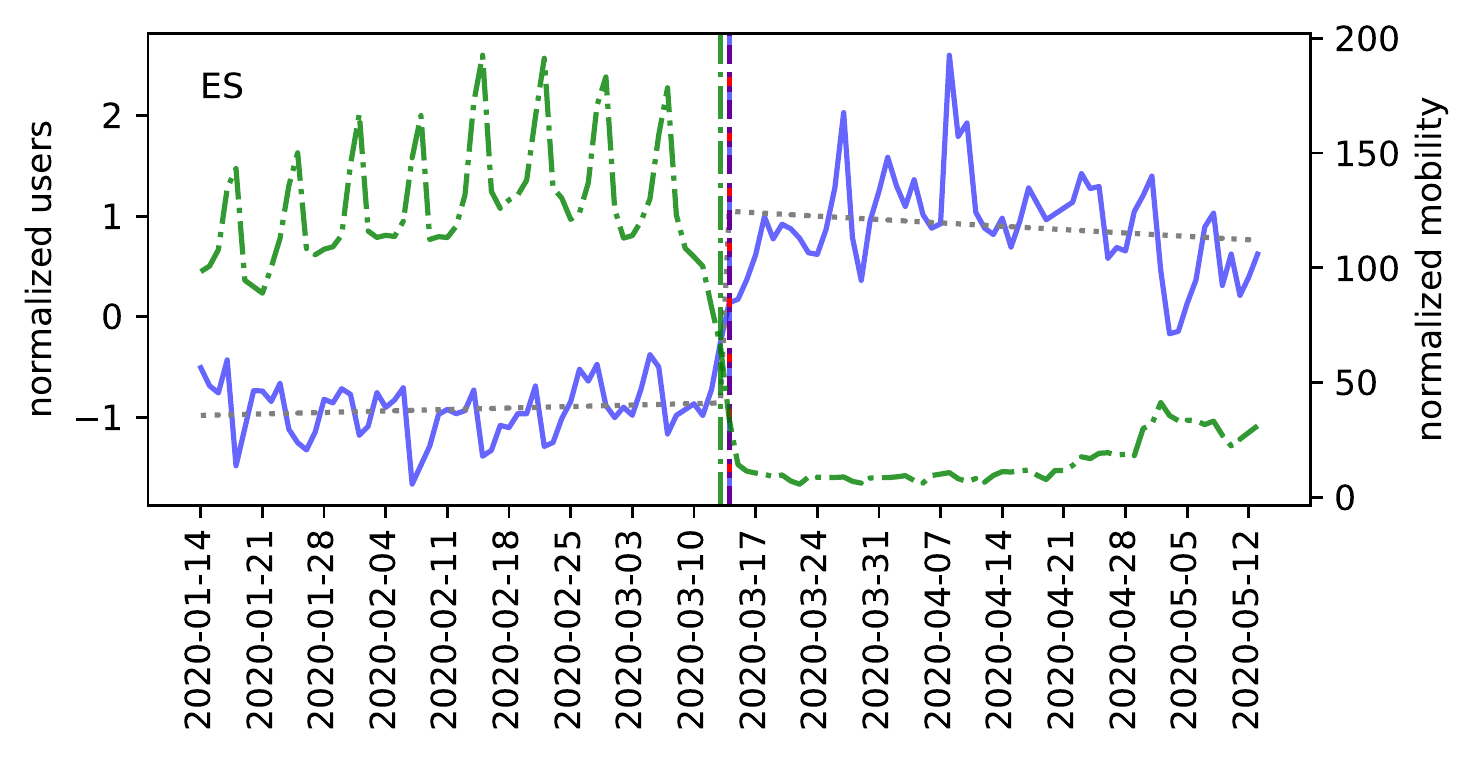}\\

\includegraphics[width=0.49\linewidth]{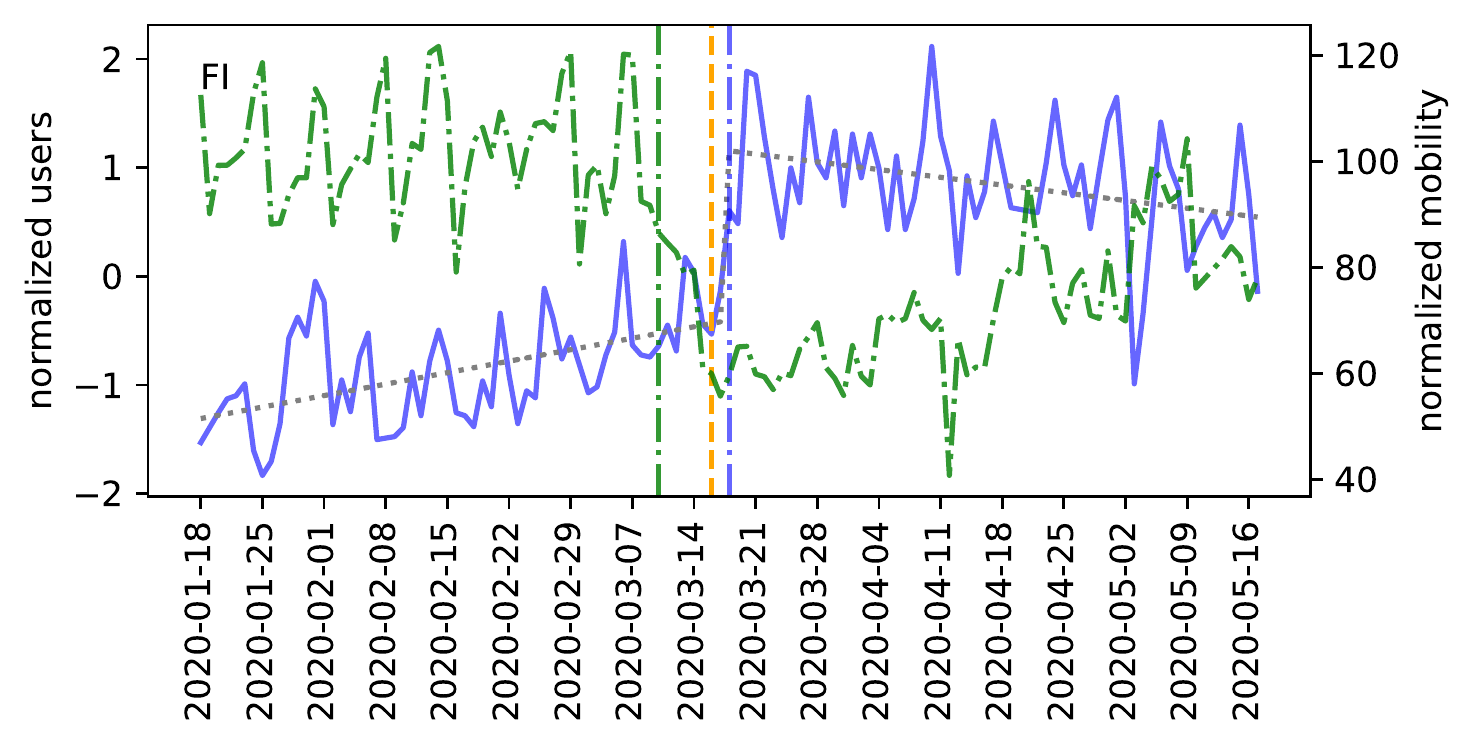}
    \vspace{0.1cm}
\includegraphics[width=0.49\linewidth]{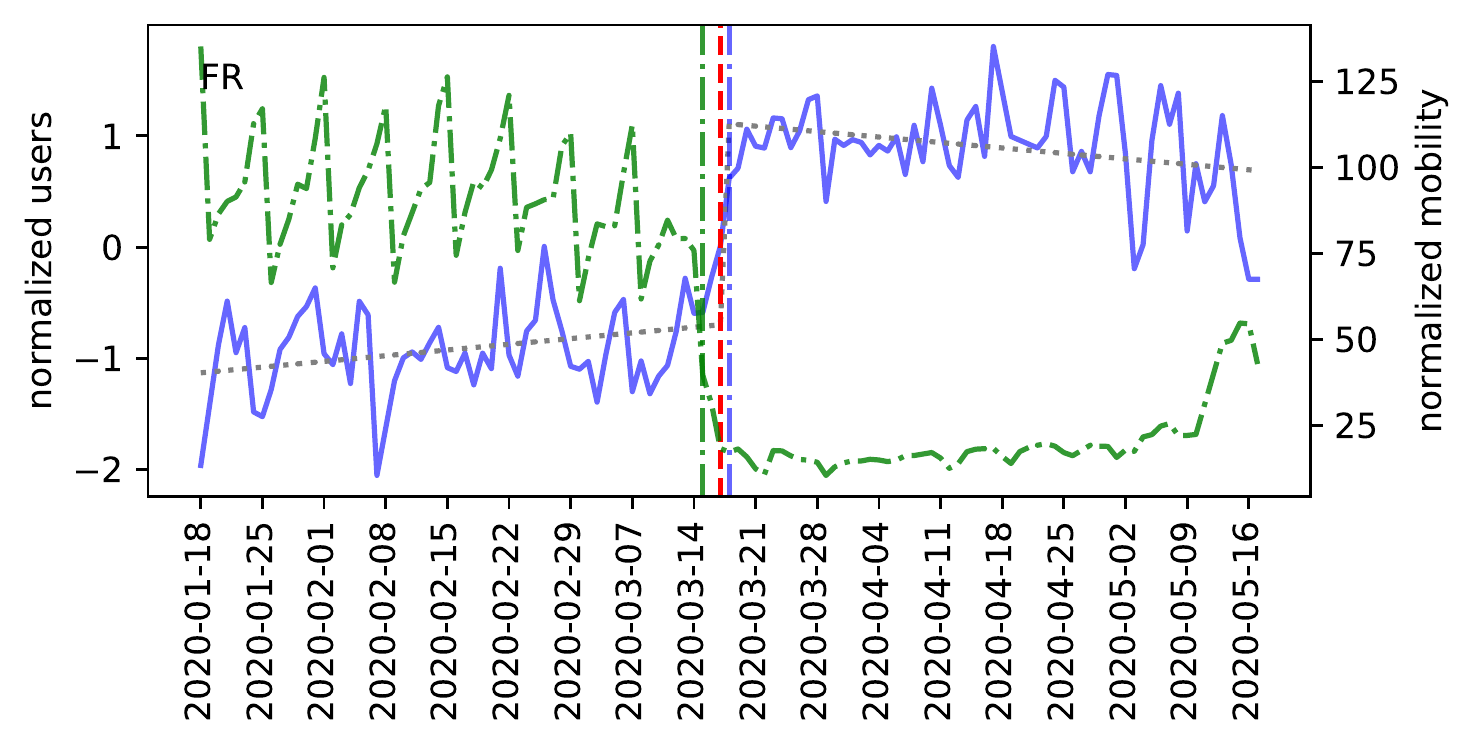}

\includegraphics[width=0.49\linewidth]{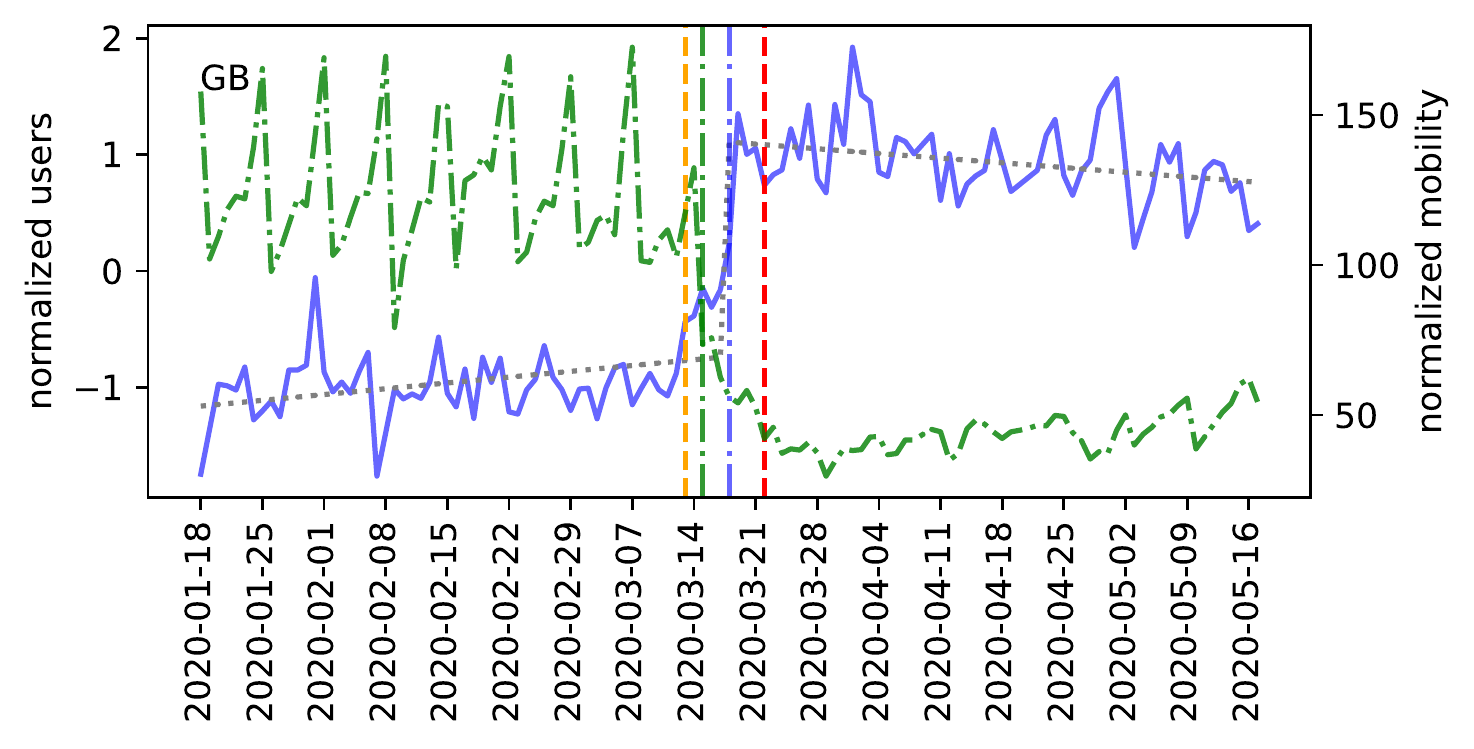}
    \vspace{0.1cm}
\includegraphics[width=0.49\linewidth]{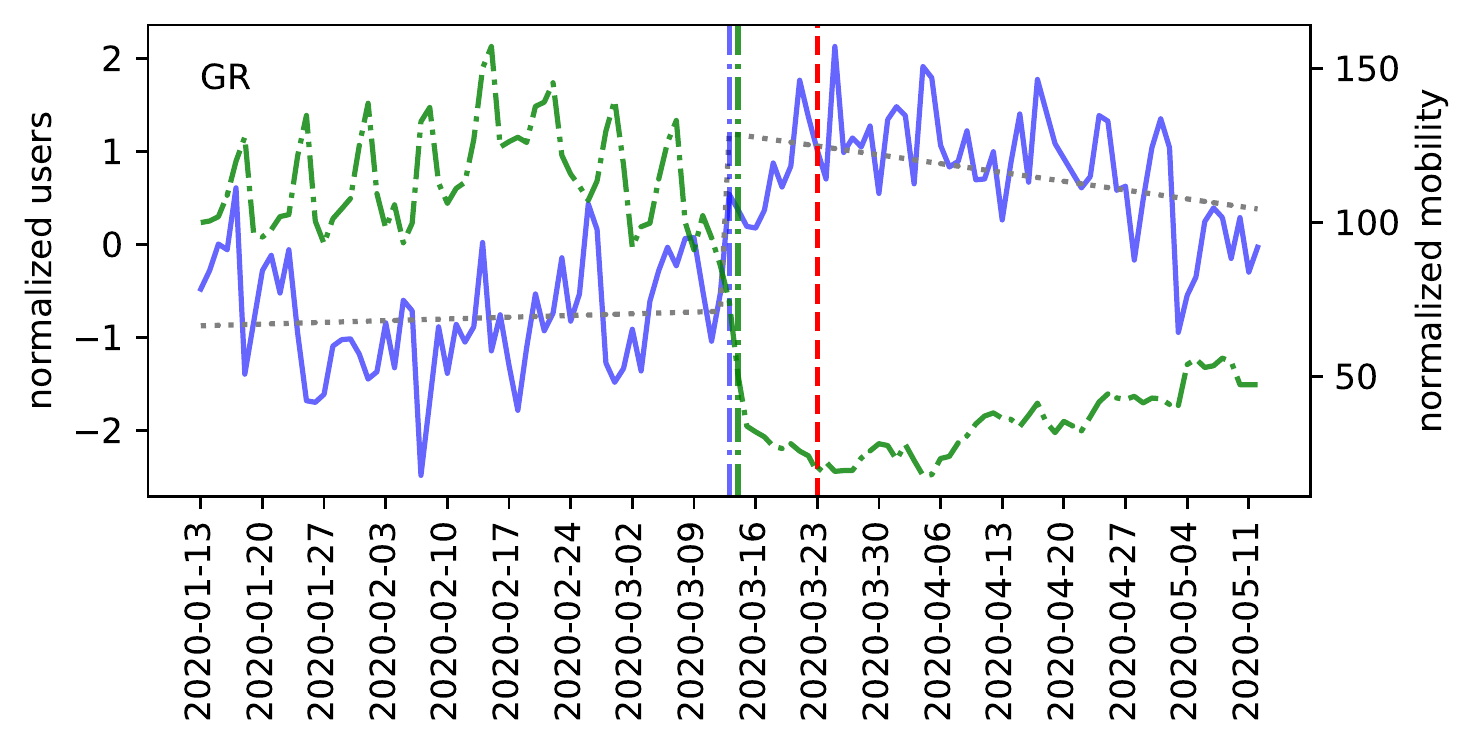} \\

\includegraphics[width=0.49\linewidth]{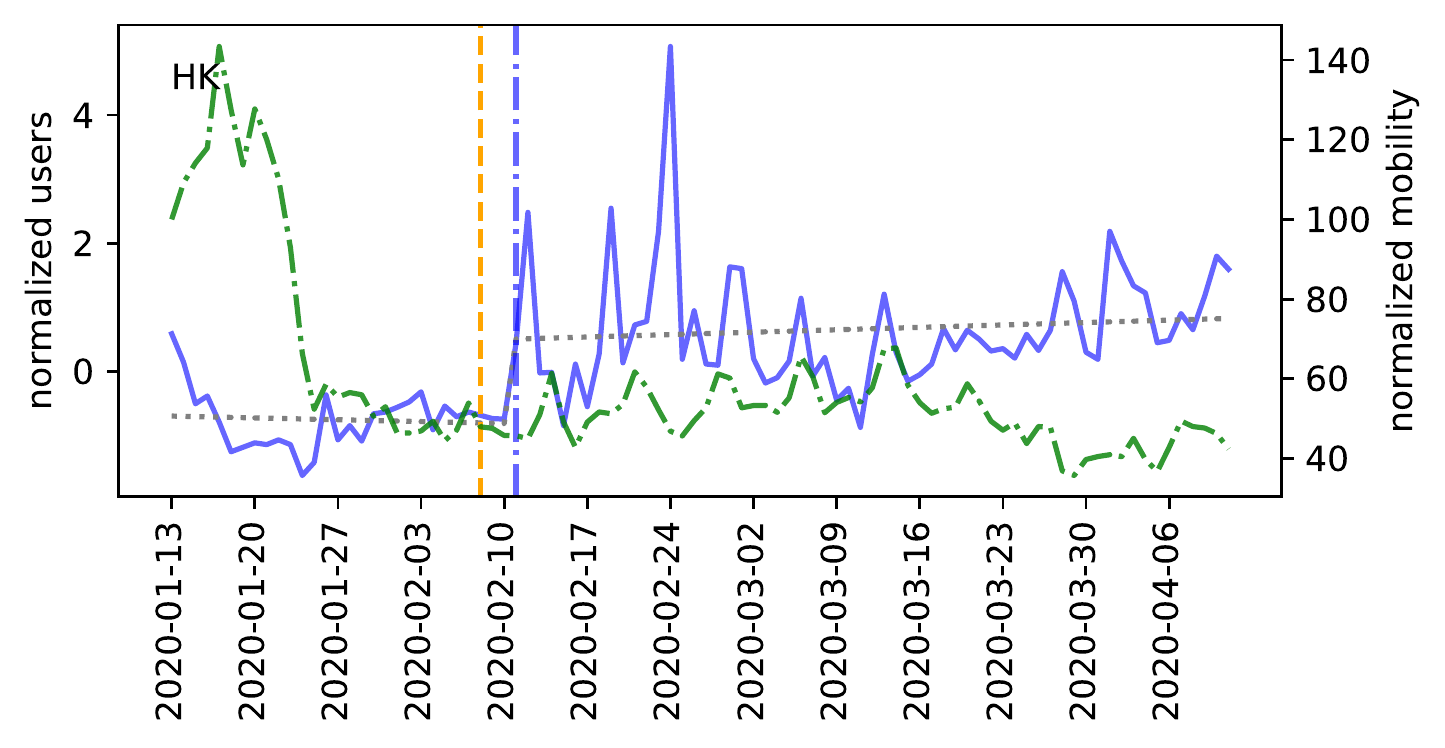}
    \vspace{0.1cm}
\includegraphics[width=0.49\linewidth]{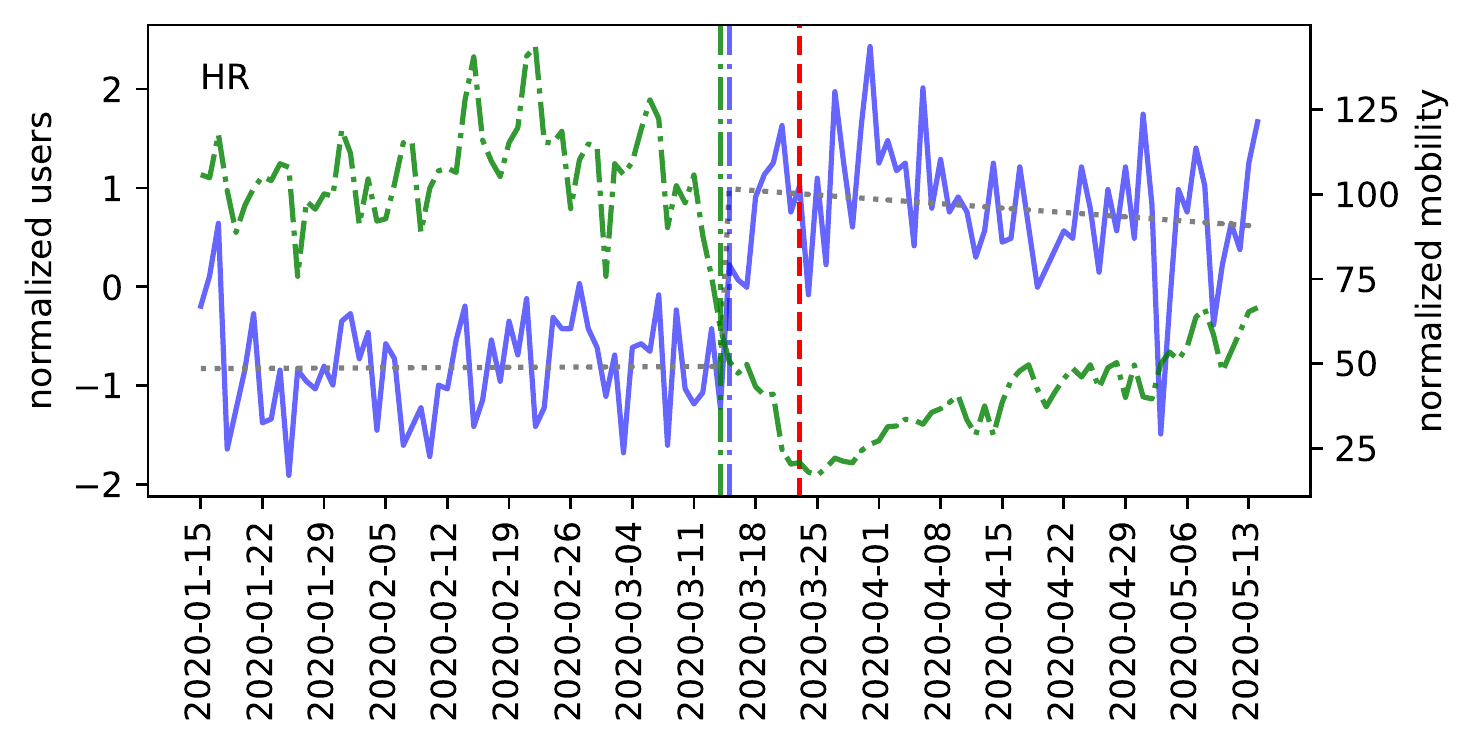}\\

\includegraphics[width=0.49\linewidth]{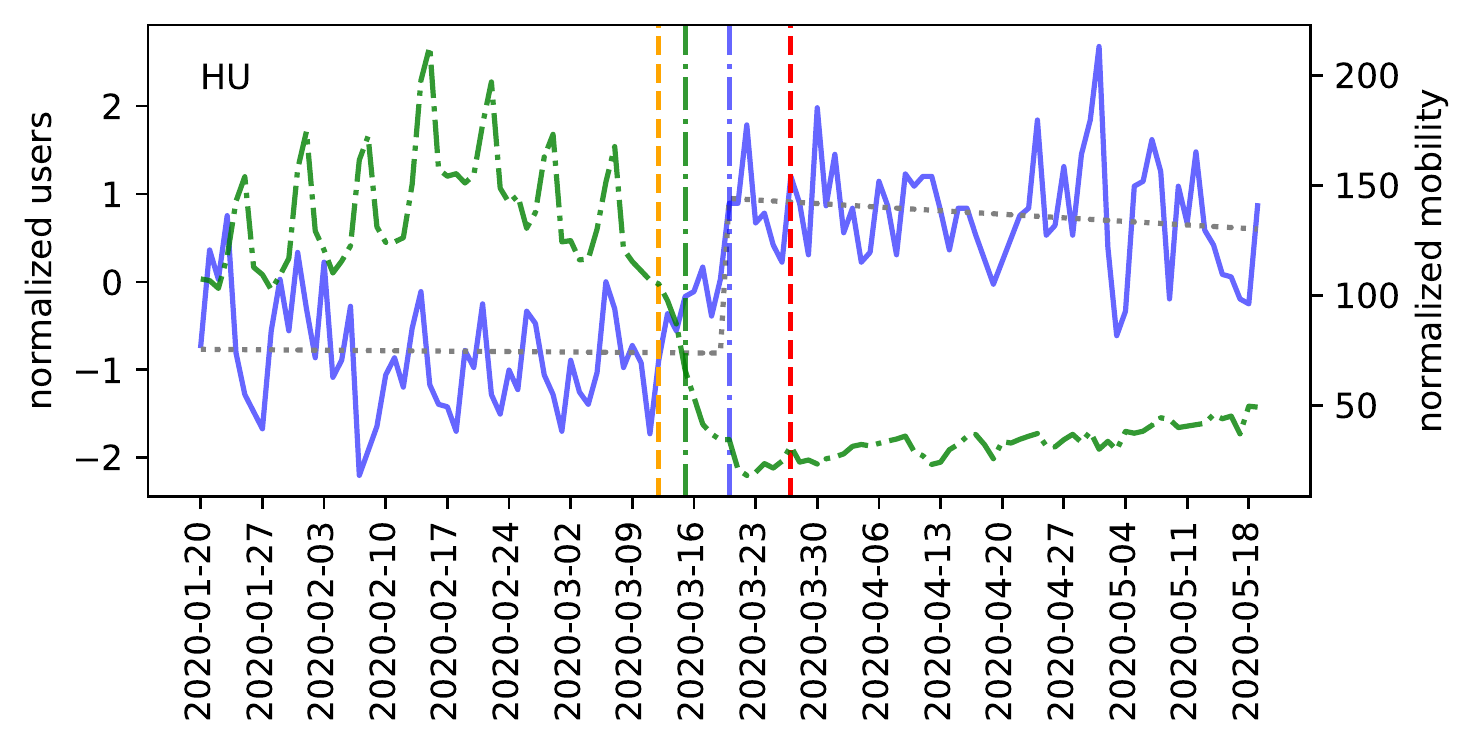}
    \vspace{0.1cm}
\includegraphics[width=0.49\linewidth]{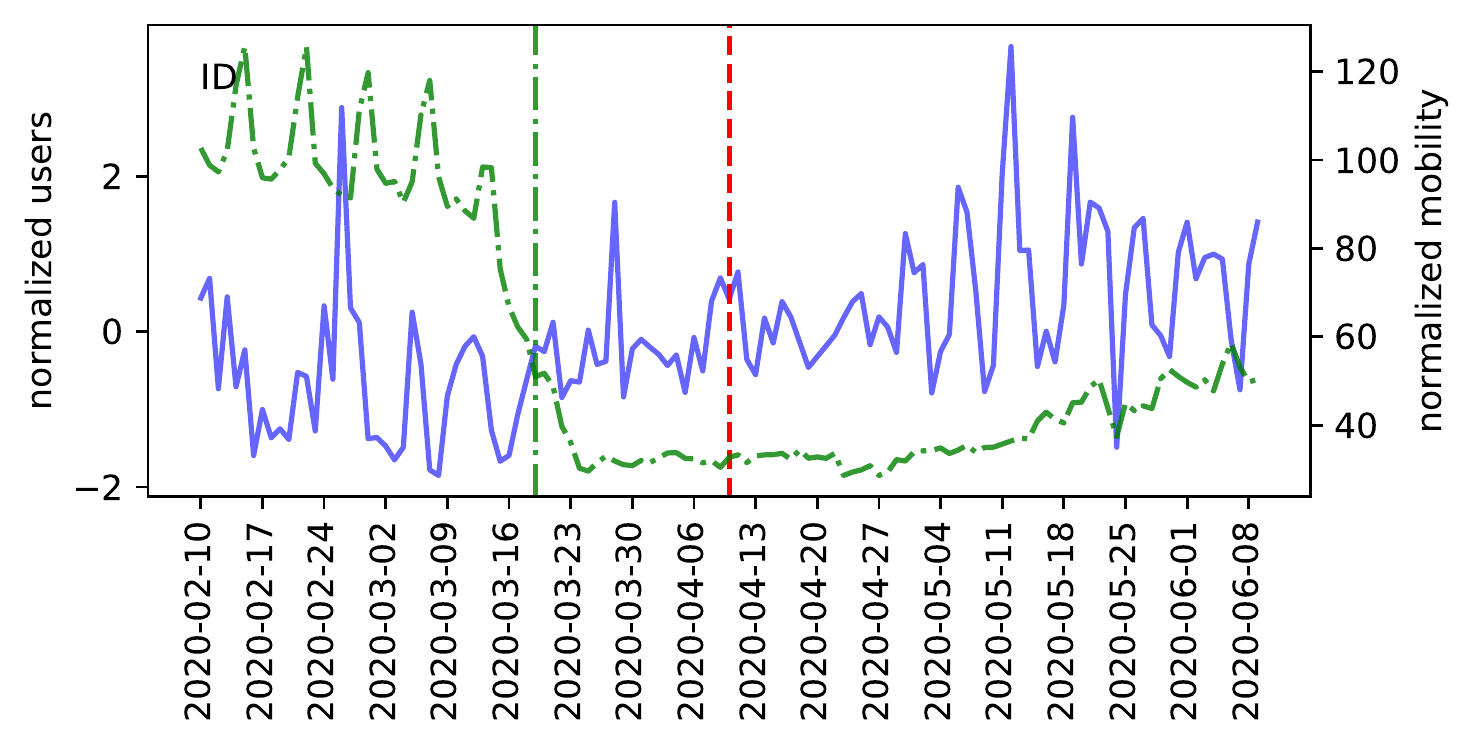}\\

\includegraphics[width=0.49\linewidth]{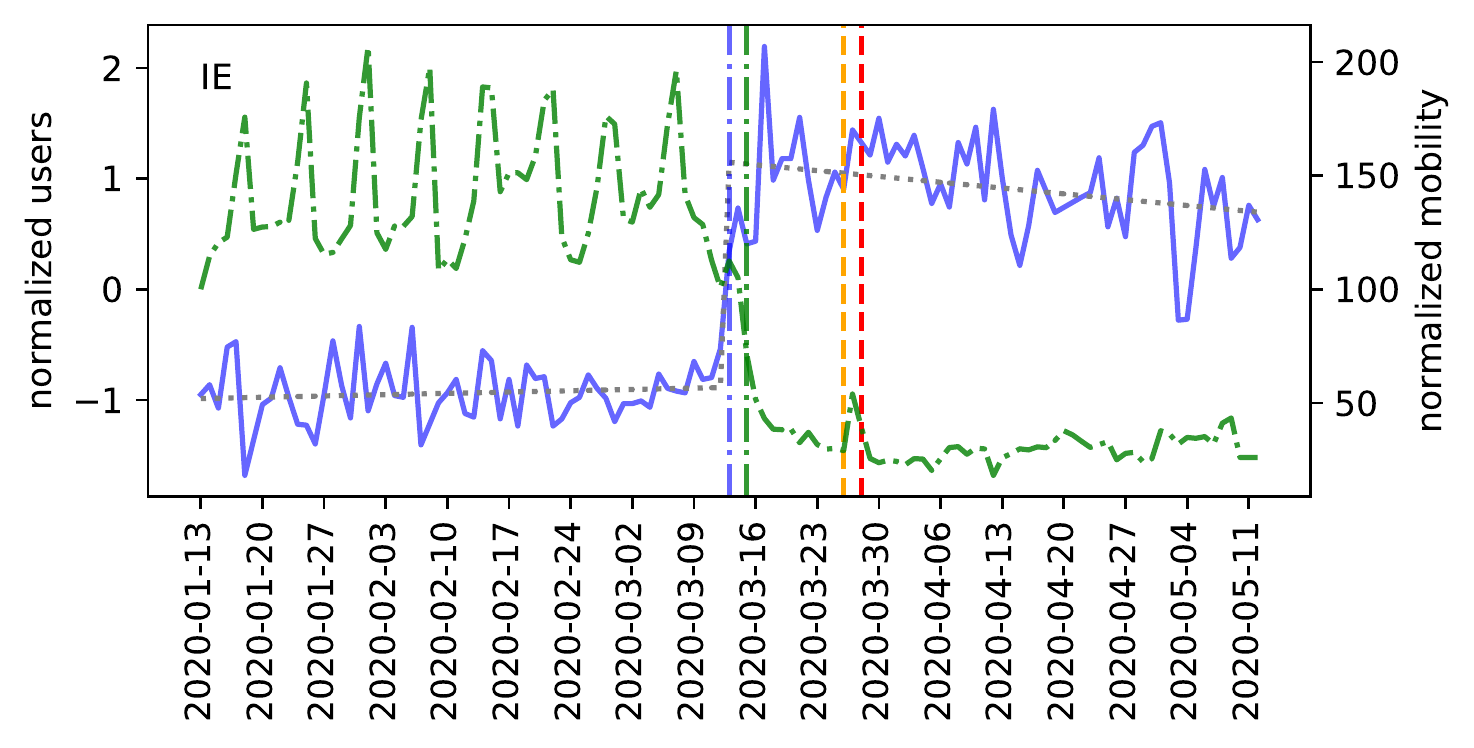}
    \vspace{0.1cm}
\includegraphics[width=0.49\linewidth]{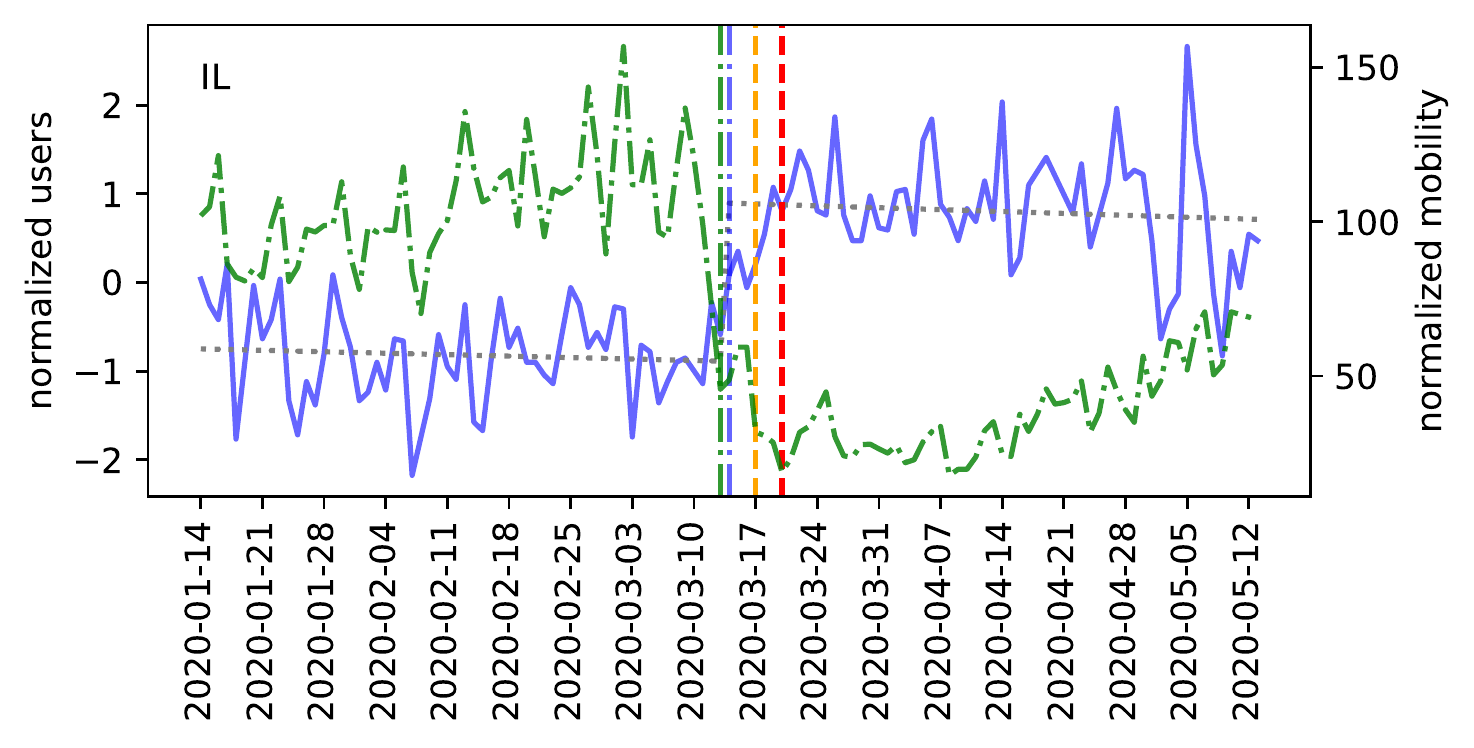}\\

\includegraphics[width=0.49\linewidth]{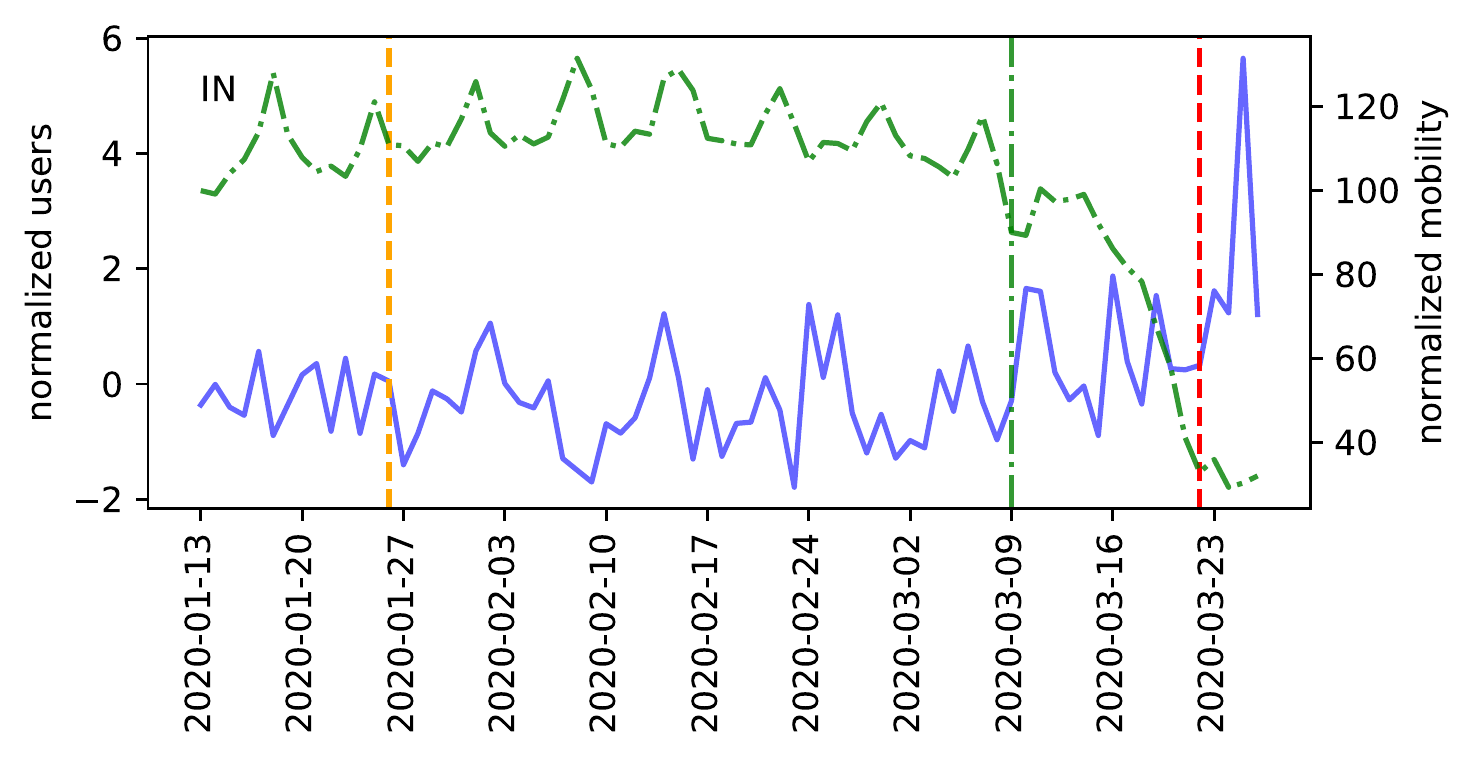}
    \vspace{0.1cm}
\includegraphics[width=0.49\linewidth]{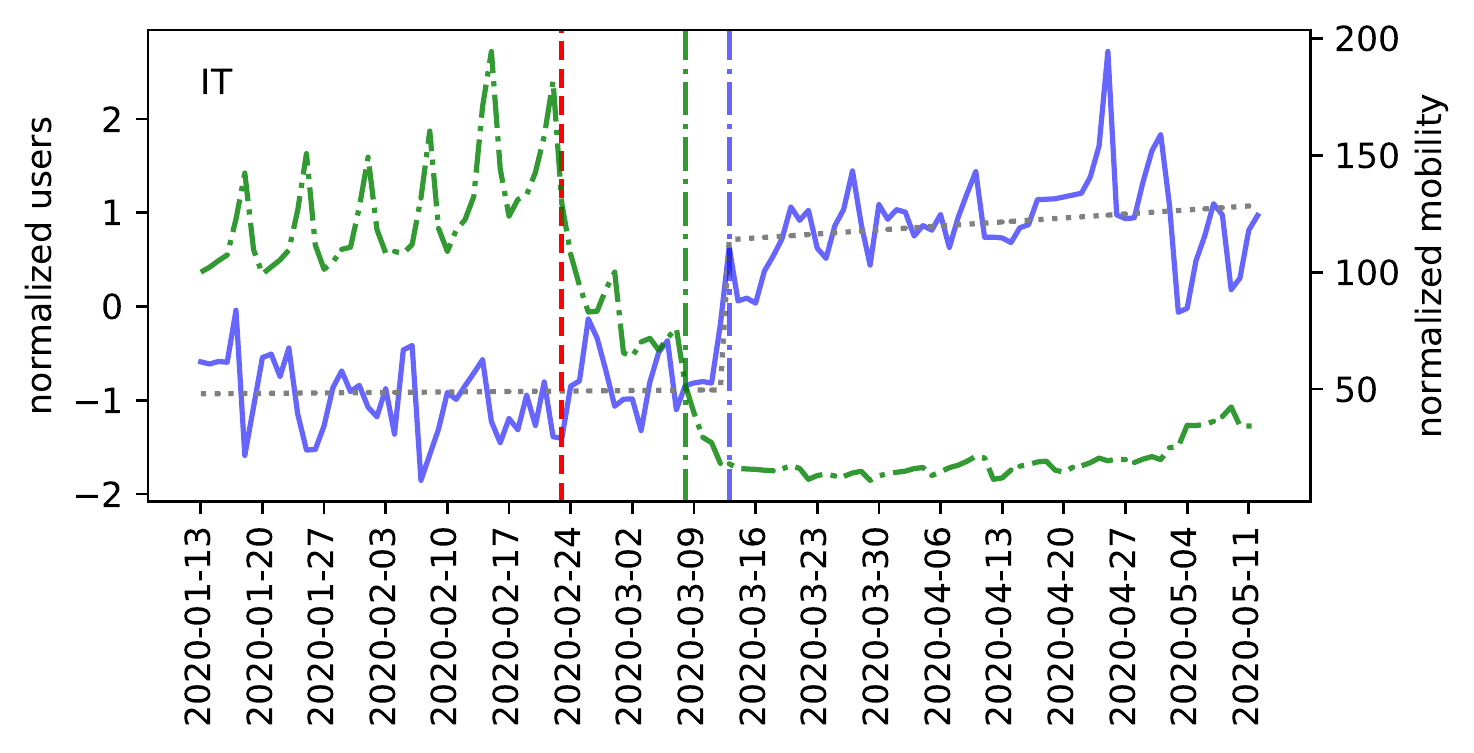}

\includegraphics[width=0.49\linewidth]{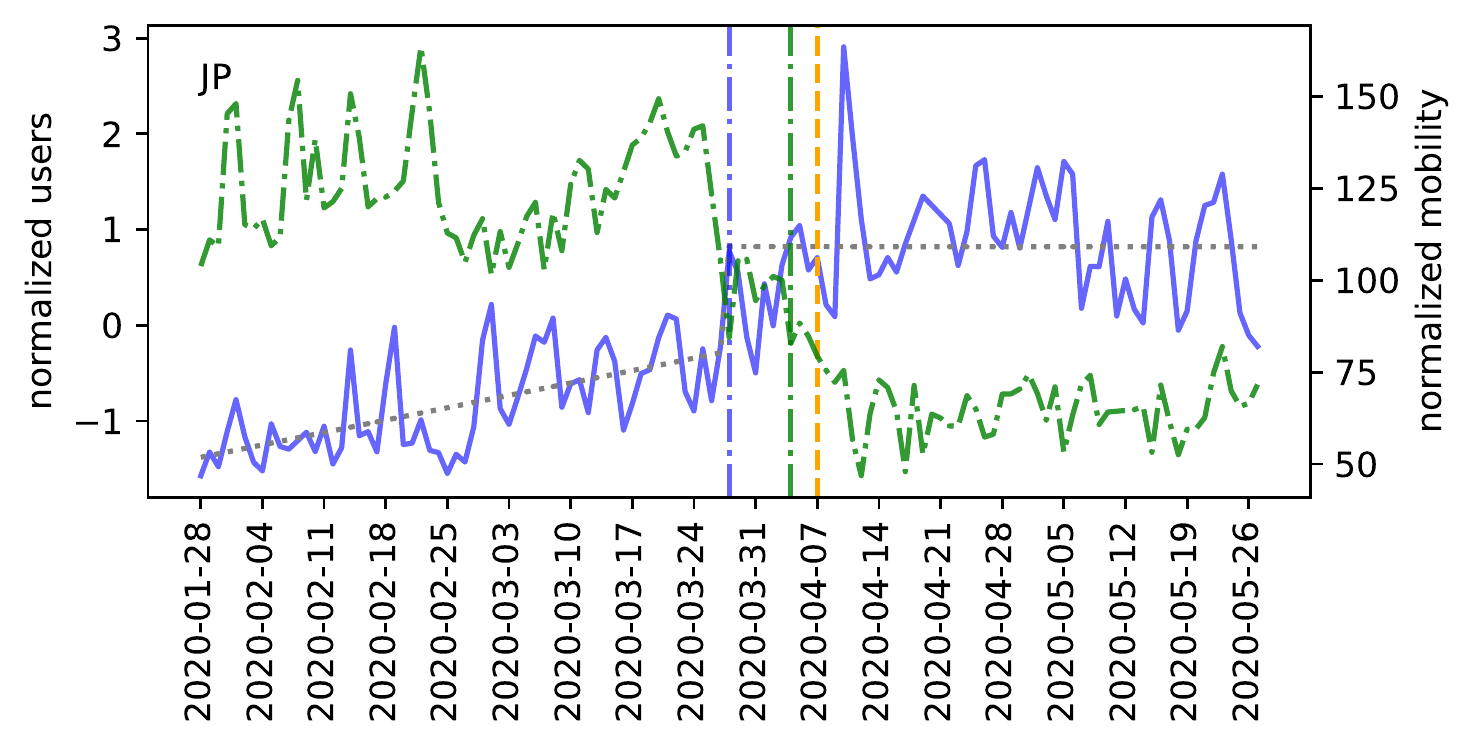}
    \vspace{0.1cm}
\includegraphics[width=0.49\linewidth]{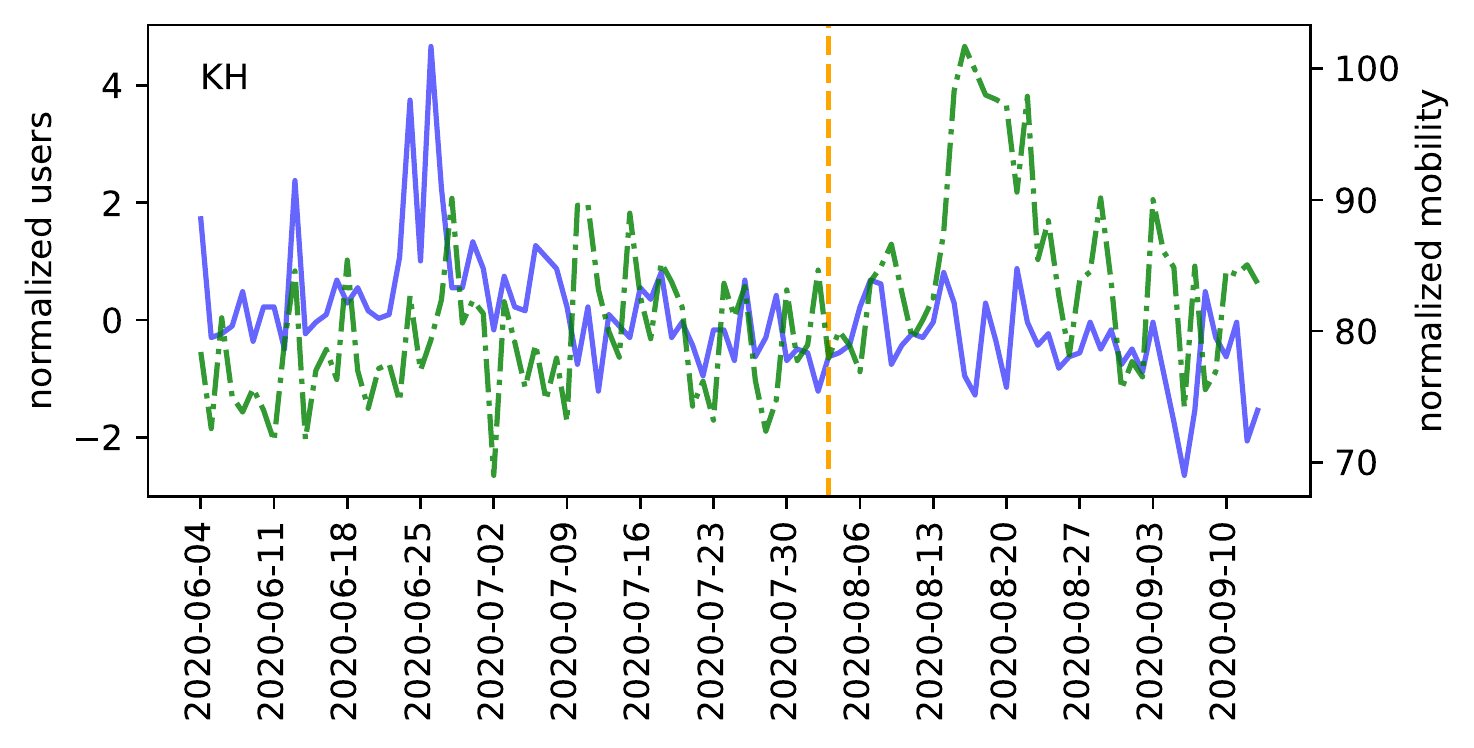} \\

\includegraphics[width=0.49\linewidth]{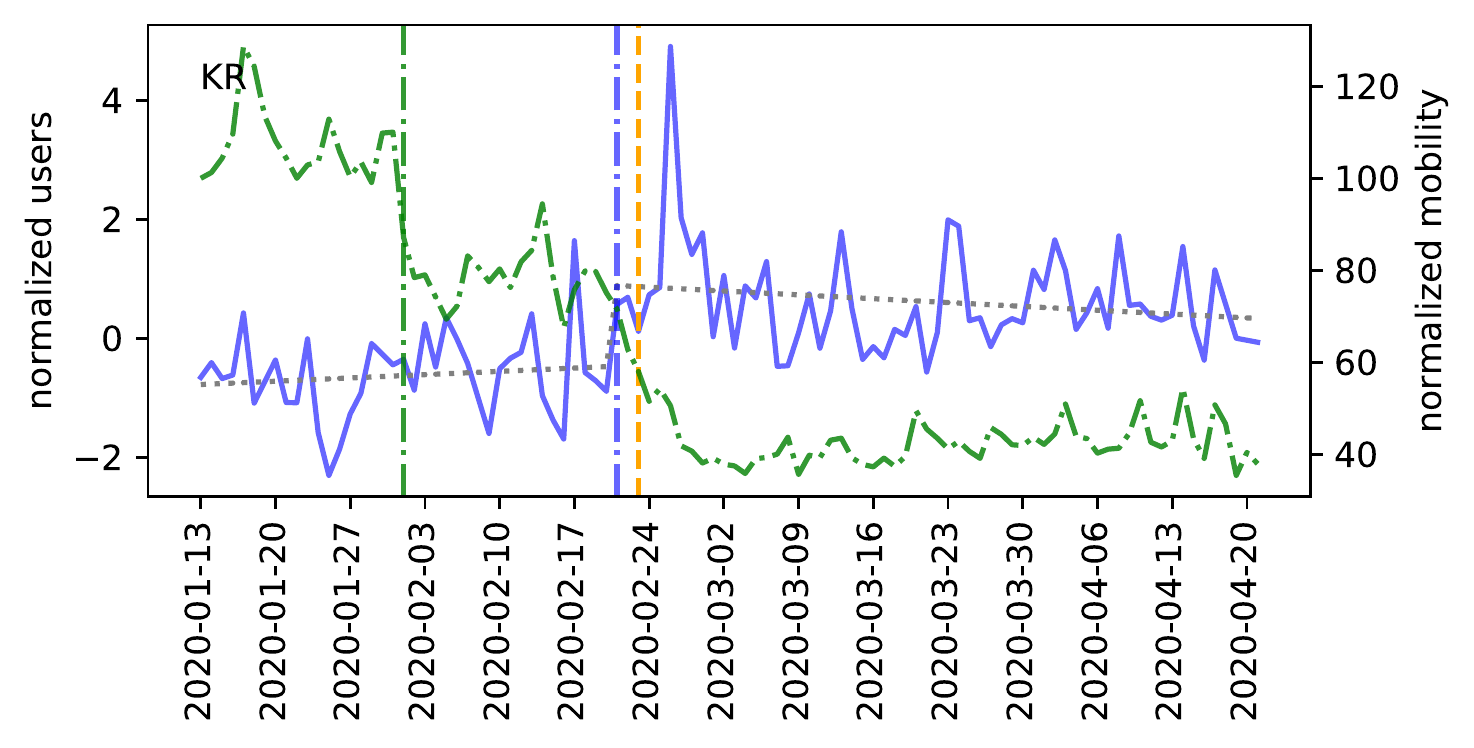}
    \vspace{0.1cm}
\includegraphics[width=0.49\linewidth]{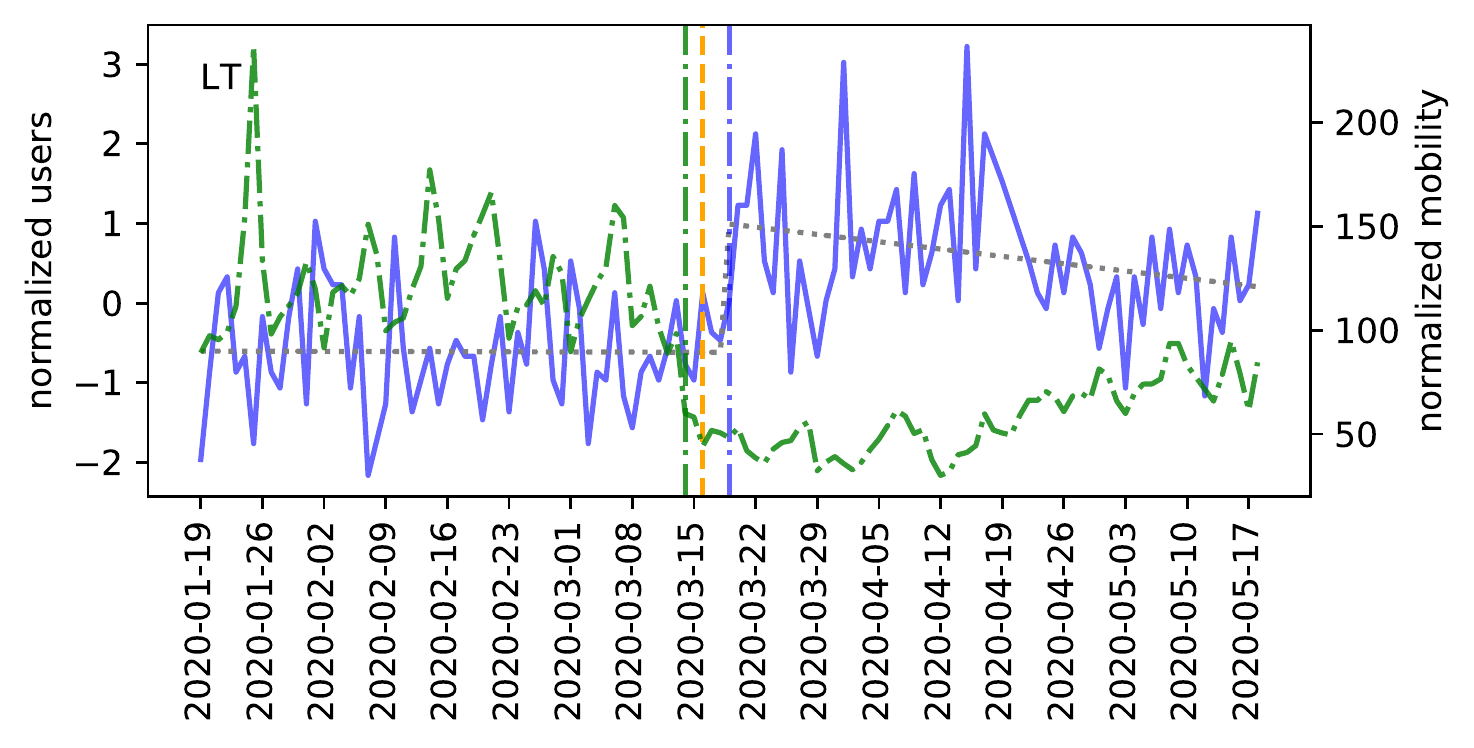}\\

\includegraphics[width=0.49\linewidth]{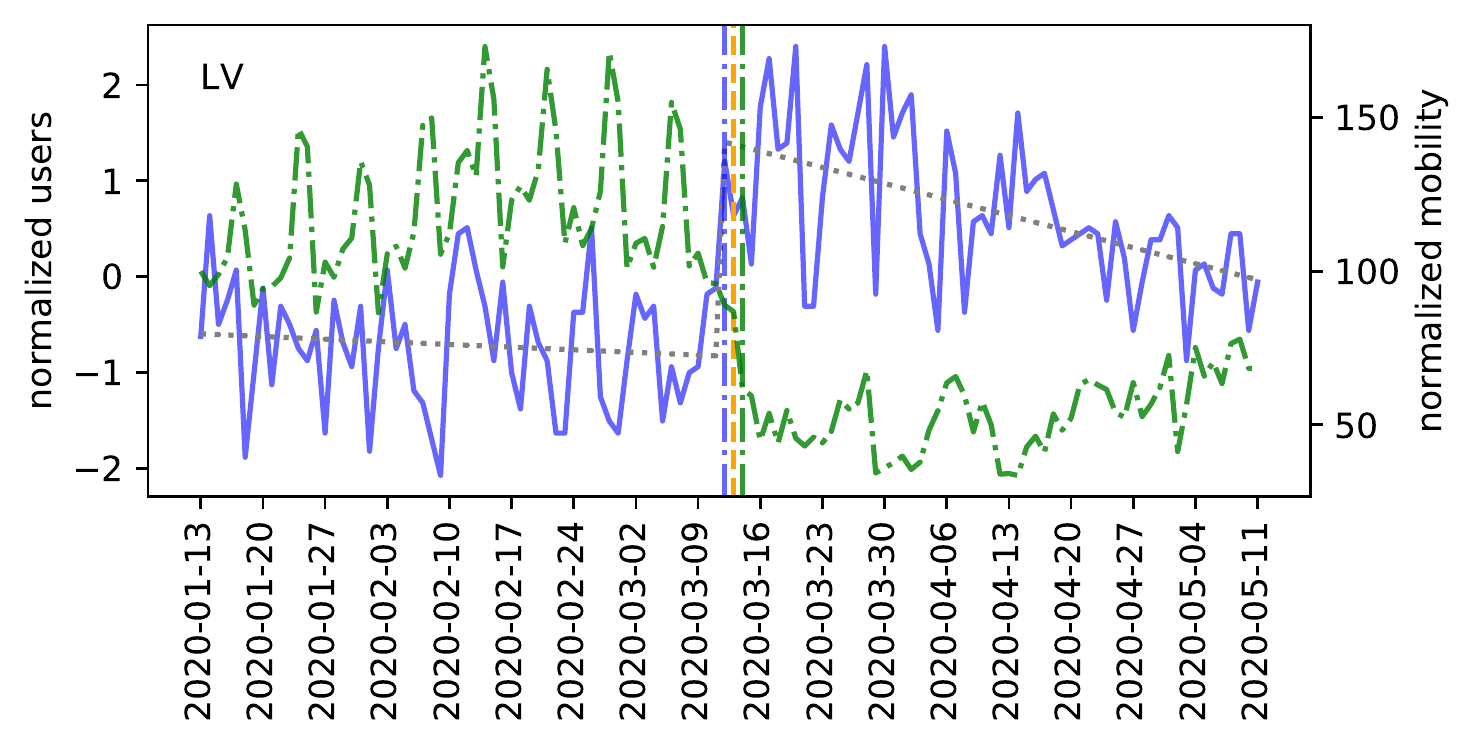}
    \vspace{0.1cm}
\includegraphics[width=0.49\linewidth]{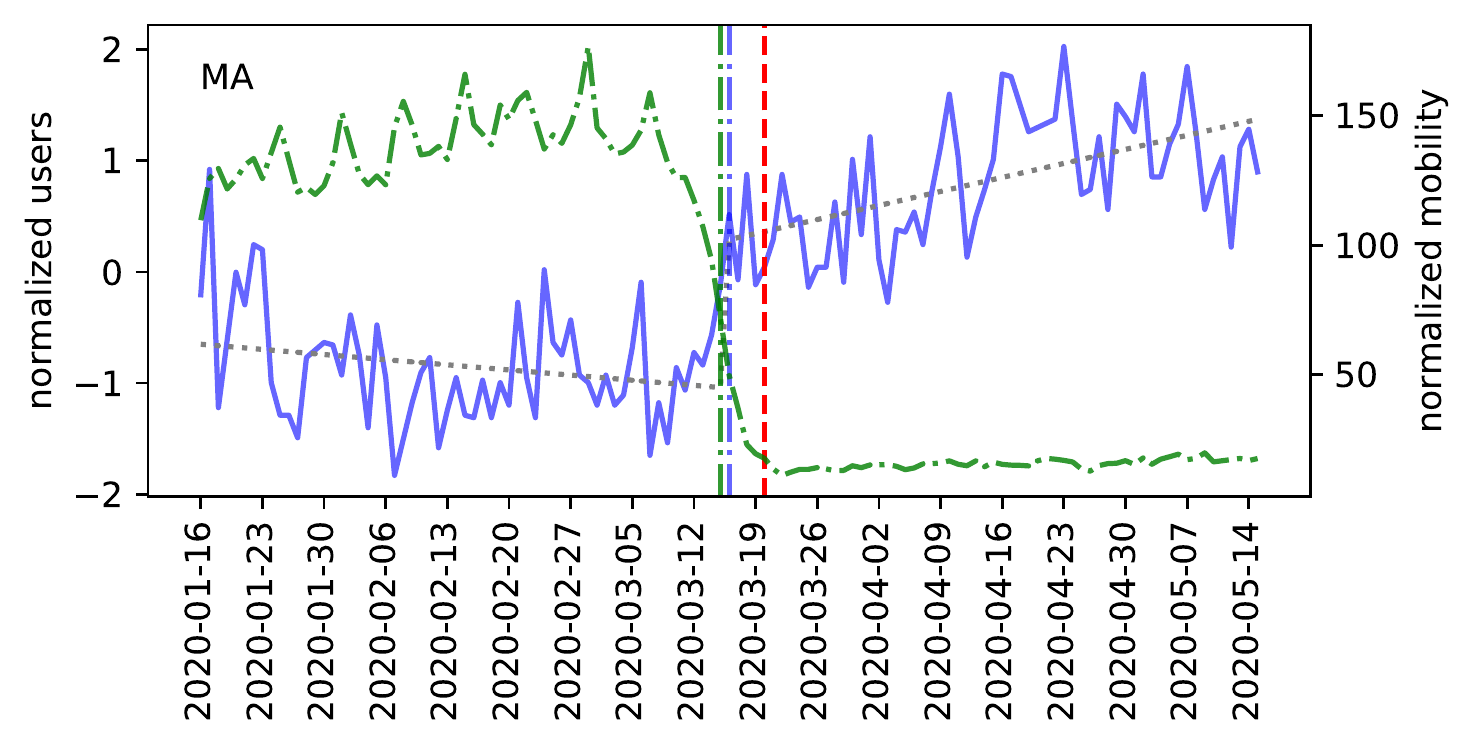}\\

\includegraphics[width=0.49\linewidth]{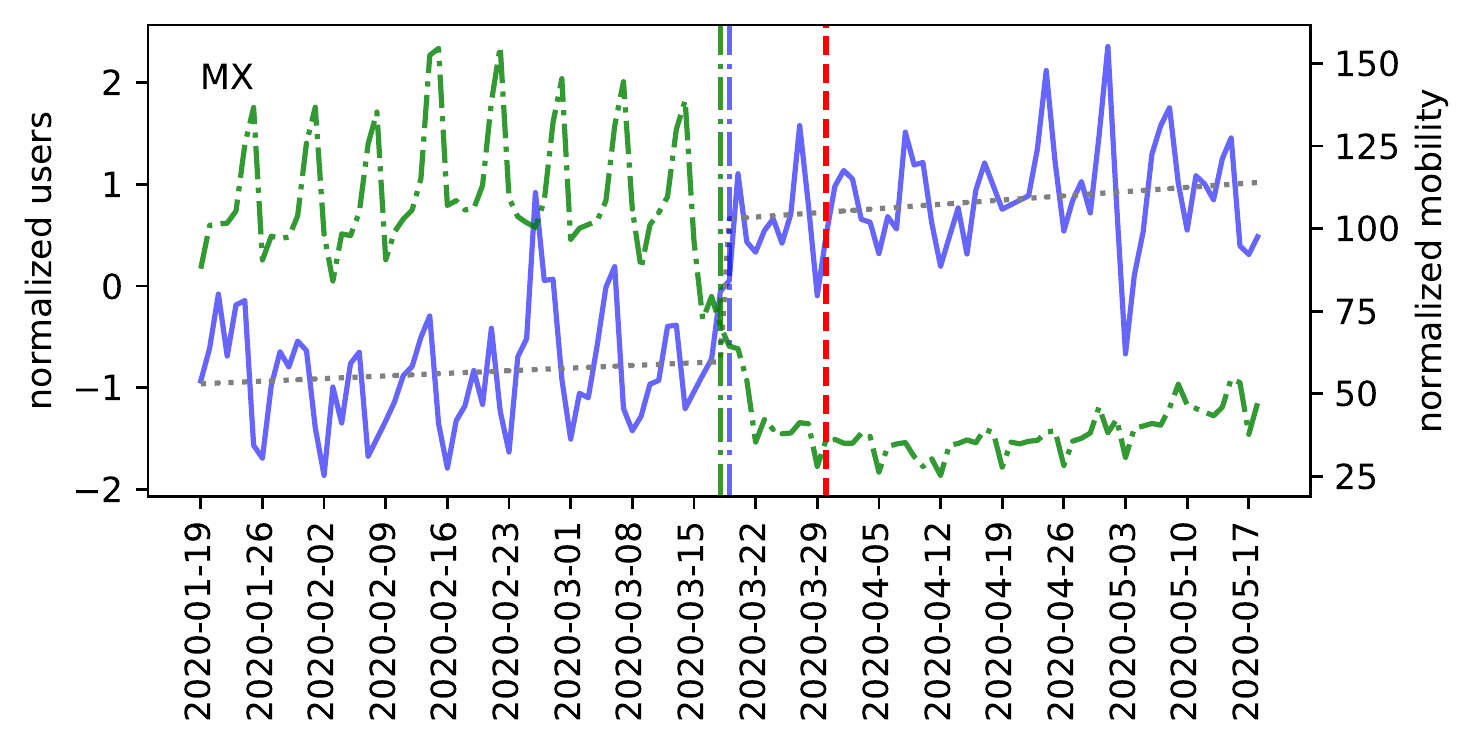}
    \vspace{0.1cm}
\includegraphics[width=0.49\linewidth]{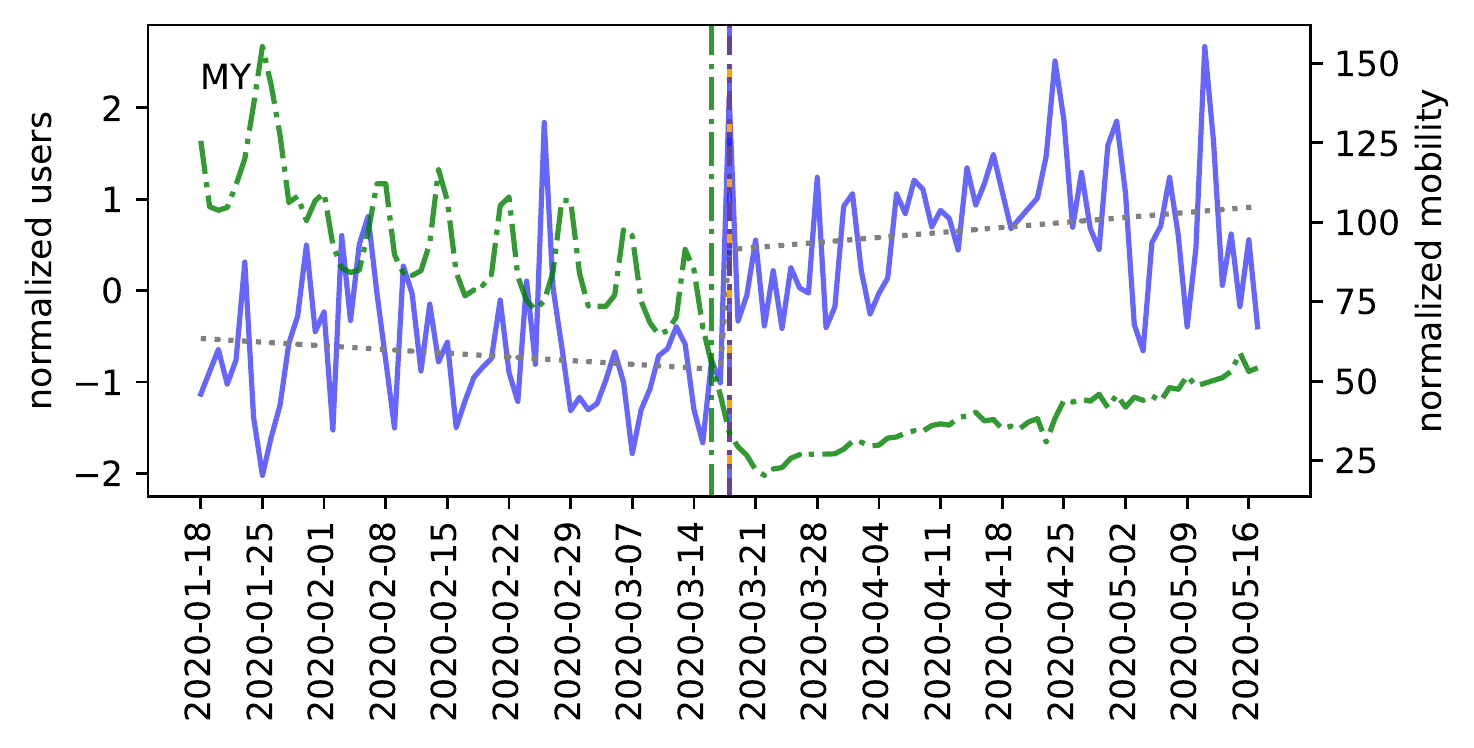}\\

\includegraphics[width=0.49\linewidth]{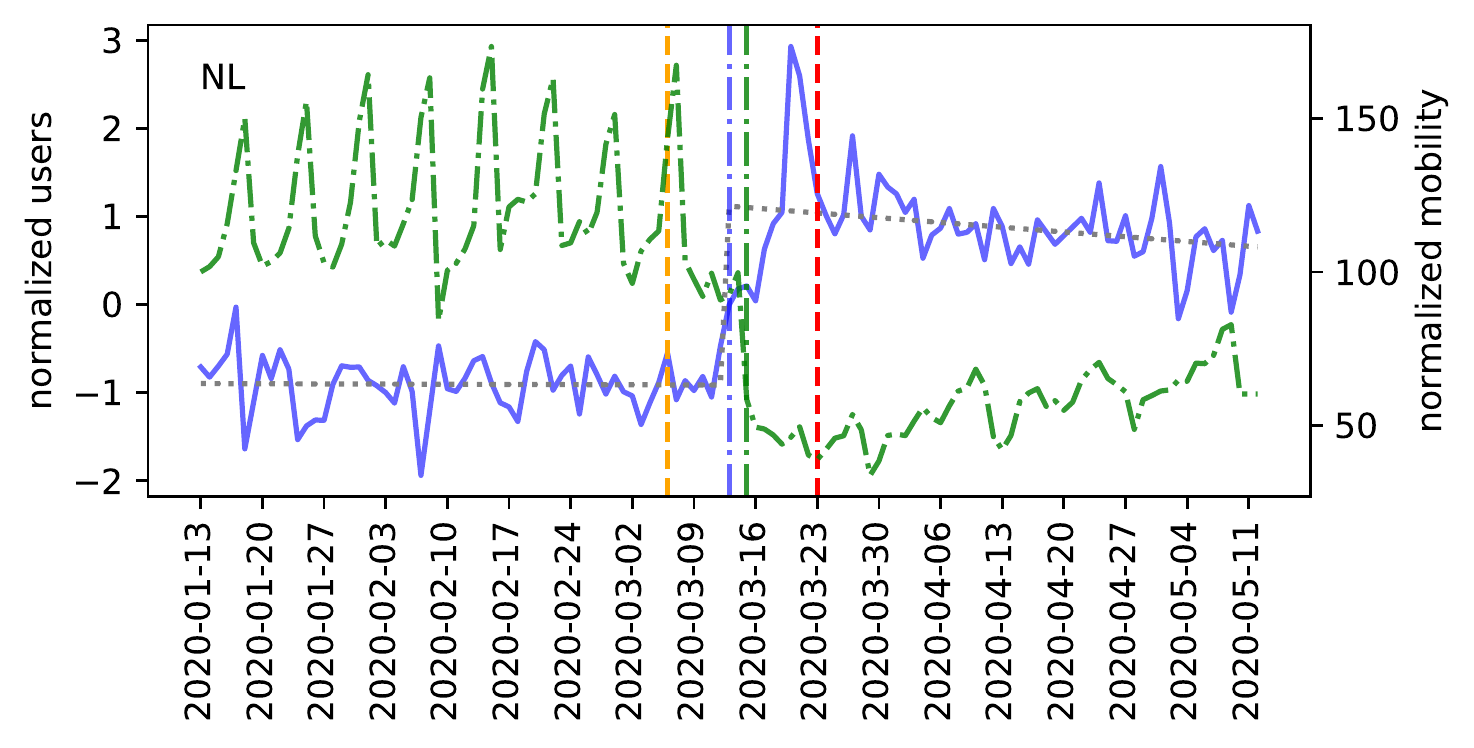}
    \vspace{0.1cm}
\includegraphics[width=0.49\linewidth]{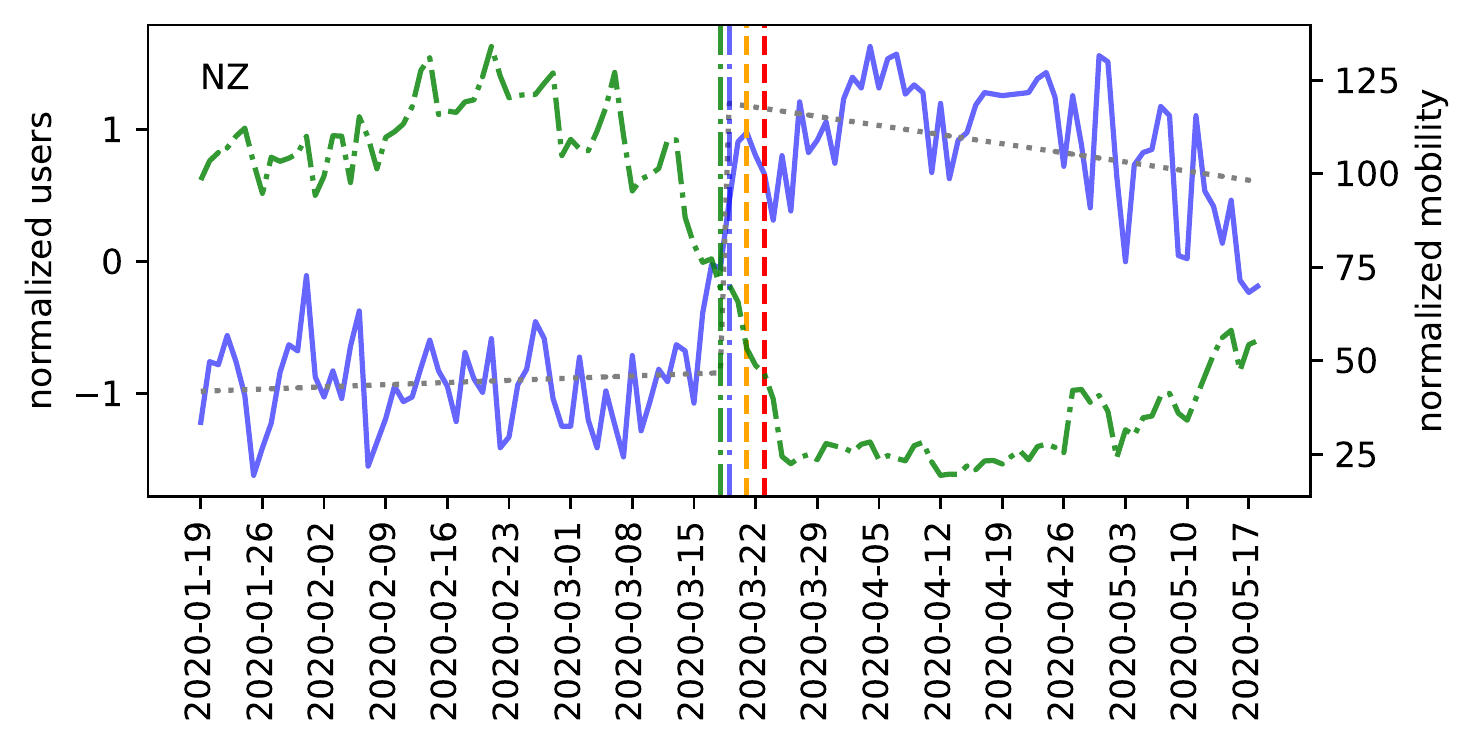}

\includegraphics[width=0.49\linewidth]{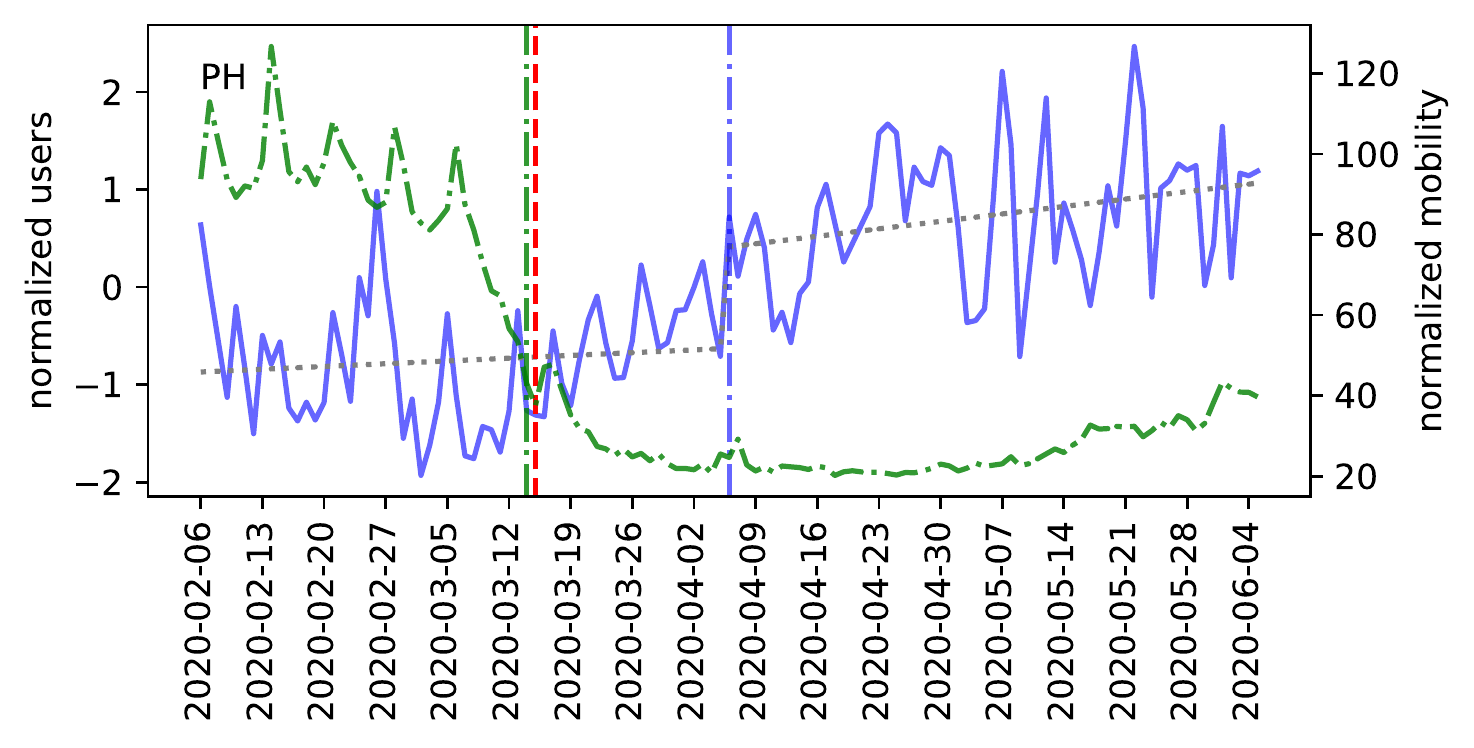}
    \vspace{0.1cm}
\includegraphics[width=0.49\linewidth]{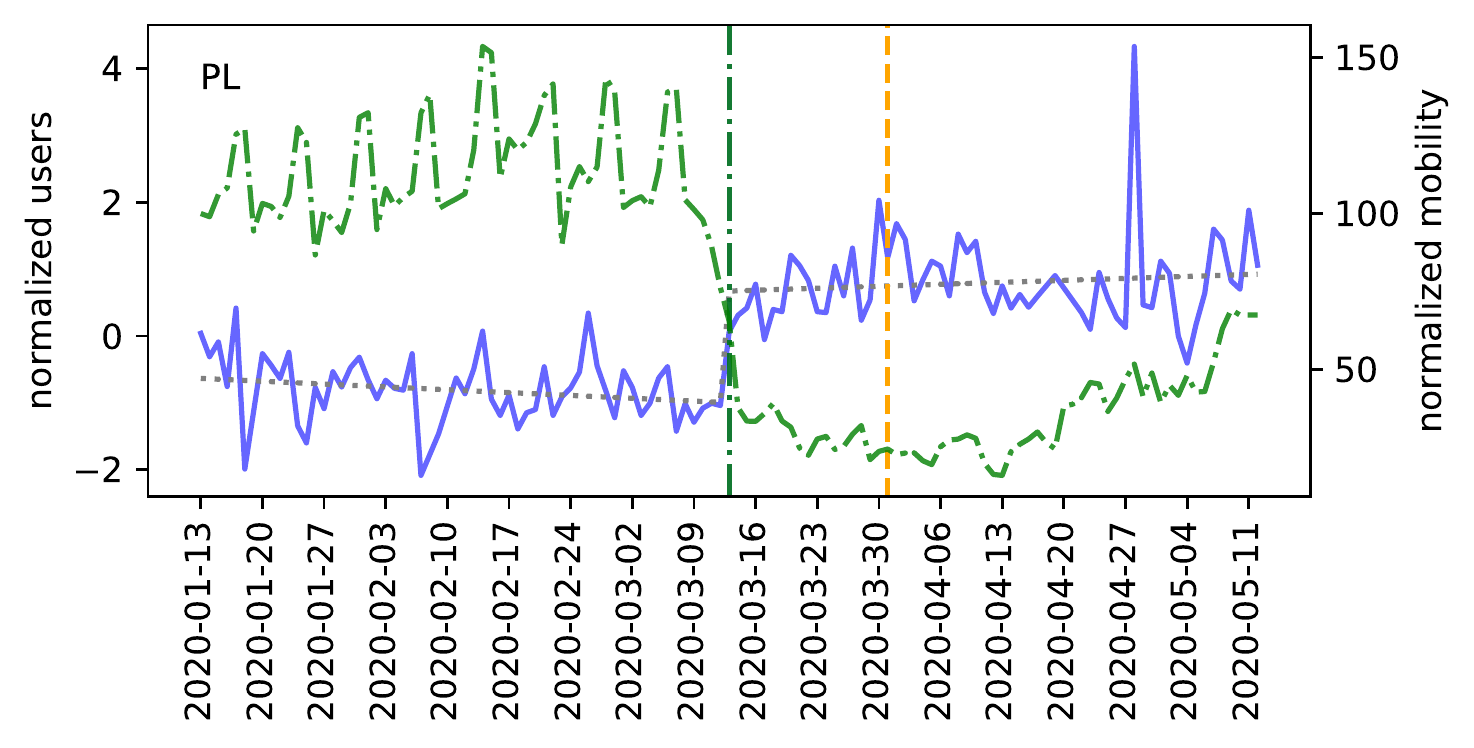} \\

\includegraphics[width=0.49\linewidth]{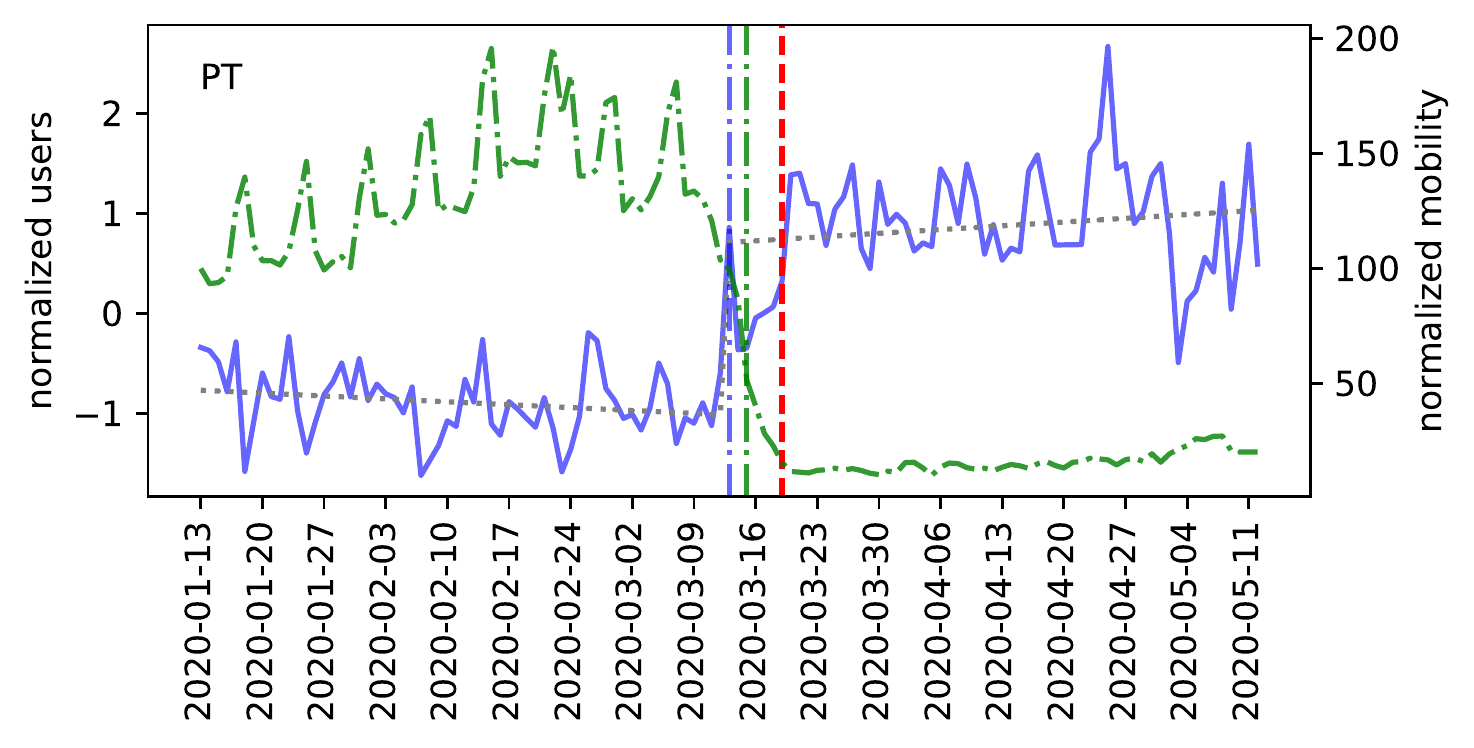}
    \vspace{0.1cm}
\includegraphics[width=0.49\linewidth]{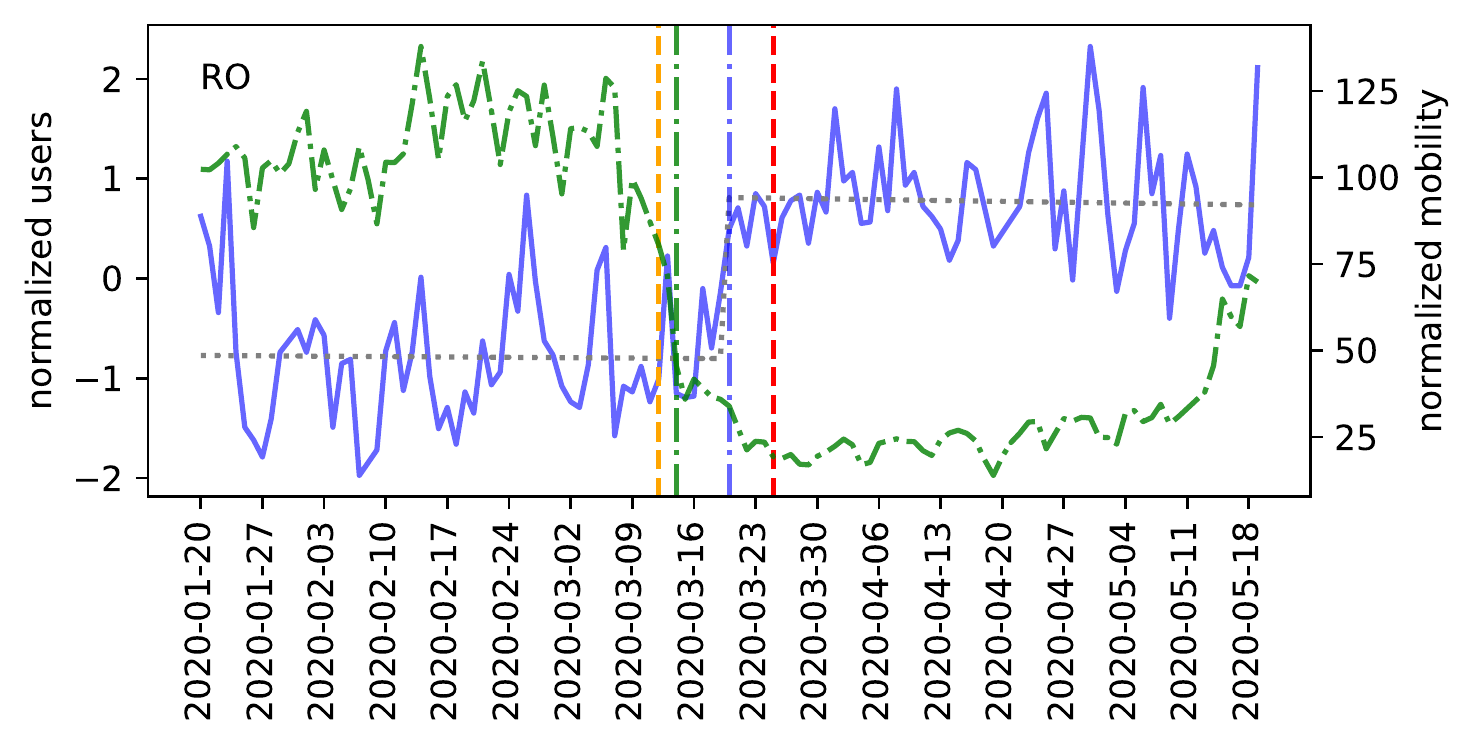}\\

\includegraphics[width=0.49\linewidth]{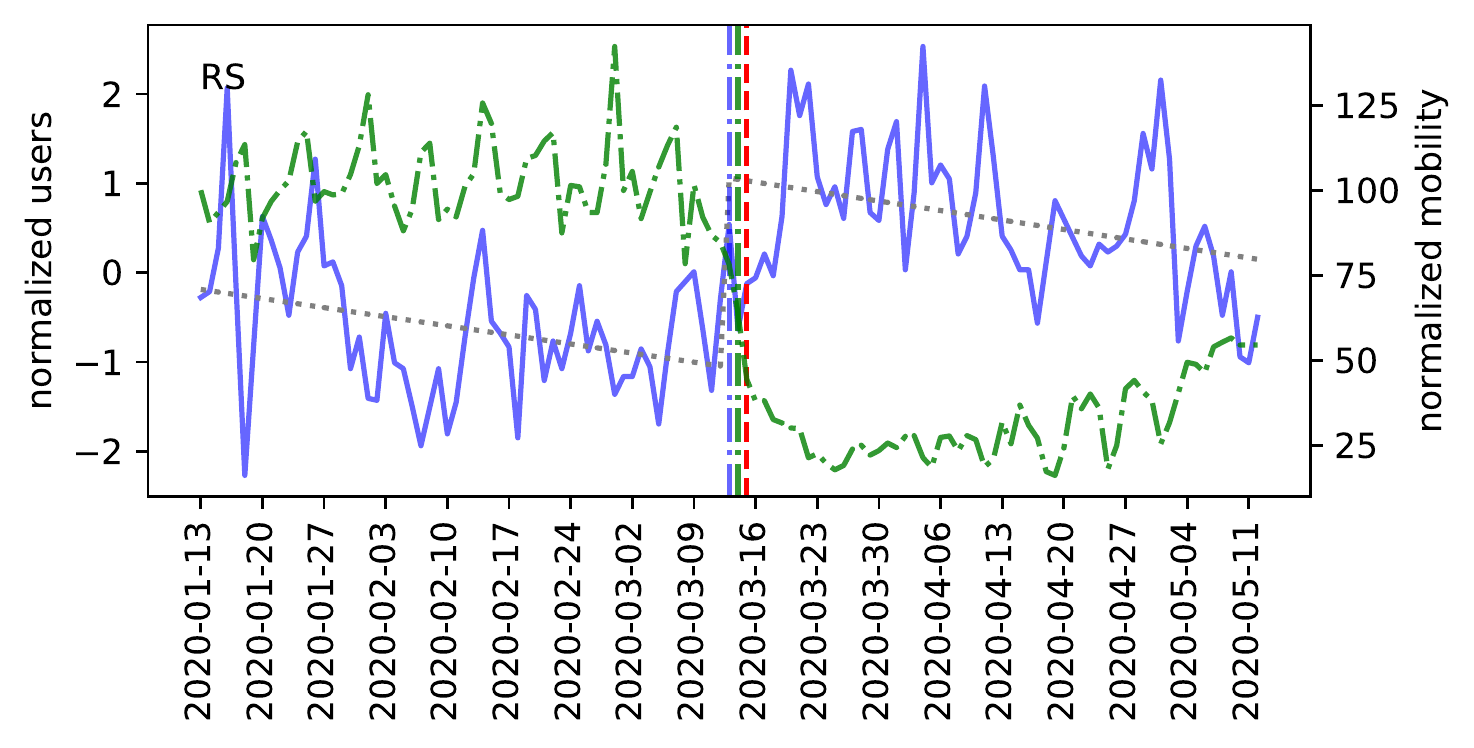}
    \vspace{0.1cm}
\includegraphics[width=0.49\linewidth]{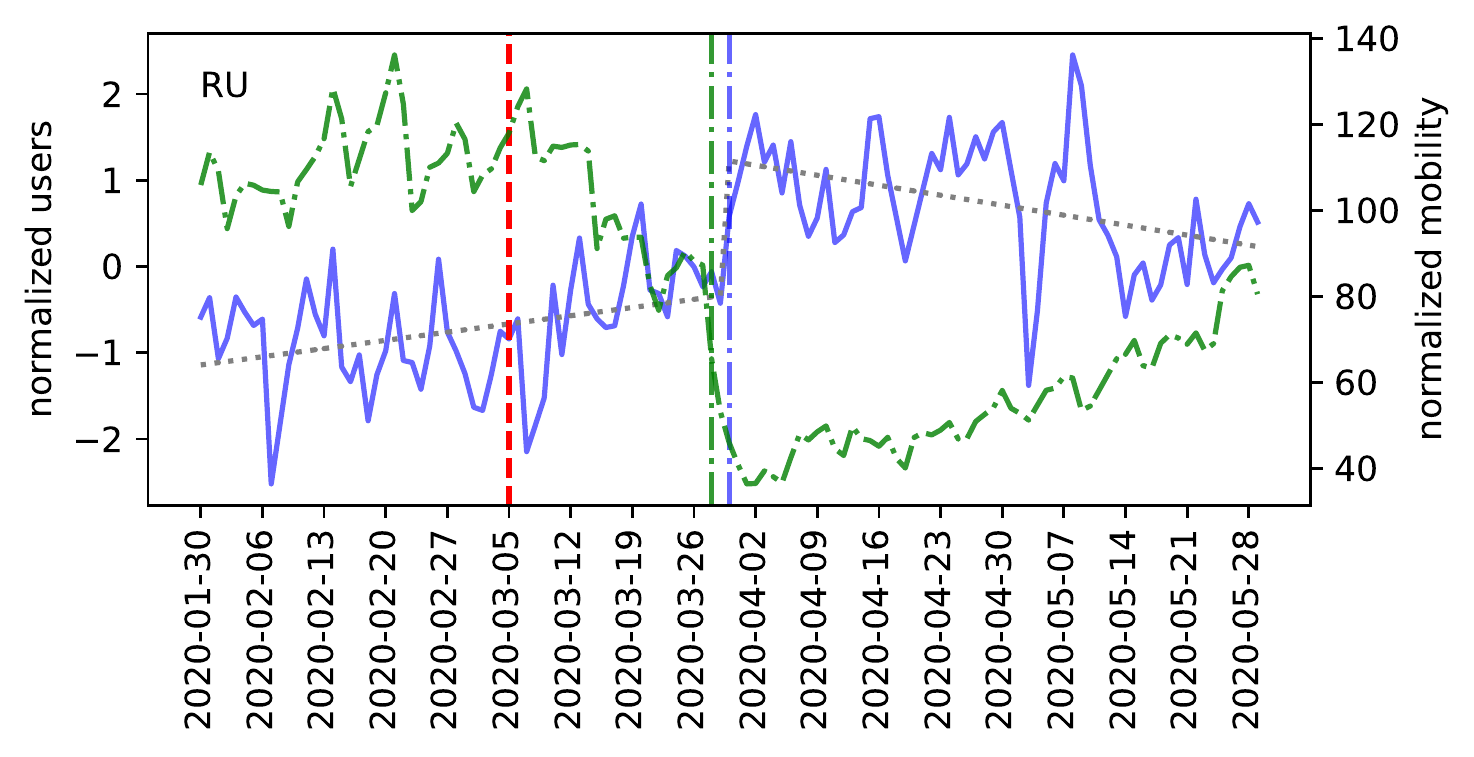}\\

\includegraphics[width=0.49\linewidth]{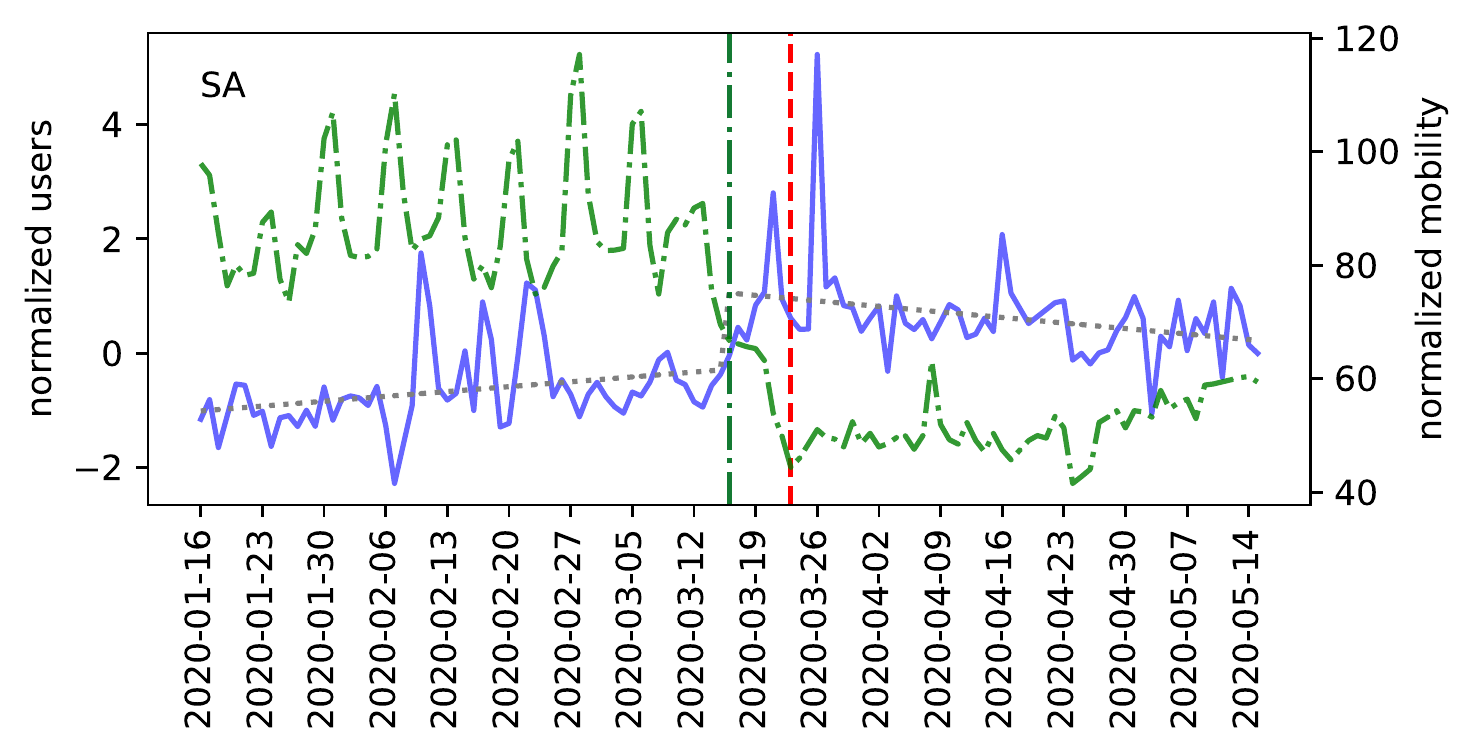}
    \vspace{0.1cm}
\includegraphics[width=0.49\linewidth]{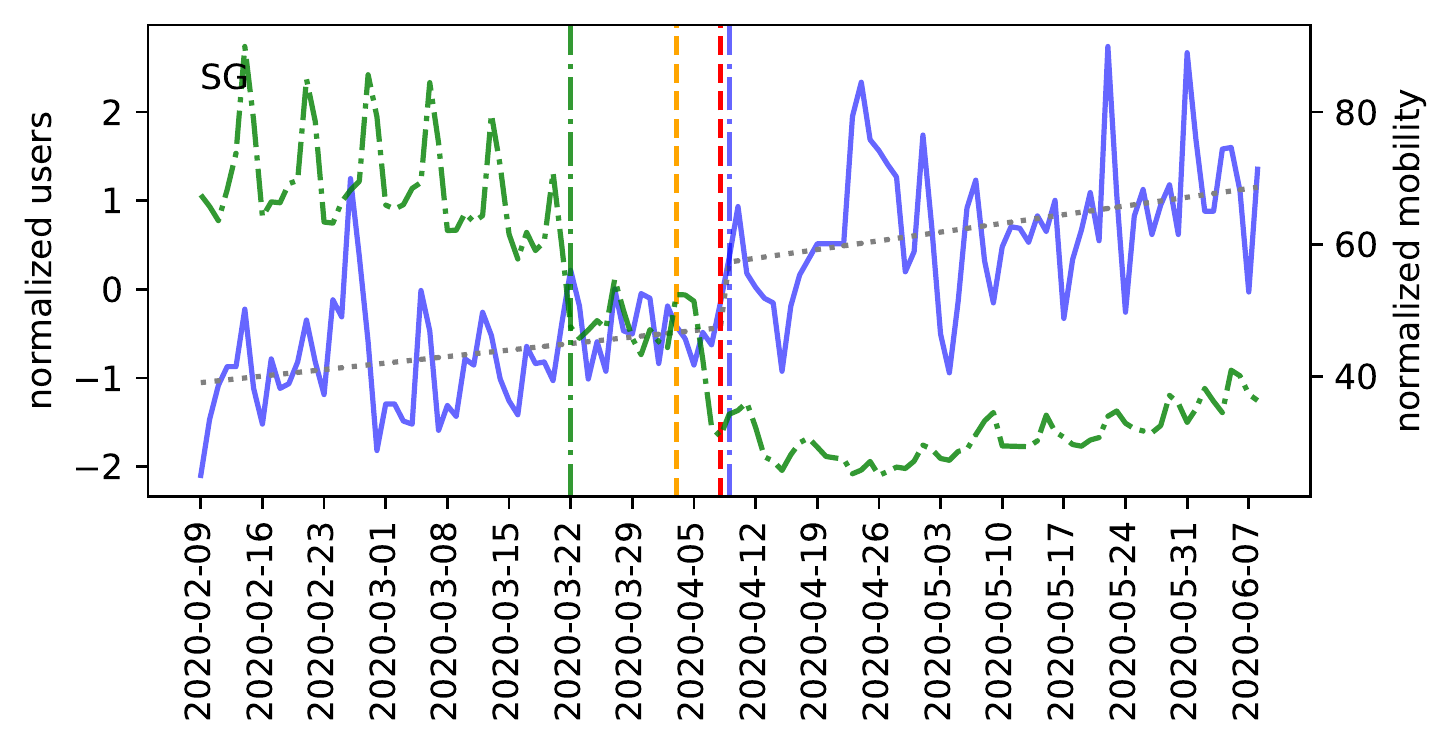}\\

\includegraphics[width=0.49\linewidth]{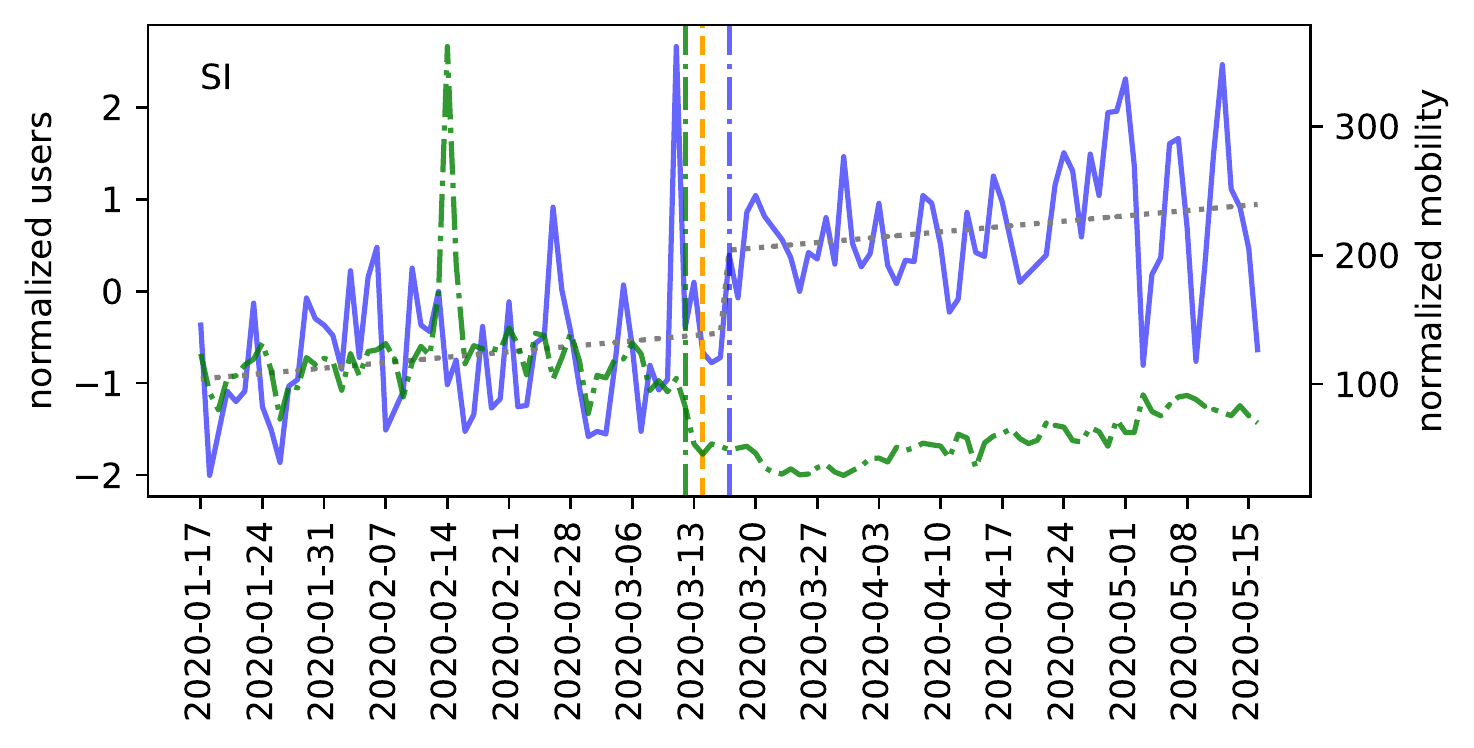}
    \vspace{0.1cm}
\includegraphics[width=0.49\linewidth]{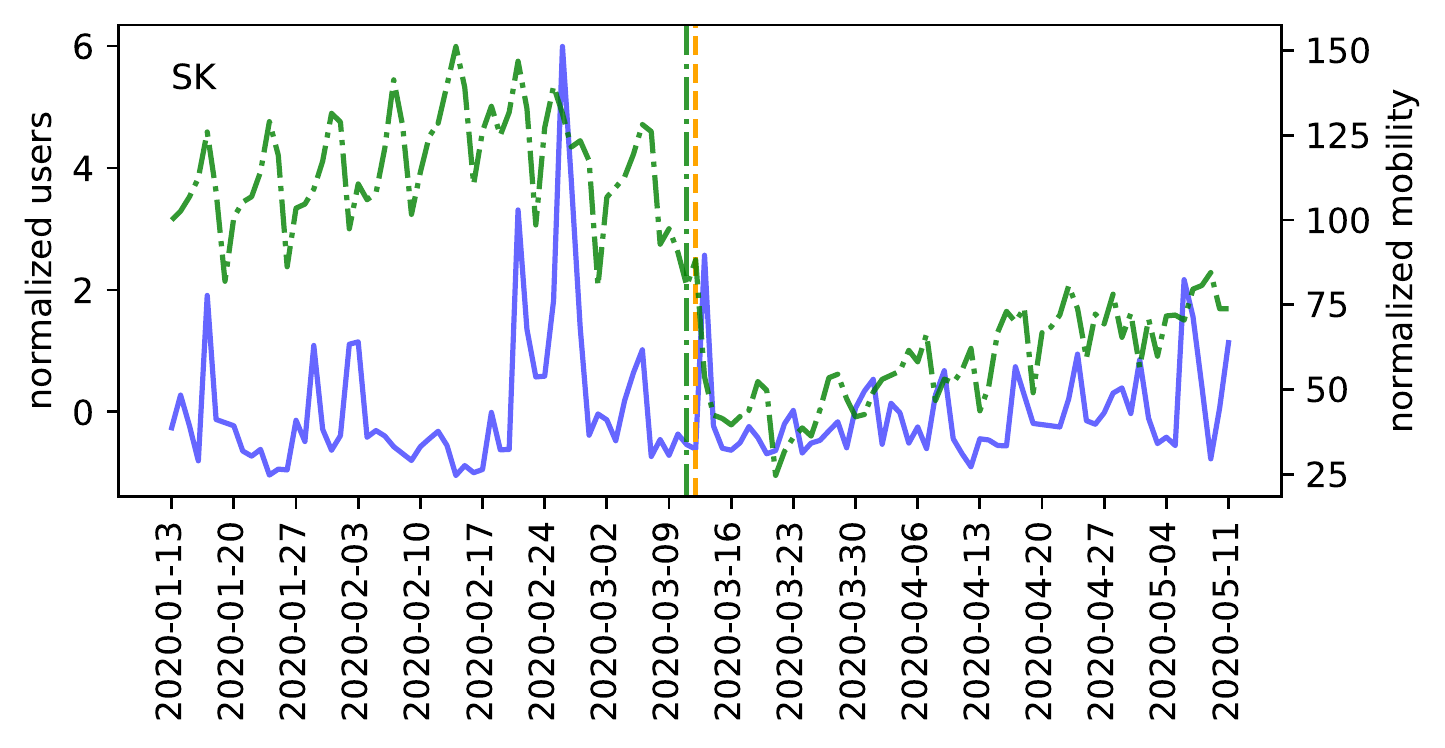}

\includegraphics[width=0.49\linewidth]{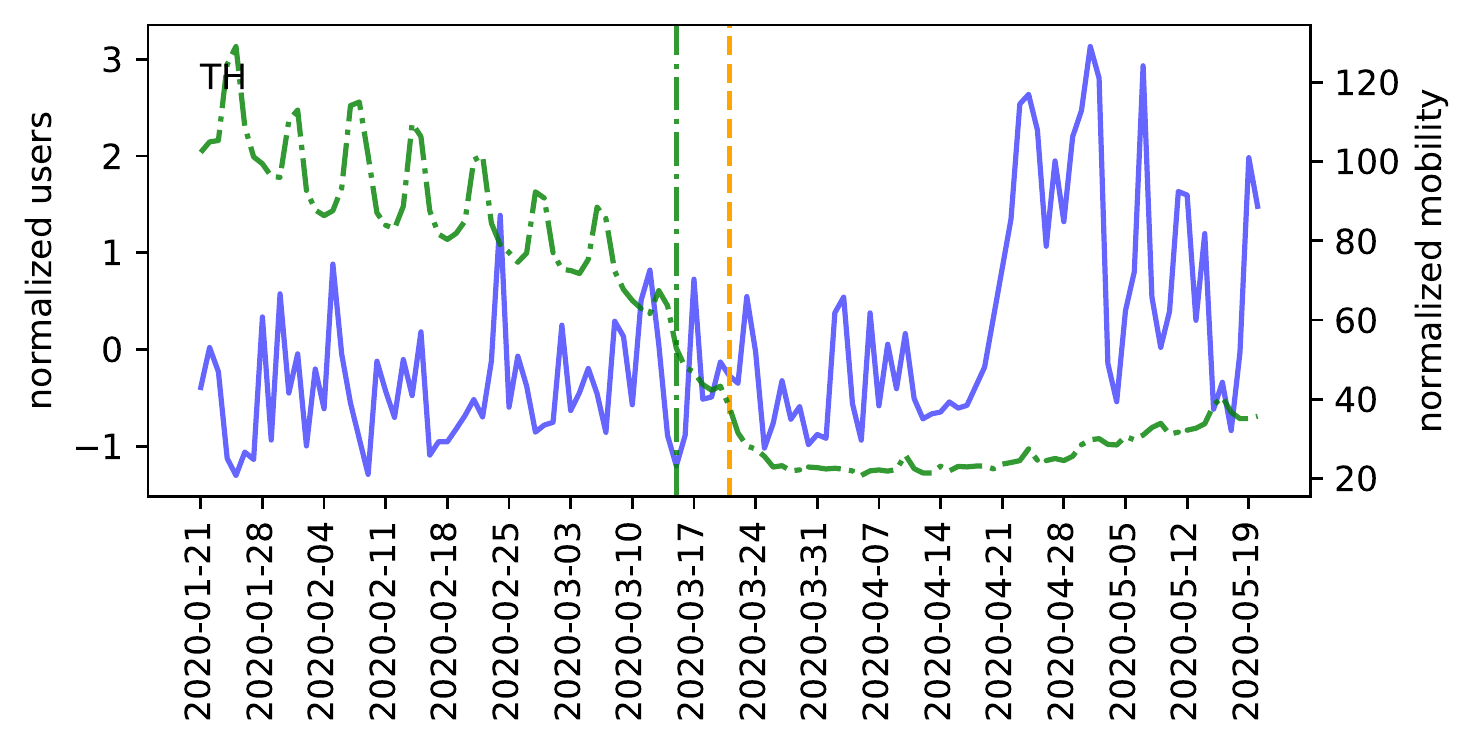}
    \vspace{0.1cm}
\includegraphics[width=0.49\linewidth]{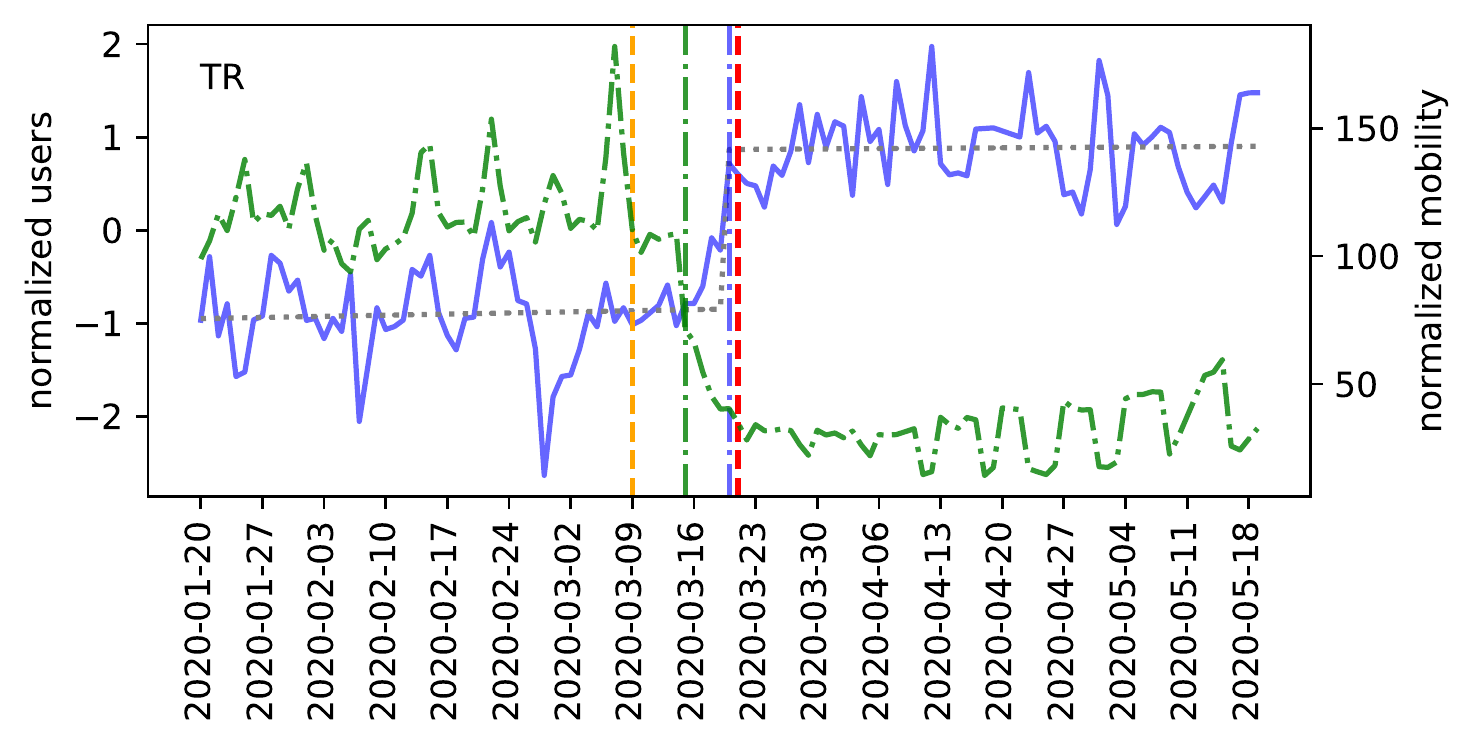} \\

\includegraphics[width=0.49\linewidth]{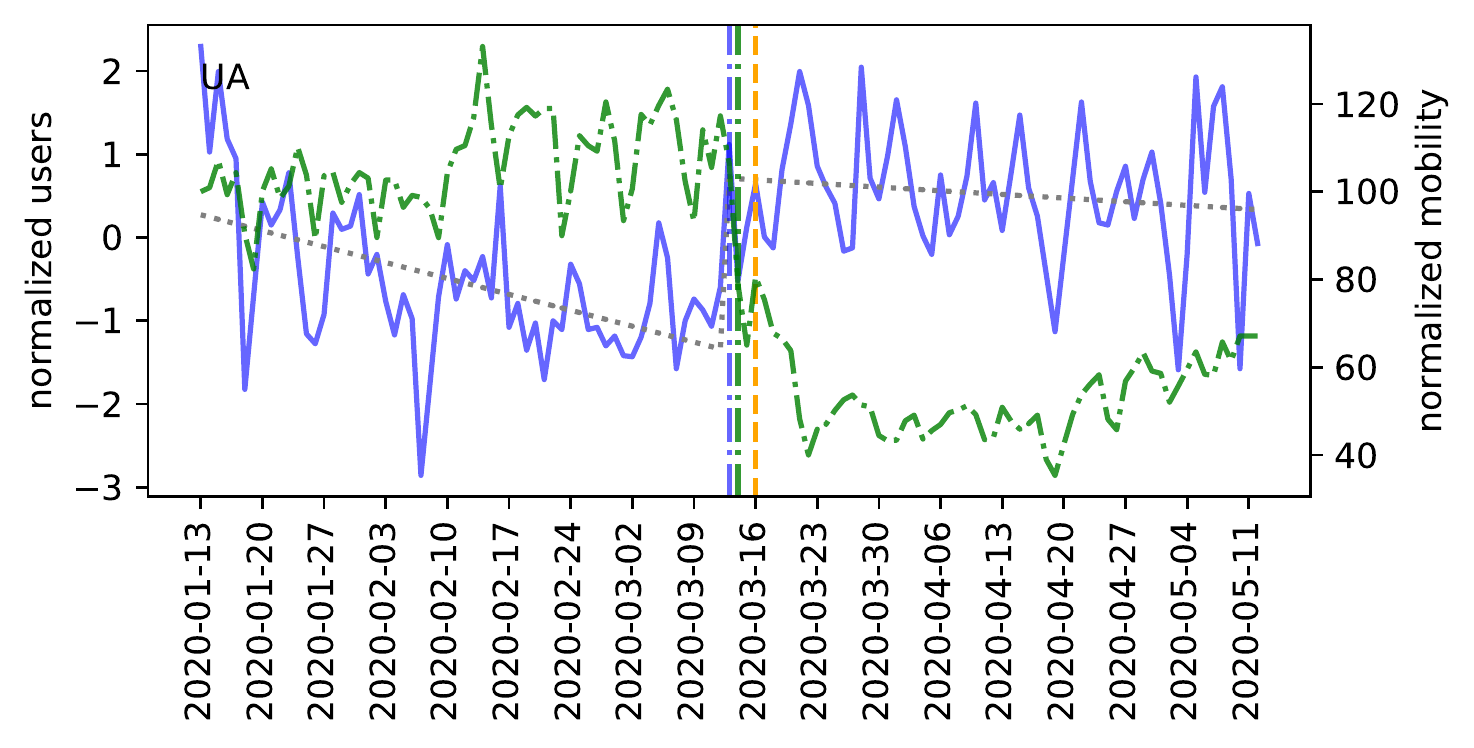}
    \vspace{0.1cm}
\includegraphics[width=0.49\linewidth]{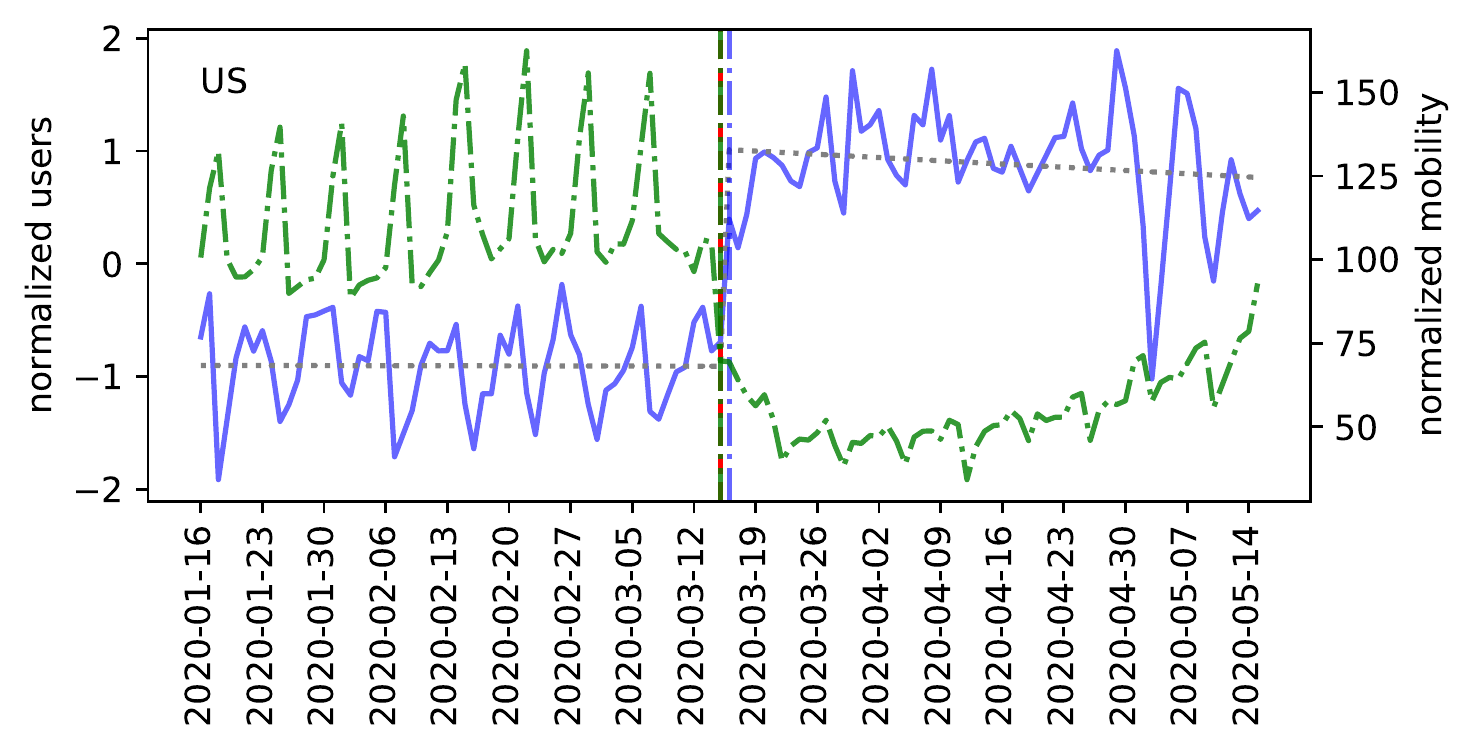}\\

\includegraphics[width=0.49\linewidth]{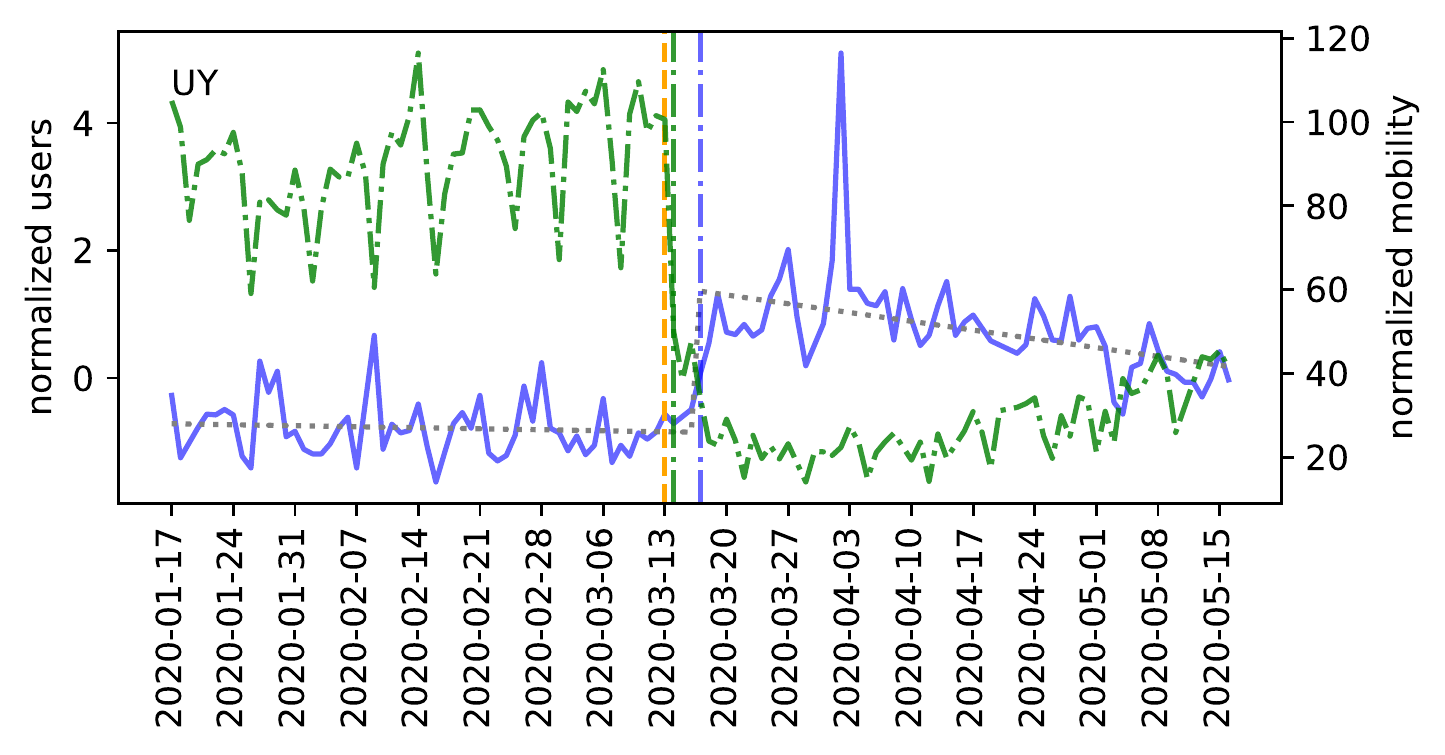}
    \vspace{0.1cm}
\includegraphics[width=0.49\linewidth]{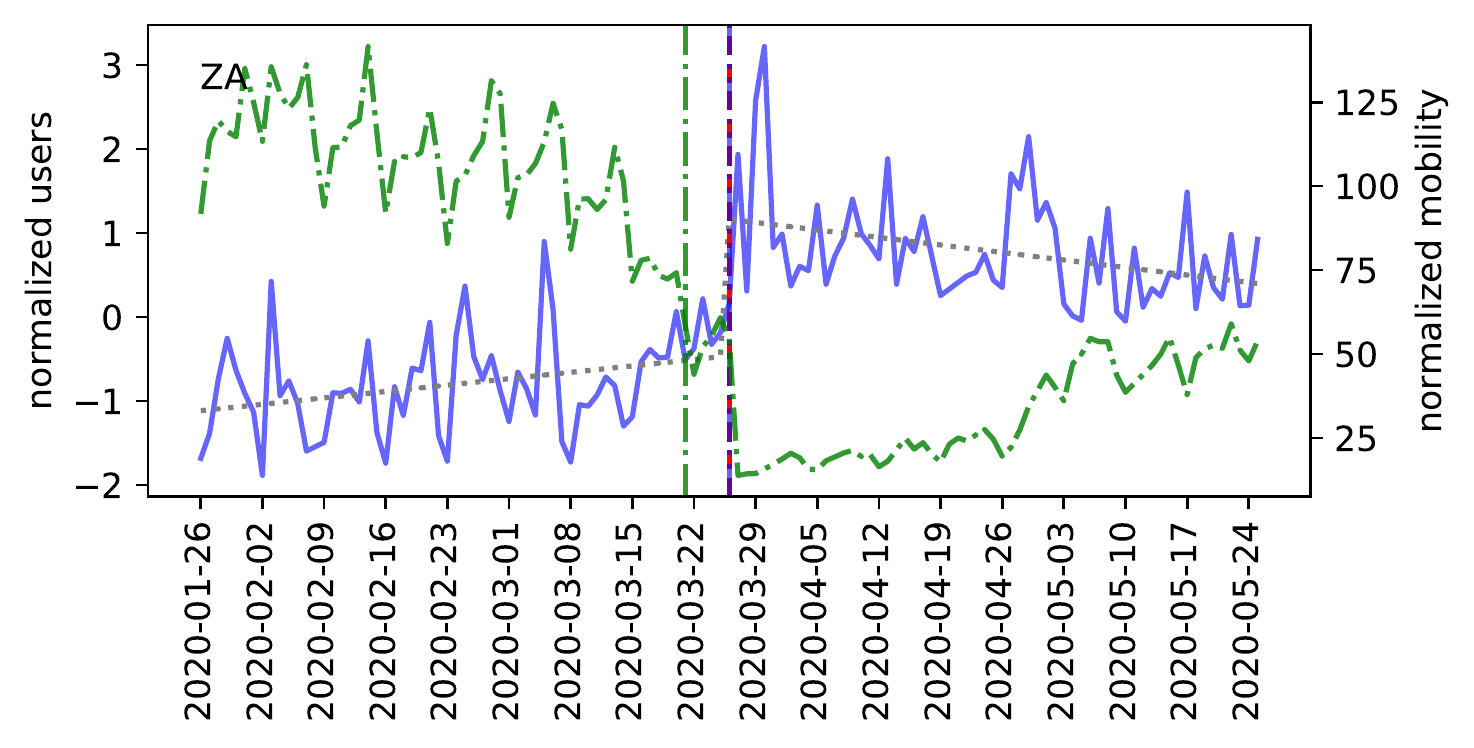}

\end{document}